\documentclass[a4paper,11pt]{article}
\pdfoutput=1 % if your are submitting a pdflatex (i.e. if you have
\usepackage{jcappub}

\usepackage{amsmath}
\usepackage{amssymb}
\usepackage{amsfonts}
\usepackage{epsfig}
\usepackage{epsfig,multicol,bbm}
\usepackage[printonlyused]{acronym}
\usepackage{subfig}
\usepackage{multirow}
\usepackage{bm}
\usepackage{float}
\usepackage{array}
\usepackage{graphicx}
\usepackage[T1]{fontenc}
\usepackage{cancel}
\usepackage{mathrsfs}
\usepackage{listings}
\usepackage{slashed}
\usepackage{url}
\usepackage[usenames,dvipsnames]{color}
\usepackage{hyperref}
\usepackage{hhline}
\usepackage{textalpha}
\usepackage{mathtools}
\usepackage{verbatim}
\newcommand{\be}{\begin{equation}}
\newcommand{\ee}{\end{equation}}

\newcommand{\App}[1]{App.~\ref{#1}}

\newcommand \trans{{\mathcal{T}}}

\newcommand{\vp}{\vec{p}}
\newcommand{\vq}{\vec{q}}

\newcommand{\vk}{\vec{k}}

\newcommand{\ttt}[1]{\texttt{#1}}

\newcommand{\Mpl}{M\unt{Pl}}

\newcommand{\gst}{g_{\star}}
\newcommand{\gsts}{g_{\star s}}

\newcommand{\mean}[1]{\left\langle #1\right\rangle}
\newcommand{\algn}[1]{\begin{aligned} #1\end{aligned}}
\newcommand{\dev}[2]{\frac{d #1}{d #2}}
\newcommand{\fracinv}[2]{\frac{#2}{#1}}

\newcommand{\unt}[1]{_{\mathrm{#1}}}
\newcommand{\upt}[1]{^{\mathrm{#1}}}

\newcommand{\obar}[1]{\mkern 1.5mu\overline{\mkern-1.5mu#1\mkern-1.5mu}\mkern 1.5mu}
\newcolumntype{?}{!{\vrule width 1pt}}
\usepackage{makecell}

\makeatletter
\lstnewenvironment{mylisting}
    { \lstset{ 
  backgroundcolor=\color{white},   % choose the background color; you must add \usepackage{color} or \usepackage{xcolor}; should come as last argument
  basicstyle=\footnotesize\ttfamily,        % the size of the fonts that are used for the code
  breakatwhitespace=false,         % sets if automatic breaks should only happen at whitespace
  breaklines=true,                 % sets automatic line breaking
  captionpos=b,                    % sets the caption-position to bottom 
  commentstyle=\color{BrickRed},    % comment style
  deletekeywords={},            % if you want to delete keywords from the given language
  extendedchars=true,              % lets you use non-ASCII characters; for 8-bits encodings only, does not work with UTF-8
  frame=true,	                   % adds a frame around the code
  keepspaces=true,                 % keeps spaces in text, useful for keeping indentation of code (possibly needs columns=flexible)
  keywordstyle=\color{Blue},       % keyword style
  language=python,             % the language of the code
  morekeywords={ exp, double },            % if you want to add more keywords to the set
  numbers=none,                    % where to put the line-numbers; possible values are (none, left, right)
  numbersep=0,                   % how far the line-numbers are from the code  
  numberstyle=\tiny\color{Magenta}, % the style that is used for the line-numbers
  rulecolor=\color{black},         % if not set, the frame-color may be changed on line-breaks within not-black text (e.g. comments (green here))
  showspaces=false,                % show spaces everywhere adding particular underscores; it overrides 'showstringspaces'
  showstringspaces=false,          % underline spaces within strings only
  showtabs=false,                  % show tabs within strings adding particular underscores
  stepnumber=2,                    % the step between two line-numbers. If it's 1, each line will be numbered
  stringstyle=\color{Violet},     % string literal style
  tabsize=2,	                   % sets default tabsize to 2 spaces
  title=\lstname                   % show the filename of files included with \lstinputlisting; also try caption instead of title
}   }
\makeatother
\title{\boldmath Lower Mass Bounds on FIMP \\ Dark Matter Produced via Freeze-In}
\author[a,b]{Francesco D'Eramo}
\author[a,b,c]{, Alessandro Lenoci}

\affiliation[a]{Dipartimento di Fisica e Astronomia, Universit\`a degli Studi di Padova, \\ Via Marzolo 8, 35131 Padova, Italy}
\affiliation[b]{Istituto Nazionale di Fisica Nucleare (INFN), Sezione di Padova, \\ Via Marzolo 8, 35131 Padova, Italy}
\affiliation[c]{DESY, Notkestrasse 85, D-22607 Hamburg, Germany}
\emailAdd{francesco.deramo@pd.infn.it}
\emailAdd{alessandro.lenoci@desy.de}
\preprint{DESY 20-219}
\abstract{Feebly Interacting Massive Particles (FIMPs) are dark matter candidates that never thermalize in the early universe and whose production takes place via decays and/or scatterings of thermal bath particles. If FIMPs interactions with the thermal bath are renormalizable, a scenario which is known as freeze-in, production is most efficient at temperatures around the mass of the bath particles and insensitive to unknown physics at high temperatures. Working in a model-independent fashion, we consider three different production mechanisms: two-body decays, three-body decays, and binary collisions. We compute the FIMP phase space distribution and matter power spectrum, and we investigate the suppression of cosmological structures at small scales. Our results are lower bounds on the FIMP mass. Finally, we study how to relax these constraints in scenarios where FIMPs provide a sub-dominant dark matter component.}

\begin{document} 

\maketitle

\flushbottom

%%%%%%%%%%%%%%%%%%%%%%%%%%%%%%%%%%%%%%%%%%%%%%%%%%%%%%%%%
%%%%%%%%%%%%%%%%%%%%%%%%% INTRO %%%%%%%%%%%%%%%%%%%%%%%%%%%%
%%%%%%%%%%%%%%%%%%%%%%%%%%%%%%%%%%%%%%%%%%%%%%%%%%%%%%%%%

\section{Introduction}
\label{sec:intro}

The microscopic nature of dark matter (DM) is still a mystery in fundamental physics~\cite{Jungman:1995df,Bertone:2004pz,Feng:2010gw}. Weakly Interacting Massive Particles (WIMPs) are theoretically motivated candidates with relic density depending on masses and couplings that we can measure in our laboratories or astrophysically~\cite{Lee:1977ua,Ellis:1983ew,Goldberg:1983nd,Scherrer:1985zt,Srednicki:1988ce,Gondolo:1990dk}. Famously, perturbative unitarity of the S-matrix puts an upper bound on the WIMP mass of approximately 100 TeV~\cite{Griest:1989wd}. If DM annihilates to visible final states via s-wave processes, it cannot be lighter than approximately 10 GeV~\cite{Leane:2018kjk} otherwise out-of-equilibrium annihilations would alter the Cosmic Microwave Background (CMB) anisotropy spectrum~\cite{Padmanabhan:2005es,Galli:2009zc,Slatyer:2009yq,Ade:2015xua}. If these dangerous processes are absent, such as for p-wave annihilations, DM lighter than approximately the MeV scale spoils the successful predictions of Big Bang Nucleosynthesis (BBN)~\cite{Sabti:2019mhn}. Although one can find some exceptions~\cite{Berlin:2017ftj}, it is fair to say that thermal relics lighter than the MeV scale are challenging to reconcile with observations. 

The lack of conclusive evidence for WIMPs, in spite of a vast and diverse experimental effort, motivates the exploration of alternative paradigms. Feebly Interacting Massive Particles (FIMPs), with couplings to the visible world way smaller than the case for WIMPs, are an appealing option. They never manage to reach thermal equilibrium through the cosmological history of our universe, but it is possible to have them around today with a cosmological abundance. Besides the obvious possibility of having them produced at very early times, such as from inflaton decays, processes among particles in the primordial thermal bath such as decays or binary collisions can produce FIMPs that free-stream subsequently. If FIMP interactions with the primordial bath are renormalizable then most DM particles are produced at low temperatures, typically around the mass of the heaviest particle participating in the production process~\cite{McDonald:2001vt,Kusenko:2006rh,Ibarra:2008kn,Hall:2009bx}, via a mechanism that has been dubbed \textit{freeze-in} in the literature~\cite{Hall:2009bx}. Using the usual jargon, we say that DM freeze-in is ``IR-dominated''. It is remarkable how freeze-in abundances depend only on quantities that we can measure today in our laboratories and/or astrophysically, and it is insensitive to unknown ``UV physics'' such as the reheating temperature. This scenario is realized in several motivated frameworks~\cite{Bernal:2017kxu}.

In this work, we set lower bounds on the mass of FIMPs. The constraints mentioned above do not apply. On one hand, FIMPs are so weakly-coupled that their out-of-equilibrium processes do not deposit any perceptible energy on the CMB. On the other hand, they are never in thermal equilibrium, and therefore their abundance at the time of BBN cannot affect the Hubble expansion rate. What sets then the lower bound on the FIMP mass? 

We observe dwarf galaxies with a size around the kpc, and the DM de Broglie wavelength cannot be larger than this value. This translates into $m_{\rm DM} \gtrsim 10^{-22} \, {\rm eV}$; when this bound is saturated and the wave nature of DM manifests itself on astrophysical scales we have fuzzy DM~\cite{Hu:2000ke,Hui:2016ltb}. The constraint is much stronger for fermionic DM candidates as a consequence of the Pauli exclusion principle which leads to $m_{\rm DM} \gtrsim {\rm keV}$, also known as the Tremaine-Gunn bound~\cite{Tremaine:1979we,Boyarsky:2008ju,DiPaolo:2017geq,Savchenko:2019qnn,Alvey:2020xsk}. These are bounds that hold independently on the production mechanism. 

Once we focus on freeze-in, there are additional complications if the DM is too light. Within such a framework, DM particles are produced via decays and collisions of thermal bath particles. As already explained above, freeze-in is ``IR-dominated'' and therefore production is most efficient when the primordial plasma has a temperature around the mass of the heaviest particle participating in the production process. This is also the typical energy of the process itself, and it sets the energy that FIMPs  inherit in the final state. The lighter the FIMP is, the larger its initial kinetic energy would be. After production, FIMPs just free-stream and if they begin their life with too much kinetic energy they erase cosmological structures on large scales. Thus light FIMPs wash out structures below some characteristic free-streaming scale $\lambda\unt{FS}$ whereas they behave as cold DM (CDM) on larger scales. If $\lambda\unt{FS}$ turns out to be larger than approximately $0.1 \, {\rm Mpc}$ then we are in conflict with observations. We provide in this paper a quantitative analysis for this qualitative statement. 

Potential hints from cosmological observations suggest that the low FIMP mass region could be of phenomenological relevance. The $\Lambda$CDM model provides a consistent picture at large length scales but there are still tensions between theory and observations at small sub-galactic scales~\cite{Weinberg:2013aya,Walker:2014isa}. Numerical simulations predict dwarf galaxies in excess with respect to the ones we observe, an issue known as the missing satellite  problem~\cite{Klypin:1999uc,Moore:1999nt}. They also predict a DM steep power-law density profile in the innermost regions of galaxies whereas observations provide an approximately constant density, a mystery dubbed as the core-cusp problem~\cite{2010AdAst2010E...5D}. Last, but not least, the too-big-to-fail problem: N-body simulations cannot reproduce the observed dynamics of massive Milky Way satellites~\cite{2011MNRAS.415L..40B,2012MNRAS.422.1203B}. The evidence is far from conclusive, and known physics such as baryonic feedback is likely to alleviate or even solve completely these problems. Nevertheless, it is worth keeping in mind that these small scale issues could be new hints from the dark sector. 

These $\Lambda$CDM shortcomings can be solved if we go beyond the CDM paradigm. Warm DM (WDM) of thermal origin and with a mass in the $m\unt{WDM} \sim 1-10 \, {\rm keV}$ range possesses a relatively large free-streaming scale suppressing structure formation. The consequent cut-off in the matter power spectrum alleviates these small-scale shortcomings. Motivated particles candidates for WDM in this mass range include sterile neutrinos~\cite{Konig:2016dzg,Schneider:2016uqi,Merle:2017jfn}, and the tensions are alleviated even in mixed (i.e., cold plus warm) frameworks~\cite{Schneider:2016ayw,Diamanti:2017xfo,Gariazzo:2017pzb}. However, the mass of thermal WDM is strongly constrained by the observation of the Lyman-$\alpha$ forest which gives a lower bound in the range $1-10$ keV, depending on the assumptions and datasets. This is exactly where the mass should lie to alleviate the excess of power on small scales giving rise to the $\Lambda$CDM model shortcomings. Nonetheless, these tight constraints apply only if the DM phase-space distribution is thermal and the tension with Lyman-$\alpha$ data might be alleviated within scenarios in which DM is produced non-thermally such as for freeze-in. 

\begin{figure}
\centering
\includegraphics[width=0.95\textwidth]{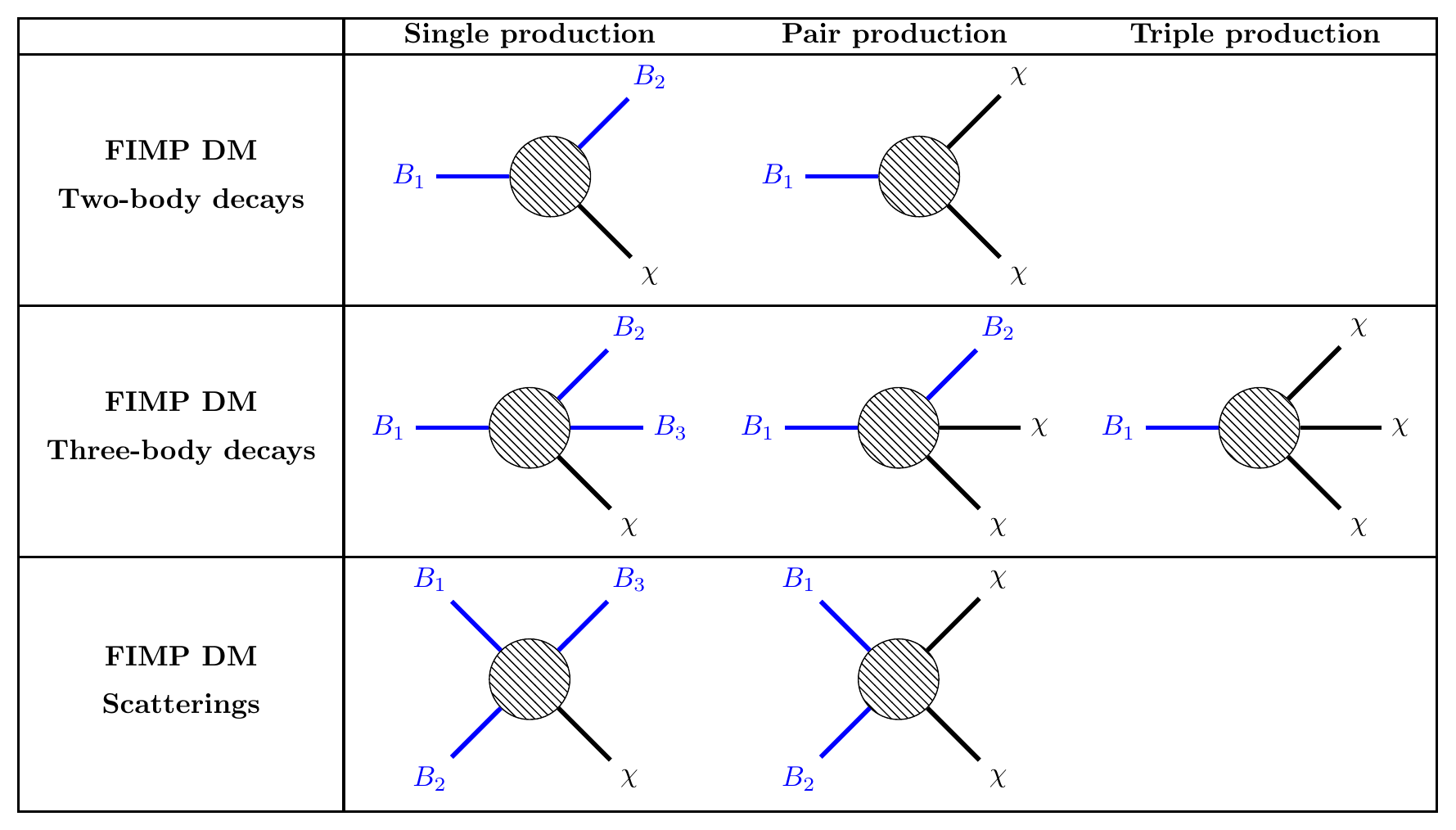}
\caption{Production of a FIMP $\chi$ via decays and scatterings of bath particles $B_i$.}
\label{fig:setup}
\end{figure}

We consider the freeze-in production mechanisms sketched in Fig.~\ref{fig:setup}: a FIMP particle $\chi$ is produced in the early universe via decays or binary collisions involving degrees of freedom $B_i$ belonging to the primordial thermal bath. We keep our analysis as general as possible and we do not commit to any specific identity of neither $B_i$ nor $\chi$. In particular, the bath particles $B_i$ can be either Standard Model (SM) fields or new degrees of freedom beyond the SM that are in thermal equilibrium in the early universe with the SM itself. For production via two-body decays, we do not need to specify the interactions to figure out the energy and momentum distribution of the final state FIMPs since two-body decays always give monochromatic final states. This is not the case for three-body decays and scatterings, and differential DM production rates depend on the microscopic theory under consideration. However, freeze-in works well in the IR and therefore the majority of FIMPs are produced in a quite narrow range of temperatures. This allows us to approximate the probability transition amplitudes, also known as the matrix elements, to a constant value and perform our model-independent analysis in full generality. The size of the matrix elements can be set to reproduce the observed abundance, or we can trade it with the fractional DM abundance if we are willing to consider the FIMP as a sub-dominant component. Related analysis within specific models can be found in Refs.~\cite{Shaposhnikov:2006xi,Biswas:2016iyh,Heeck:2017xbu,Bae:2017dpt,Boulebnane:2017fxw,Huo:2019bjf,Kamada:2019kpe,Baumholzer:2019twf,Dvorkin:2020xga,Hager:2020une}.

We review the Boltzmann equation formalism to derive the momentum distribution of FIMPs in Sec.~\ref{sec:PSFI}, and we apply it to the specific channels shown in Fig.~\ref{fig:setup} in Sec.~\ref{sec:PSD}. We illustrate how the warmness of DM particles constraints their mass in Sec.~\ref{sec:warmness}. Upper bounds on quantities such as the free-streaming length and the velocity dispersion imply lower bounds on the FIMP mass. We also improve this analysis by computing the linear matter power spectrum and comparing the suppression on small scales with the one in the WDM case. Finally, we derive bounds from the comparison between the predicted Milky Way satellites and the observed ones. We present our results in Sec.~\ref{sec:results}. We constrain the FIMP mass under the assumption that FIMPs reproduce the whole present relic density in Sec.~\ref{sec:FIMPDM}, and we relax this assumption in Sec.~\ref{sec:CDMWDM} considering mixed FIMP/CDM scenarios. We summarize our findings in Sec.~\ref{sec:conclusions}. Readers interested in technical details can refer to our appendices.

%%%%%%%%%%%%%%%%%%%%%%%%%%%%%%%%%%%%%%%%%%%%%%%%%%%%%%%%%
%%%%%%%%%%%%%%%%%%%%%%%% BE GENERAL %%%%%%%%%%%%%%%%%%%%%%%%%%
%%%%%%%%%%%%%%%%%%%%%%%%%%%%%%%%%%%%%%%%%%%%%%%%%%%%%%%%%

\section{Freeze-in in phase space}
\label{sec:PSFI}

We set up the formalism to study how the momentum distribution for the FIMP, always denoted with the symbol $\chi$, evolves through the history of the universe. We analyze freeze-in in a Friedmann-Robertson-Walker (FRW) expanding universe with its energy content dominated by a thermal bath of relativistic particles. Homogeneity and isotropy of the FRW metric ensure that the phase space distribution (PSD) $f_\chi$ can only depend on the cosmic time $t$ and the modulus of its physical momentum $p(t)$, which is also time dependent. In what follows, we omit the explicit time dependence and we just write $f_\chi(p)$. We summarize in App.~\ref{app:useful} our remaining conventions as well as our notation and results useful to our analysis. 

The Boltzmann equation describing the PSD time evolution takes the general form
\be
L[f_\chi(p)] = C[f_\chi(p)] \ .
\ee
On the left-hand side, the Liouville operator $L[f_\chi(p)]$ describes the PSD variation due to the space-time geometry whereas the collision operator $C[f_\chi(p)]$ on the right-hand side accounts for processes changing the net number of $\chi$ particles. We write down the explicit expression of the Liouville operator in a FRW background~\cite{Bernstein:1988bw,Kolb:1990vq,Dodelson:2003ft}, and the Boltzmann equation reads 
\be
g_\chi \frac{df_\chi(p)}{dt} = g_\chi \frac{C[f_\chi(p)]}{E} \ .
\label{eq:BEgeneral}
\ee
Here, $g_\chi$ is the number of internal degrees of freedom and the energy follows from the dispersion relation $E^2 = p^2 + m_\chi^2$. The time derivative acts on the explicit time dependence of $f_\chi$, which we do not write explicitly, as well as on the implicit time dependence through the physical momentum $p(t)$. In the most general case, this is an integro-differential equation and it is part of a Boltzmann system with one equation for each particle in the framework.

The most general process producing (and, for the opposite reaction, destroying) a net number $n$ of $\chi$ particles reads
\be
\underbrace{B_1(K_1)+\dots+B_\ell(K_\ell)}_{\ell} \longleftrightarrow \underbrace{B_{\ell+1}(K_{\ell+1})+\dots+B_{\ell+m}(K_{\ell+m})}_m+\underbrace{\chi(P)
+\dots+\chi(P_n)}_{n} \ .
\label{eq:genprocess}
\ee
A number of $\ell$ bath particles $B_i$ collides and produces a final state with $m$ bath particles in addition to $n$ FIMPs. Each particle has a four-momentum as indicated between parenthesis. Here and below, we denote four-momenta by an uppercase character, e.g. $K$, and the modulus of the associated spatial momentum by a lowercase character, e.g. $k$. 

\begin{figure}
\centering
\includegraphics[width=0.5\textwidth]{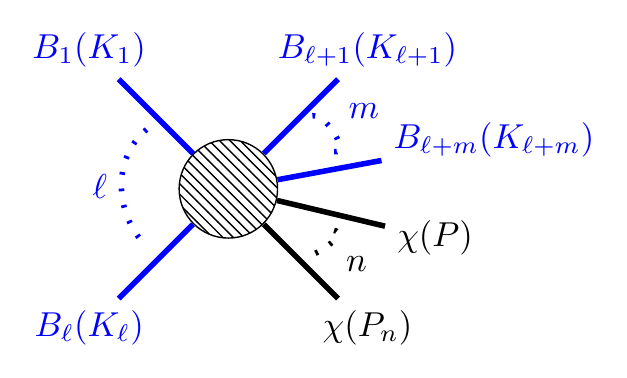}
\caption{Most general FIMP production process: $\ell$ bath particles collide and give rise to $m$ bath particles and $n$ FIMPs. Production via decays corresponds to the $\ell=1$ case.}
\label{fig:general_process}
\end{figure}

We isolate one FIMP and we write down the collision operator applied to its PSD. Without loss of generality, we choose it to be the first in Eq.~\eqref{eq:genprocess} with four-momentum $P$ as shown in Fig.~\ref{fig:general_process}. The collision operator, for which we provide a derivation in \App{app:Cderivation}, reads
\be
\algn{
C[f_\chi(p)] =  & \, n \, \times \, \frac{1}{2}   \int \prod_{i=1}^{\ell+m} d\mathcal{K}_i\prod_{i=2}^{n} d\Pi_i \; (2\pi)^4 \, \delta^{(4)}(P_\text{f} - P_\text{i}) \, \times \\ & \times 
\left[ \obar{|\mathcal{M}_{\rightarrow }|^2} \times \prod_{i=1}^{\ell} f_i(k_i) \times  (1\pm f_\chi(p)) \prod_{i=2}^{n}(1\pm f_\chi(p_i)) \prod_{i=\ell + 1}^{\ell+m}(1\pm f_i(k_i))  + \right. \\ &  \left. \qquad - \obar{|\mathcal{M}_{\leftarrow }|^2} \times f_\chi(p) \prod_{i=2}^{n}  f_\chi(p_i)  \prod_{i=\ell + 1}^{\ell+m} f_i(k_i) \times \prod_{i=1}^{\ell}(1\pm f_i(k_i)) \right] \ ,}
\label{eq:Cfgeneral}
\ee
with $P_\text{i} = \sum_{i=1}^\ell K_i$ and $P_\text{f} = \sum_{i=\ell+1}^{\ell+m} K_i + P + \sum_{i=2}^{n} P_i$ for the ease of notation. The overall factor of $n$ accounts for the net number of $\chi$'s produced by each process. The integration, with Lorentz invariant phase space measures given in Eqs.~\eqref{eq:LIPS1} and \eqref{eq:LIPS2}, is over the momenta of the other particles participating in the process, namely the $\ell+m$ bath degrees of freedom and the remaining $n-1$ DM particles. The $\ell+m+n-1$ integration momenta are not independent but they need to be consistent with energy and momentum conservation as ensured by the four-dimensional Dirac delta function. The two terms inside the square brackets quantify probabilities for the process and its inverse, respectively, and they are proportional to the squared matrix elements (averaged over both \textit{initial} and \textit{final} states and with appropriate symmetry factors $1/r!$ for $r$ identical particles in the initial or final states). If interactions preserve $CP$ (or time inversion $T$) the two squared matrix elements are equal $\obar{|\mathcal{M}_{\leftarrow}|^2} = \obar{|\mathcal{M}_{\rightarrow}|^2} = \obar{|\mathcal{M}|^2}$. Each probability is also proportional to the PSD's for initial state particles, and we account for Bose enhancement ($+$ sign) or Pauli blocking ($-$ sign) in the final state.

Within the freeze-in paradigm, we consider only the production process in Eq. (\ref{eq:genprocess}) and safely neglect quantum degeneracy effects for final state particles. The resulting expression for the collision operator simplifies significantly
\be
\mathcal{C}(T,p)  \equiv \frac{n}{2}\int \prod_{i=1}^{\ell+m} d\mathcal{K}_i \, \prod_{i=2}^{n} d\Pi_i \, (2\pi)^4  \, \delta^{(4)}(P_\text{f} - P_\text{i}) \,  \obar{|\mathcal{M}|^2} \prod_{i=1}^{\ell} f_i(k_i)  \ . 
\label{eq:collterm}
\ee
Notice how the PSD $f_\chi$ of the particle $\chi$ under investigation does not appear on the right-hand side. In other words, once we focus on freeze-in, the collision operator is not actually an operator but rather a function that we dub \textit{collision term} and we denote with the symbol $\mathcal{C}$. Such a function depends only on the cosmic time (through the bath particles PSD's $f_i(k_i)$), or equivalently on the bath temperature $T$, and the momentum of the DM particle $p$.

We introduce a dimensionless ``time variable'', $x \equiv M / T$, and it is convenient to set $M$ to the mass of the heaviest particle involved in the process since most FIMPs are produced at that temperature. We write the Boltzmann equation by using this new evolution variable. We trade the time derivative with temperature derivative, see Eq.~\eqref{eq:t_to_T}, and we use Eq.~\eqref{eq:T_to_x} to switch from $T$ to $x$. The resulting Boltzmann equation reads
\be
g_\chi \dev{f_\chi(p)}{\log x} = \frac{1}{H(x)} \bigg(1-\frac{1}{3}\dev{\log\gsts}{\log x}\bigg) g_\chi \frac{\mathcal{C}(x,p)}{E} \ .
\label{eq:BE_FI_x} 
\ee
This is the master equation for our analysis. 

The Hubble parameter, for a radiation dominated universe, as a function of $x$ reads
\be
H(x) = \frac{\pi \gst^{1/2}(T)}{3 \sqrt{10}} \frac{M^2}{\Mpl} \, x^{-2} \ ,
\label{eq:Hubblevsx}
\ee
with $\Mpl$ the reduced Planck mass and $\gst(T)$ the effective number of relativistic degrees of freedom. When we integrate Eq.~\eqref{eq:BE_FI_x} over $x$, the support of the integral is concentrated around $x \simeq 1$. For $x \ll 1$, namely $T$ much higher than $M$, freeze-in is not efficient because the universe is not old enough to give an appreciable amount of FIMPs. For temperatures much lower than $M$, $x \gg 1$, the right-hand side of the Boltzmann equation is exponentially suppressed because at least one particle participating in FIMP production is too heavy to be around. In order to find the functional form for the PSD $f_\chi(p)$ at late times, $x_{\rm fin} \gg 1$, we integrate the Boltzmann equation from $x_{\rm in} \rightarrow 0$ to $x_{\rm fin}$. 

The procedure described above is rather inconvenient. While it is certainly true that freeze-in is inactive at temperatures below $M$, i.e. $C(x,p) \simeq 0$ for $x \gtrsim 1$, FIMP physical momenta keep changing with time due to the Hubble expansion. The evolution of the DM momentum is straightforward: it red-shifts with the scale factor $a$ as $p \propto a^{-1}$. In other words, the PSD maintains its shape but the scale of momenta changes because of the cosmological red-shift. We isolate this effect by introducing the dimensionless comoving momentum
\be
q \equiv \frac{p}{M} \frac{a(T)}{a(M)} \ ,
\label{eq:q}
\ee
with $a(T)$ the value of the scale factor when the bath temperature was $T$. This momentum variable is not altered in the absence of number changing processes because $p \, a = {\rm const}$ for free streaming particles. The explicit solution at late times reads
\be
g_\chi{f_\chi}(q)=\int_0^\infty d\log x\,\frac{1}{H(x)}\bigg(1-\frac{1}{3}\dev{\log\gsts}{\log x}\bigg) g_\chi\frac{\mathcal{C}(x,q)}{E} \ .
\label{eq:PSD}
\ee

Even though FIMP particles produced via freeze-in are never in thermal equilibrium, it is convenient to introduce a DM ``temperature'' as follows
\be
T_\chi \equiv \frac{p}{q} = \bigg(\frac{\gsts(T)}{\gsts(M)}\bigg)^{1/3} \, T \ .
\label{eq:Tchi}
\ee
The ratio of the entropic degrees of freedom arises after we impose entropy conservation. This parametrization is commonly adopted in the literature to characterize the PSD of non-thermally produced DM candidates.

%%%%%%%%%%%%%%%%%%%%%%%%%%%%%%%%%%%%%%%%%%%%%%%%%%%%%%%%%
%%%%%%%%%%%%%%%%%%%%%%% BE FIMPS %%%%%%%%%%%%%%%%%%%%%%%%%%%%
%%%%%%%%%%%%%%%%%%%%%%%%%%%%%%%%%%%%%%%%%%%%%%%%%%%%%%%%%

\section{Phase space distributions for FIMPs}
\label{sec:PSD}

For the different topologies illustrated in Fig. \ref{fig:setup}, we can compute the associated PSD from Eq.~\eqref{eq:PSD}. The relevant collision term is different for each case, and here we report results valid when bath particles involved in the production are characterized by the Maxwell-Boltzmann (MB) classical statistics. We show later on in this section how quantum effects, with general expressions given in App.~\ref{app:Ctopologies}, give negligible corrections. Approximating both fermions and bosons with a MB distribution offers the possibility to work in a model-independent framework where we do not need to specify the thermal bath particles statistics. 

\begin{description}

\item[Two-body decays.] The general collision term for two-body decays is given in Eq.~\eqref{eq:Cdecayseq}. We trade the squared matrix element with the decay width of the bath particle $B_1$
\be
\Gamma_1 = \frac{g_{\mathcal{Q}_2} g_\chi}{16\pi m_1} \obar{|\mathcal{M}_{2}|^2}  \, y_{\mathcal{Q}_2} \ ,
\ee
with $g_{\mathcal{Q}_2}$ equal to $g_2$ (single production) or $g_\chi$ (double production), and we define
\be
y_{\mathcal{Q}_2} \equiv \sqrt{\lambda \left(1, \frac{M_2}{m_1}, \frac{m_\chi}{m_1} \right)} \sim\mathcal{O}(1) \ ,
\ee
a known order one factor depending on the mass spectrum with the function $\lambda(x,y,z)$ defined in Eq. \eqref{eq:lambda}, and $M_2 = (m_2, m_\chi)$ (single, double). The collision term reads 
\be
g_\chi  \frac{\mathcal{C}_2(T,p)}{E} = {n} \frac{g_1 \Gamma_1 m_1}{y_{\mathcal{Q}_2}} \frac{T \, e^{-E/T}}{Ep}\bigg\{e^{-\mathcal{E}_2^-/T}-e^{-\mathcal{E}^+_2/T}\bigg\} \ ,
\label{eq:CdecayseqMB}
\ee 
with the functions $\mathcal{E}_2^{\pm}$ given by Eq.~\eqref{eq:E2+-}.

\item[Three-body decays.] The three-body decay collision term is given in Eq.~\eqref{eq:C3decayseq}. We work in the approximation in which the squared matrix elements are constant; while this is exact for two-body decays, in general it is not the case for three-body decays and scatterings. However, freeze-in is ``IR-dominated'' and the squared matrix elements relevant for production are dominated at temperatures around the mass scale $m_1$ of the decaying particle. We present an explicit example for scattering in the next paragraph and we show how the squared matrix element can be taken to be constant. Thus we trade again the squared matrix element with the decay width of $B_1$, and we have
\be
\Gamma_1 = \frac{g_{\mathcal{Q}_2} g_{\mathcal{Q}_3} g_\chi}{256\pi^3} m_1 \obar{|\mathcal{M}_{3}|^2} \, y_{\mathcal{Q}_2 \mathcal{Q}_3 \chi} \ ,
\ee
where
\be
y_{\mathcal{Q}_2 \mathcal{Q}_3\chi} \equiv\int_{(r_\chi + r_{\mathcal{Q}_3})^2}^{(1-r_{\mathcal{Q}_2})^2} \frac{d\xi}{\xi}\sqrt{\lambda(\xi,1,r_{\mathcal{Q}_2})\lambda(\xi,r_{\mathcal{Q}_3},r_\chi)} \sim \mathcal{O}(1) 
\ee
is an order one dimensionless factor depending on the mass spectrum of the theory. Similarly to what we have done before, we define $(g_{\mathcal{Q}_2}, M_2) = (g_2, m_2)$ for single and double production and $(g_{\mathcal{Q}_2}, M_2) = (g_\chi, m_\chi)$ for triple production. Likewise, we define $(g_{\mathcal{Q}_3}, M_3) = (g_3, m_3)$ for single production and $(g_{\mathcal{Q}_3}, M_3) = (g_\chi, m_\chi)$ for double and triple production. The mass ratios follow consequently: $r_{\mathcal{Q}_2} = M_2/m_1$, $r_{\mathcal{Q}_3} = M_3/m_1$ and $r_\chi = m_\chi / m_1$. Within our approximations, the collision term results in  
\be\label{eq:C3decayseqMB}
g_\chi \frac{\mathcal{C}_3(T,p)}{E} =n   \frac{g_1\Gamma_1}{ y_{\mathcal{Q}_2 \mathcal{Q}_3\chi}m_1 } \frac{T}{pE} \int_{(M_2+M_3)^2}^{(m_1-m_\chi)^2} \frac{ds}{{ s}}\sqrt{\lambda (s,M_2,M_3)} \bigg\{e^{-E_1^-/T}-e^{-E^+_1/T}\bigg\} \ ,
\ee
where the $E_1^\pm$ are functions given by Eqs. (\ref{eq:E1dec3}).

\begin{figure}
\centering
\includegraphics[width=0.65\textwidth]{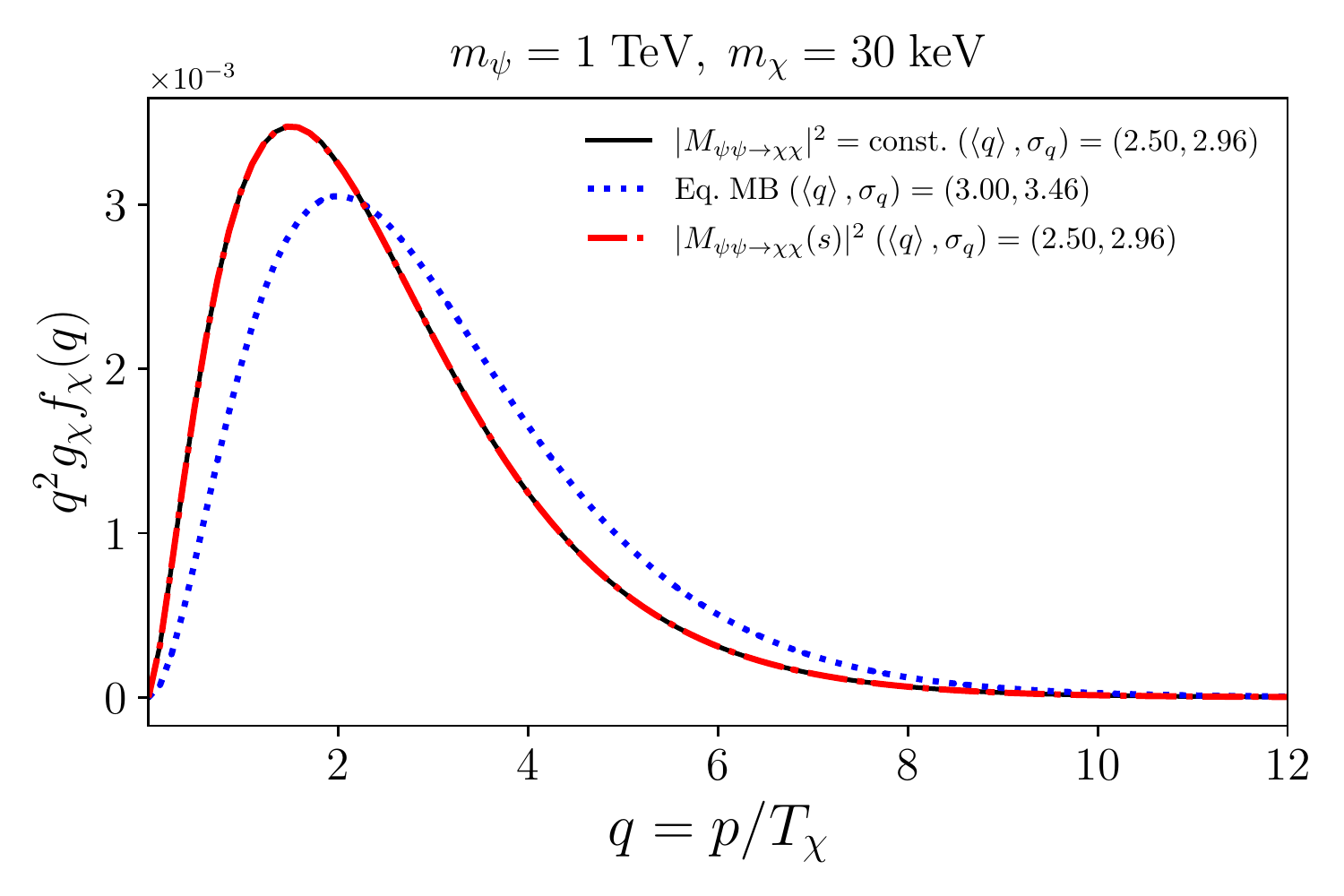}
\caption{FIMPs pair-production: comparison between the complete theory and a constant squared matrix element. We consider fermion FIMPs $\chi$'s produced via a binary collisions of fermions $\psi$'s belonging to the thermal bath, and the process is mediated by a scalar particle.}
\label{fig:PSDcomparison}
\end{figure} 

\item[Scatterings.] The general collision term for binary collision is given in Eq.~\eqref{eq:Csca_gen}. We trade the dependence on the squared amplitude with the relevant observable for this process: the Lorentz-invariant scattering cross-section. This is possible if the squared matrix element is (or is approximated as a) constant. We work within this approximation in which the Lorentz-invariant cross-section reads
\be
\sigma_{\mathcal{Q}_3\chi}(s) = \frac{g_{\mathcal{Q}_3}g_\chi}{16\pi s} \obar{|\mathcal{M}_{s}|^2}\sqrt{\frac{\lambda (s,M_3,m_\chi)}{\lambda (s,m_1,m_2)}} \ .
\ee
Here $g_{\mathcal{Q}_3}=g_3$ (single production) or $g_\chi$ (double), and $M_3=m_3$ (single) or $m_\chi$ (double). Unlike for decays, the relevant observable has a dependence on the initial state particle energy, i.e. on $s$. We evaluate the cross sections for $s=M^2 $ where $M$ is the relevant energy scale for the FIMP production, $M=\max\{m_1, m_2, M_3, m_\chi\}$. The choice is reasonable since FIMP production is IR-dominated. Fig.~\ref{fig:PSDcomparison} illustrated with one explicit example how good this approximation is.\footnote{There can be exceptions in peculiar parameter space region such as around the mediator resonance similarly to the well known case for WIMPs~\cite{Griest:1990kh}.} We consider a fermion FIMP candidate $\chi$ which is pair-produced via scatterings of thermal bath fermions $\psi$'s. The process is mediated by the virtual exchange of a scalar particle $\phi$ interacting with the FIMP via a Yukawa interaction $\phi \bar{\chi} \chi$ as well as with the thermal bath fermions $\phi \bar{\psi} \psi$. We present the resulting PSD obtained by the numerical integration of the Boltzmann equation within the full theory and for a constant squared matrix element. The black and red lines are on top of each other, and the moments of the two distributions are identical. Thus we parameterize the strength of the interaction with the cross-section computed at the freeze-in energy scale
\be
\sigma\upt{FI}_{\mathcal{Q}_3\chi}\equiv\sigma_{\mathcal{Q}_3\chi}(M^2)=\frac{g_{\mathcal{Q}_3}g_\chi}{16\pi M^2}  \obar{|\mathcal{M}_{s}|^2}y_{\mathcal{Q}_3\chi}\ ,
\ee
where
\be
y_{\mathcal{Q}_3\chi}^s=\sqrt{\frac{\lambda(M,M_3,m_\chi)}{\lambda(M,m_1,m_2)}} 	\sim \mathcal{O}(1)
\ee
is a dimensionless order one factor depending on the mass spectrum. Therefore the collision term for binary collision reads
\be\label{eq:CscaeqMB}
g_\chi  \frac{\mathcal{C}_s(T,p)}{E} =\frac{n}{16 \pi^2 }  \frac{g_1 g_2 M^2\sigma\upt{FI}_{\mathcal{Q}_3\chi}}{ y_{\mathcal{Q}_3\chi}^s}  \frac{Te^{-E/T}}{pE}\int_{ s\unt{min}}^{{\infty}}  \frac{ds}{ s} \sqrt{\lambda(s,m_1,m_2) }\bigg\{e^{-\mathcal{E}_3^-/T}-e^{-\mathcal{E}^+_3/T}\bigg\}
\ee
where $\mathcal{E}_3^\pm$ are functions given by Eqs. (\ref{eq:E3sca}) and $s\unt{min}=\max\left\{(m_{1}+M_{2})^2,(M_{3}+m_\chi)^2\right\}$.  
\end{description}

\begin{figure}
\centering
\includegraphics[width=0.99\textwidth]{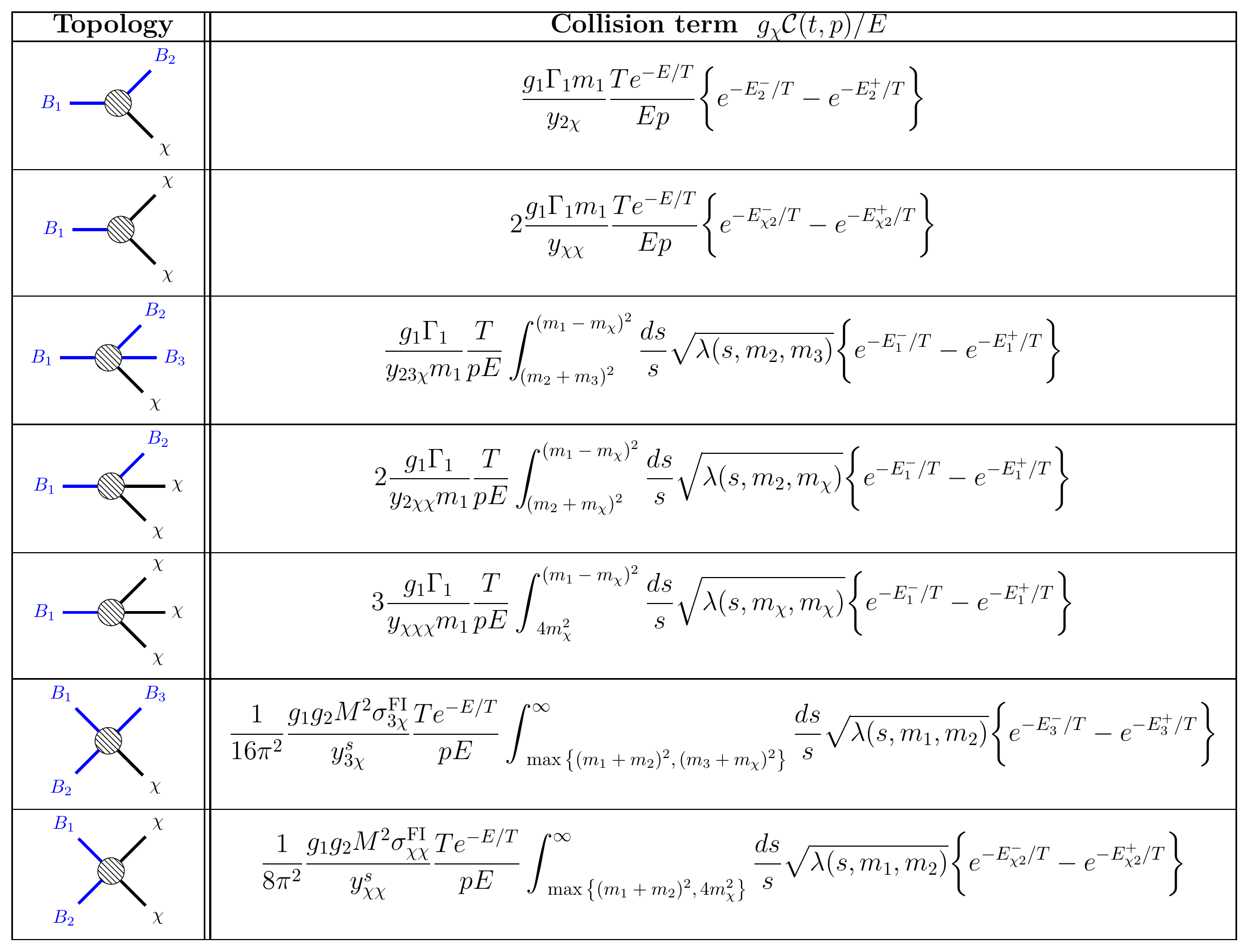}
\caption{Collision terms for each topology and for MB statistics (see text for definitions of $E_i^\pm$ and $y_i$). Results are exact for two-body decays, for three-body decays and scatterings they are valid if the matrix element is approximately constant as it is the case for freeze-in.}
\label{fig:gCE}
\end{figure}

\begin{figure}
\centering
\includegraphics[width=0.75\textwidth]{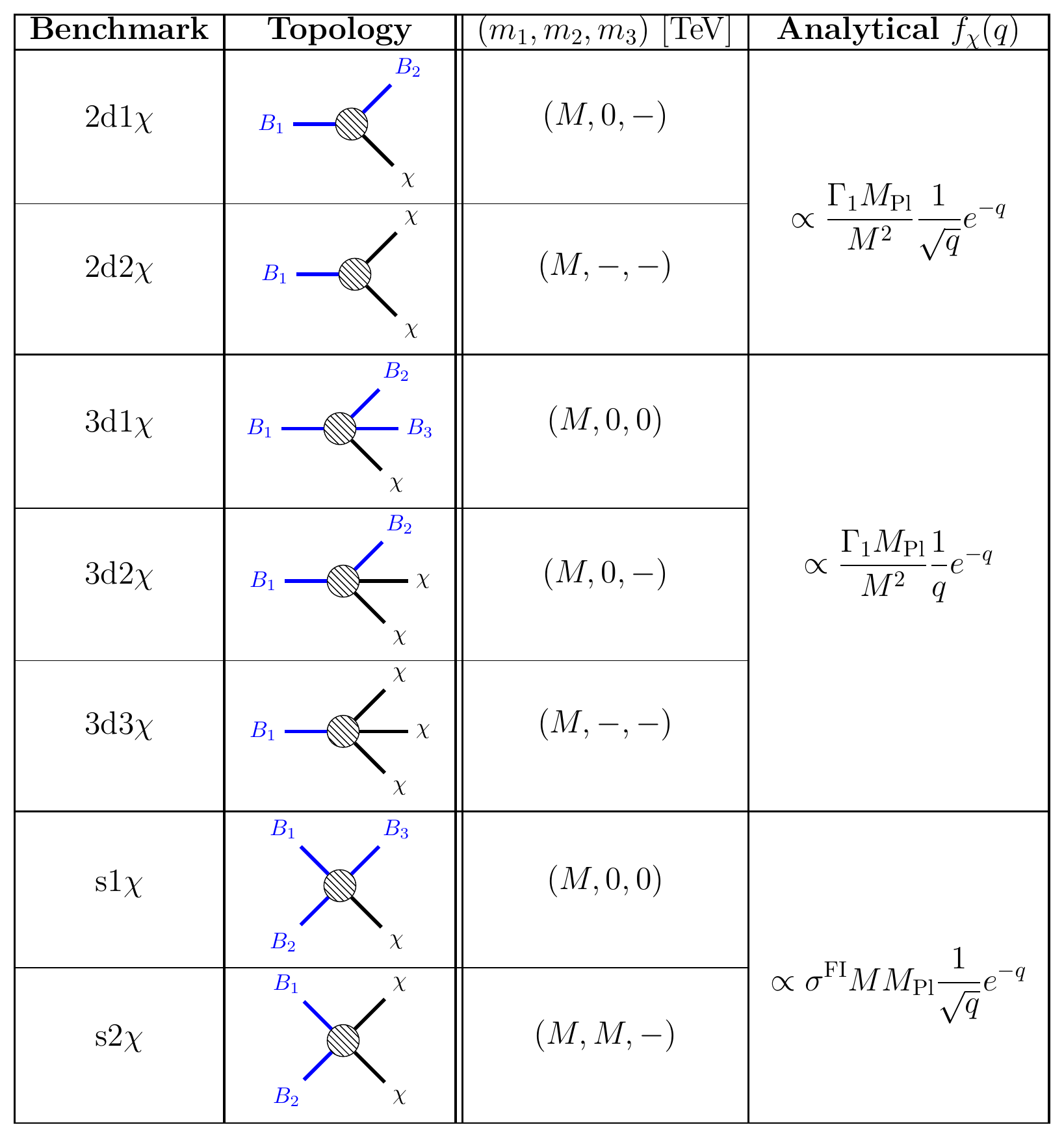}
\caption{ Illustration of the chosen benchmarks, one for each topology. We show the associated topology, the mass spectrum together with the analytical approximaton for the PSD $f_\chi(q)$ obtained in App. \ref{app:PSD_an} to highlight the dimensional dependences and the functional shape.}
\label{fig:benchmarks}
\end{figure} 

We present a compact summary in Fig.~\ref{fig:gCE} where we provide the collision term for each topology. In order to show numerical results, we choose a benchmark spectrum for each topology and we show them in Fig.~\ref{fig:benchmarks}. We label our benchmarks with names denoting the process producing DM (``2d'' for two-body decay, ``3d'' for three-body decays and ``s'' for scatterings) and the number of $\chi$ particles produced (``$1\chi$'', ``$2\chi$'' or ``$3\chi$''). We are always interested in FIMPs behaving similarly to WDM, and therefore with a mass $m_\chi$ much lighter than the one of the heaviest bath particle participating in the process, which we call $M$. In order to compute the resulting PSD, we can take the limit $m_\chi \ll M$ and neglect the DM mass. Remarkably, for these benchmarks we can obtain an approximated analytical solution for the resulting PSD computed after solving the Boltzmann equation. We refer to App. \ref{app:PSD_an} for the derivations of these analytical expressions for the PSD with the appropriate numerical factors, and we only show the 
dimensional and functional dependences in Fig.~\ref{fig:benchmarks}. As we can see from the explicit expressions, multiple DM production does not change the functional form of the PSD but only its normalization. Thus each benchmark is defined by the overall mass scale $M$, and we set the masses of the other particles to be much smaller than $M$ such that they are irrelevant to the calculations, as it is often the case in concrete microscopic models. The remaining physical quantities such as the DM mass $m_\chi$ and the interaction strength, quantified by the decay width or the scattering cross section, are fixed to reproduce the DM relic density (or a fraction $F < 1$ if we consider mixed FIMP/CDM frameworks).

\begin{figure}
\centering
\includegraphics[width=0.95\textwidth]{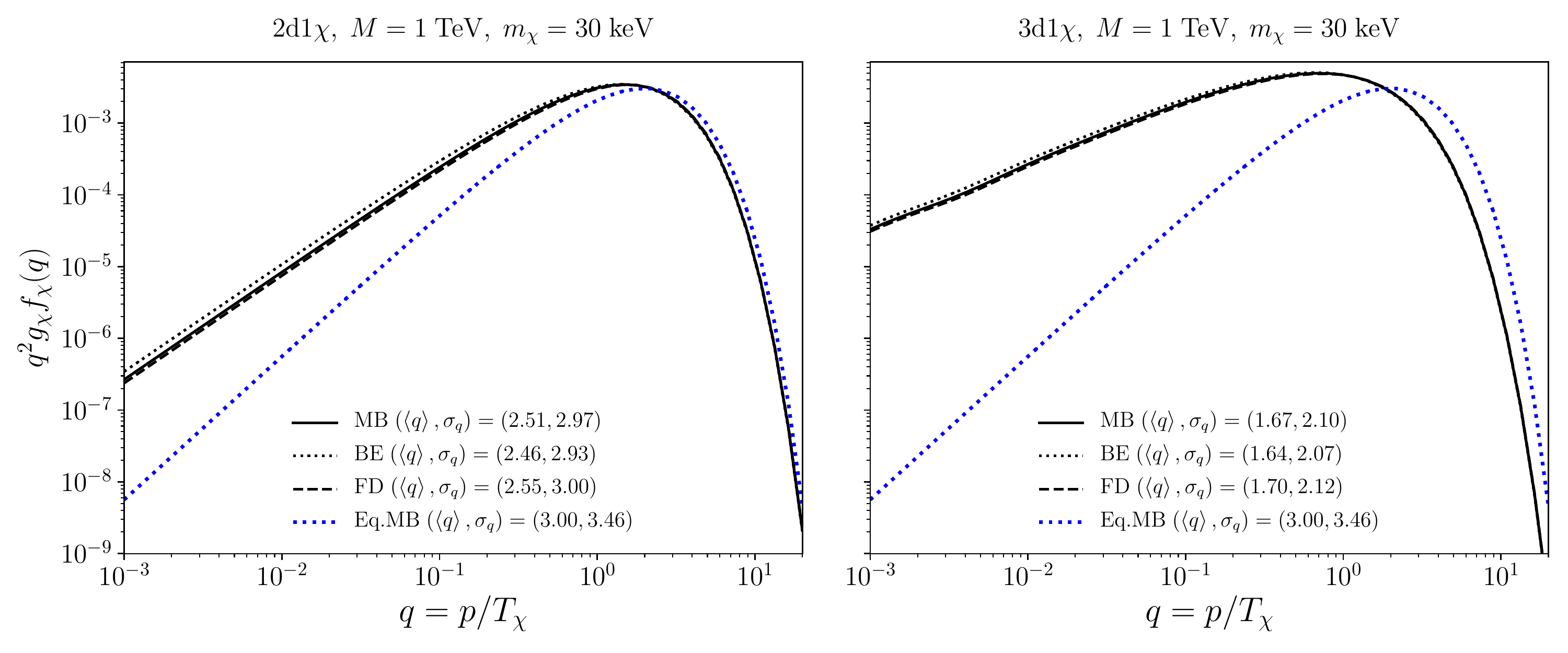}
\caption{Numerical PSDs for single DM production benchmarks via two-body decays (left) and three-body decays (right). Black lines correspond to different statistics for the decaying  particle: Maxwell-Boltzmann (MB, solid), Bose-Einstein (BE, dotted), and Fermi-Dirac (FD, dashed); the differences are tiny. We report for comparison also the equilibrium MB expression for the PSD (dotted blue). All curves are normalized in order to reproduce the DM relic density for that given value of $m_\chi$. We also provide the values of $\mean{q}$ and $\sigma_q$.}
\label{fig:PSD_bench1}
\end{figure}

We show in Fig.~\ref{fig:PSD_bench1} numerical results for the PSD obtained after integrating the Boltzmann equation. Focusing on single DM production, we consider both two-body (left panel) and three-body (right panel) decays. The PSDs for the other benchmarks (double production and triple production) are identical in shape to the ones shown and only differ in the normalization constant. We consider different statistics fot the decaying particle: boson (BE), fermion (FD) and also a MB; the difference is almost imperceptible. Furthermore, we compare our results with a thermal MB equilibrium distribution. For each case, we characterize the PSD by its first two moments: the average comoving momentum and the comoving momentum dispersion
\begin{align}
\mean{q} = & \, \frac{\int d q\, q^3 f_\chi(q)}{\int d q\, q^2 f_\chi(q)} \ , \\
\label{eq:sigmaq} \sigma_q^2 = & \, \frac{\int d q\, q^4 f_\chi(q)}{\int d q \,q^2f_\chi(q)} \ .
\end{align}
The latter quantity is a measure of the DM warmness as we discuss in Sec. \ref{sec:dispersion}.

\begin{figure}
\centering
\includegraphics[width=0.95\textwidth]{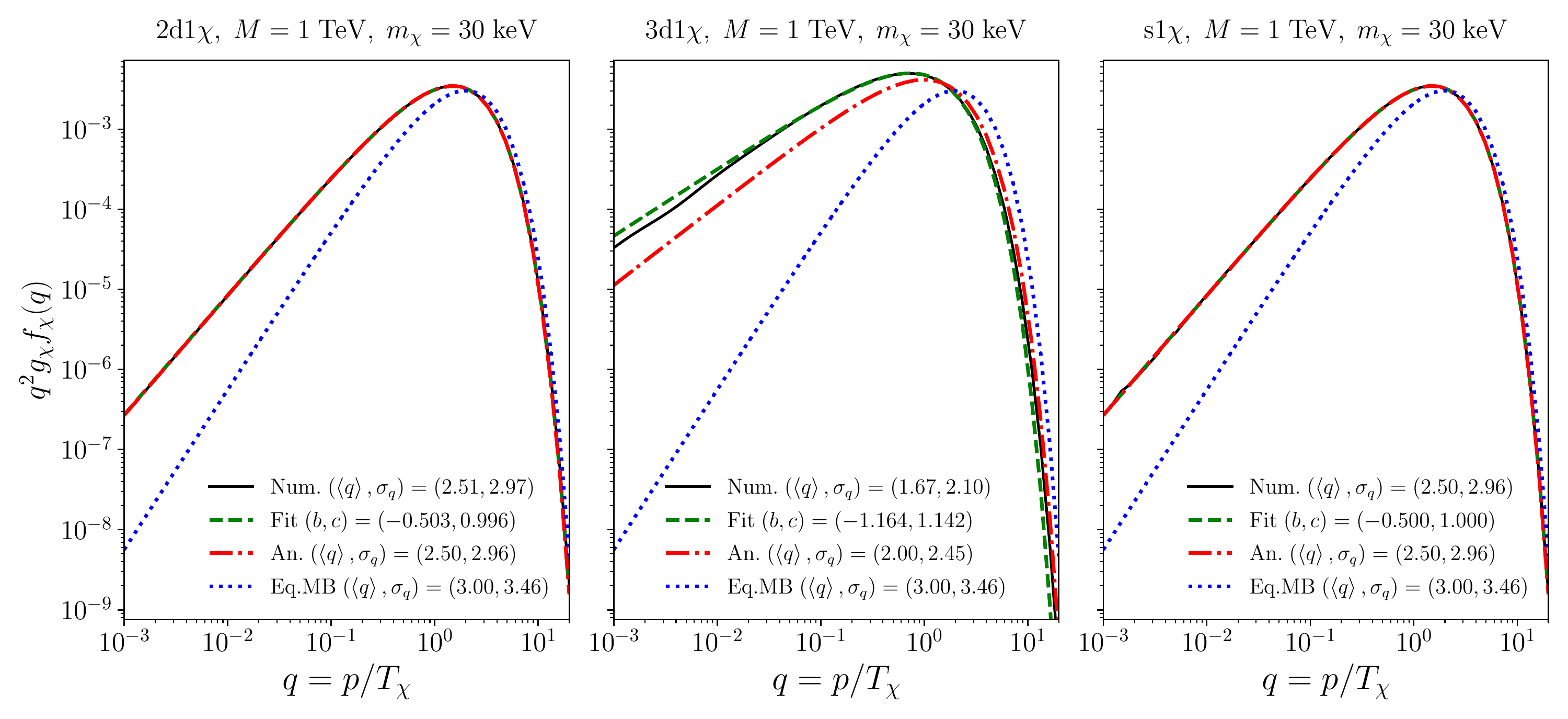}
\caption{PSD for single DM production benchmarks via two-body decays (left), three-body decays (center) and scatterings (right). We compare numerical solutions (solid black lines) with the analytical estimate (dot-dashed red), the fit of the numerical MB solution (dashed green) with Eq. (\ref{eq:fit_easy}) and the MB equilibrium distribution (dotted blue). All curves are normalized to reproduce the DM relic density, and we also show the values of $\mean{q}$ and $\sigma_q$.}
\label{fig:PSD_bench2}
\end{figure}

Quantum corrections to the bath particles statistics give negligible corrections, and from now on we take the MB distribution for all of them. We show in Fig.~\ref{fig:PSD_bench2} numerical results for the PSD for the single production benchmarks via two-body decays (left panel), three-body decays (center panel) and scatterings (right panel). We superimpose in the figure also our analytical solutions. The analytical estimates of the PSD are well suited for two-body decays and scattering while they are not that accurate for three-body decays. The reason is the rough saddle point approximation exploited to evaluate analytically the integral over the Mandelstam variable $s$, as explained in App. \ref{app:PSD_an}. We compare also with the equilibrium MB distribution, and we observe how our distributions are ``colder'': they have a smaller average comoving momentum and dispersion than the equilibrium ones. 

Finally, we fit the numerical solutions for the PSD with the expression
\be
g_\chi q^2 f_\chi = \mathcal{N}_F q^{2+b}e^{-cq} \ .
\label{eq:fit_easy}
\ee
This general form for the PSD, adopted by Ref.~\cite{Bae:2017dpt}, works well when DM is produced via a single process and it can be generalized when production is controlled by two or more competiting processes. The fit parameters $b,c$ leads to 
\be
\mean{q} = \frac{b+3}{c} \ , \qquad \qquad \sigma_q = \frac{[(b+3)(b+4)]^{1/2}}{c} \ .
\ee
The observed DM relic density can be reproduced upon choosing an appropriate value for the normalization constant $\mathcal{N}_F$. We see a perfect agreement of the fit in all three cases.

%%%%%%%%%%%%%%%%%%%%%%%%%%%%%%%%%%%%%%%%%%%%%%%%%%%%%%%%%
%%%%%%%%%%%%%%%%%%%%%%% WARMNESS %%%%%%%%%%%%%%%%%%%%%%%%%%%
%%%%%%%%%%%%%%%%%%%%%%%%%%%%%%%%%%%%%%%%%%%%%%%%%%%%%%%%%
 
\section{Warmness bounds}
\label{sec:warmness}

Light FIMPs, even if they are produced non-thermally, suppress cosmological structures at small scales similarly to what WDM does. The proper methodology to constrain the FIMP mass would be to assess its impact with a full analysis of structure formation. An important observables is the absorption feature of light produced by the inhomogeneous distribution of the Intergalactic Medium (IGM) along different lines of sight to distant quasars, known as Lyman-$\alpha$ forest. It provides an observable at the smallest scales available, effectively in the range $[0.5,100]$ Mpc$/h$~\cite{Viel:2013fqw,Baur:2015jsy,Irsic:2017ixq}. In principle, the most suitable approach would be to analyse Lyman-$\alpha$ data using the computed model-dependent phase-space distribution for any FIMP model and then compute the observable of interest, i.e. the flux along the line of sight. However, this is highly time-consuming and it is not a viable approach.

We rely on the existent bounds on the WDM mass and we present different methods to constrain the FIMP mass. We use the results found by Ref.~\cite{Irsic:2017ixq}, a \textit{conservative} and a \textit{stringent} bound from Lyman-$\alpha$ forest data of $m\unt{WDM}>3.5$ keV and $m\unt{WDM} > 5.3$ keV, respectively. We list four different methodologies and we show how to employ them on our benchmarks. In the next section, we apply them to the general FIMP parameter space.

\subsection{Free-streaming}

The first quantity we rely on is the free-streaming horizon $\lambda\unt{FS}$. After production, FIMPs propagate along FRW geodesics from overdense to underdense regions, and they erase cosmological perturbations on the length scale they are able to travel until the time of matter/radiation equality. When the bath temperature is $T = M / x$, a FIMP particle with comoving momentum $q$ travels with velocity  
\be
v(q,x)=\frac{p}{\sqrt{p^2+m_\chi^2}}= \bigg[1+\bigg(\frac{m_\chi}{q \, T_\chi(x)}\bigg)^2\bigg]^{-1/2} \ .
\ee
We remind the definition in Eq.~\eqref{eq:Tchi} of the FIMP ``temperature''. Thus, at a given bath temperature $x$, FIMPs have an average velocity
\be
\mean{v(x)} = \frac{\int dq \;q^2 v(q,x) f_\chi(q)}{\int   dq\; q^2 f_\chi(q)} \ .
\ee
We define the free-streaming horizon as follows
\begin{equation}
\lambda\unt{FS} = \int_{t\unt{prod}}^{t\unt{equality}}dt\frac{\mean{v(t)}}{a(t)} = \int_{1}^{x\unt{equality}}\frac{d x}{x} \; \frac{\mean{v(x)}}{H(x)}\bigg(1+\frac{1}{3}\frac{d\log \gsts}{d\log x}\bigg) \frac{T_\chi(x)}{a(M)M}  \ ,
\label{eq:lambdaFS}
\end{equation}
where we set the value of the scale factor today $a(t_0) = 1$. Moreover, we account for the fact that the linear growth of cosmological perturbations begins at matter/radiation equality when the photon temperature is $T\unt{equality} = M/x\unt{equality} \simeq {\rm eV}$. The time of FIMP production corresponds to much higher temperatures, $x\unt{prod} \simeq 1$. Finally, we use Eqs.~\eqref{eq:t_to_T} and \eqref{eq:T_to_x} to change integration variable in the second equality.

We provide an analytical estimate for the free-streaming horizon by identifying two different regimes for the FIMP velocity
\begin{equation}
v(q,x)  \simeq \begin{dcases} 
1  &  \hspace{1cm} \text{ if } x\ll x\unt{NR} \\  
q\frac{T_\chi(x)}{m_\chi} &  \hspace{1cm} \text{ if } x\gg x\unt{NR}
\end{dcases} \ ,
\end{equation}
where the time when DM becomes non-relativistic corresponds to $x\unt{NR} \simeq M/m_\chi$. Upon inserting these limiting expressions into the general definition we obtain the approximate expression for the free-streaming horizon
\be
\begin{split}
\lambda\unt{FS} \simeq & \, \frac{3\sqrt{10}}{\pi}\frac{\Mpl}{a(M)M^3}\bigg[\int_{1}^{x\unt{NR}} dx\; x \bigg(1-\frac{1}{3}\frac{d\log \gsts}{d\log x}\bigg)\frac{T_\chi(x)}{\gst^{1/2}(x)}\\
&\hspace{3cm}+ \frac{\mean{q}}{m_\chi} \int_{x\unt{NR}}^{x\unt{equality}} dx \; x \bigg(1-\frac{1}{3}\frac{d\log \gsts}{d\log x}\bigg)\frac{T_\chi(x)^2}{\gst^{1/2}(x)}\bigg] \\ & \simeq \frac{3\sqrt{10}}{\pi \gsts^{1/3}(T_0) }\frac{\Mpl}{T_0}\bigg(\frac{\gsts^{1/3}(m_\chi)}{\gst^{1/2}(m_\chi)}\bigg)\frac{1}{m_\chi}\bigg[1-\frac{m_\chi}{M}+  \mean{q} \bigg(\frac{\gsts(m_\chi)}{\gsts(M)}\bigg)^{1/3}\log\bigg(\frac{m_\chi}{T\unt{eq}}\bigg)\bigg].
\end{split}
\ee
In the first equality, we use the Hubble parameter for a radiation dominated universe given in Eq.~\eqref{eq:Hubblevsx} and we split the integral over the relativistic and non-relativistic regimes. We perform additional simplifications in the second equality by neglecting the temperature dependence of $\gsts$ and by realizing how both integrals are dominated around $x\sim x\unt{NR}$. With this in mind, we evaluate the integrals by fixing the value of $\gsts$ at $T\simeq m_\chi$ and using the scaling $T_\chi$ as in Eq.~\eqref{eq:Tchi}. In the limit we are interested in, $m_\chi \ll M$, we find, neglecting the content of the square brackets 
\be
\lambda\unt{FS}\simeq 0.1\;\text{Mpc }\bigg(\frac{1\text{ keV}}{m_\chi}\bigg)\ .
\ee

The analytical estimate is useful to find the scaling of $\lambda\unt{FS}$ with the DM mass. However, it is not very precise because the integral is dominated around the time of transition for the DM from the relativistic to the non-relativistic regime. At that time, the full form of the PSD is relevant but the approximate analytical expression does not depend significantly on the actual shape of the PSD; there is just a weak dependence through the factor of $\mean{q}$ multiplied by the logarithmic factor, which for $m_\chi=$1 keV is roughly 7. 

\begin{table}
\centering
\begin{tabular}{?c?c?c?c?}\Xhline{2\arrayrulewidth} 
Benchmark &$f_\chi(q)$ &conservative $m_\chi\upt{min}$  [keV] &stringent $m_\chi\upt{min}$  [keV]\\\Xhline{2\arrayrulewidth} 
2d1$\chi$& $\propto q^{-0.503}e^{- 0.996 q}$  & 3.44 & 5.97 \\
3d$1\chi$ & $\propto q^{-1.164}e^{-1.142 q}$ & 2.22 & 3.89 \\
s$1\chi$ & $\propto q^{-0.500}e^{-1.000 q}$  & 3.44 & 5.97 \\
 \Xhline{2\arrayrulewidth} 
\end{tabular}\caption{FIMP mass bounds from the free-streaming horizon for single DM production benchmarks with $M=1$ TeV. For each case, we give the fit behavior of the PSD and the smallest value of $m_\chi$ for which the conservative and stringent bounds on $\lambda\unt{FS}$ are satisfied.}
 \label{tab:lambdaFS}
\end{table}

We impose our bounds by evaluating the full, PSD-dependent, free-streaming horizon defined in Eq.~\eqref{eq:lambdaFS} which takes into account carefully the form of the PSD during the transition from the relativistic to the non-relativistic regime. The greater the free-streaming horizon, the tighter the constraints from structure formation. Length scales around $0.1$ Mpc, which are typical of dwarf galaxies, mark the border between hot DM (suppressing too much power on small-scales compared to CDM) and models which are only in tension with data on structure formation, like WDM. In order to be more quantitative, we compute the free-streaming horizon for WDM candidates with mass $m\unt{WDM} = 3.5 \, {\rm keV}$ and $m\unt{WDM} = 5.3 \, {\rm keV}$, corresponding to the conservative and stringent bounds, respectively. In order to perform this calculation, we use Eq.~\eqref{eq:lambdaFS} with the WDM temperature scaling as follows
\begin{equation}
T\unt{WDM}(x) = \frac{M}{x} \, \frac{T_\nu}{T_0} \, \left( \frac{\gsts(M/x)}{2}\right)^{1/3} \, \bigg(\frac{93.14\text{ eV}}{m\unt{WDM}}\bigg)^{1/3} \ ,
\end{equation}
where $T_\nu/T_0 = 0.71611$~\cite{Lesgourgues_2011}. We find the values
\be
\lambda\unt{FS}^{\rm WDM} =  
\begin{dcases} 
0.070 \, {\rm Mpc} & \hspace{2cm} m\unt{WDM} = 3.5 \; \mathrm{keV}  \\
0.041 \, {\rm Mpc} & \hspace{2cm} m\unt{WDM} = 5.3 \; \mathrm{keV}
\end{dcases} \ .
\ee
We impose the bound $\lambda\unt{FS} < \lambda\unt{FS}^{\rm WDM}$ on our single production benchmarks with $M = 1 \, {\rm TeV}$ with results shown in Tab.~\ref{tab:lambdaFS}. Notice how scatterings and two-body decays have the strongest bounds whereas three-body decays have a weaker constraints being considerably colder.
 
\subsection{Momentum dispersion}
\label{sec:dispersion}

A more refined methodology, still computationally quite simple, relies on the introduction of a warmness quantity defined from the second moment of the PSD
\be
W_\chi = \frac{\sqrt{\mean{p^2}}}{m_\chi} = \sigma_q\frac{T_\chi}{m_\chi} \ ,
\ee
with $\sigma_q$ defined in Eq.~\eqref{eq:sigmaq}. As done above, we compute such a warmness for our FIMPs and we impose that it cannot exceed the values for WDM with masses equal to the ones correspondent to the bounds given in Ref.~\cite{Irsic:2017ixq}. This is very efficient to obtain rather quickly FIMP mass bounds for simple scenarios in which, under some approximations, one can compute analytically the PSD. For example, Refs.~\cite{Bae:2017dpt,Kamada:2019kpe} apply this method to establish whether a 7 keV mass DM candidate, in some benchmark models, is compatible with Lyman-$\alpha$ constraints.

The bound $W_\chi < W\unt{WDM}$ translates into the inequality
\be
\sigma_q \frac{T\unt{\chi}}{m\unt{\chi}} <\sigma_q\upt{WDM}  \frac{T\unt{WDM}}{m\unt{WDM}} \ .
\ee
The WDM temperature is fixed by the relation
\be\label{eq:TWDM}
\frac{T\unt{WDM}}{T_0}=\frac{T_\nu}{T_0}\bigg(\frac{93.14\text{ eV}}{m\unt{WDM}}\bigg)^{1/3} (\Omega\unt{WDM}h^2)^{1/3} \ ,
\ee
where $T_\nu/T_0 = 0.71611$~\cite{Lesgourgues_2011}. The problem is shifted into computing $\sigma_q$ from the model-dependent PSD. For a thermal fermion candidate we have ${\sigma}_q\upt{WDM} \simeq 3.6$. 

If the observed DM density is accounted for by our FIMPs, the bound on $m_\chi$ reads
\begin{equation}
m_\chi > 19\;\mathrm{keV} \bigg(\frac{m\unt{WDM}}{5.3\;\mathrm{keV}}\bigg)^{4/3} \bigg(\frac{\sigma_q}{3.6}\bigg) \bigg(\frac{106.75}{\gsts(M)}\bigg)^{1/3},
\end{equation}
for the stringent Lyman-$\alpha$ constraint $m\unt{WDM}>5.3$ keV. We list the results for the minimum mass allowed for each benchmark in Tab.~\ref{tab:mom_disp}. Notice how the momentum dispersion constraints are a factor of 3 stronger than the ones obtained with the free-streaming bounds. 
\begin{table}
\centering
\begin{tabular}{?c?c?c?c?c?c?}
\Xhline{2\arrayrulewidth} 
Benchmark & $f_\chi(q)$& Num. $\sigma_q$ &  conservative $m_\chi\upt{min} $ [keV]&stringent $m_\chi\upt{min} $[keV]  \\\Xhline{2\arrayrulewidth} 
2d1$\chi$& $\propto q^{-0.503}e^{- 0.996 q}$ & $2.97$  &9.01 &15.67 \\
3d$1\chi$ & $\propto q^{-1.164}e^{-1.142 q}$ & $2.10$  &6.37  &11.08 \\
s$1\chi$ & $\propto q^{-0.500}e^{-1.000 q}$ & $2.96$ &8.98 &15.62\\
 \Xhline{2\arrayrulewidth} 
\end{tabular}
\caption{Minimum  FIMP mass allowed by the warmness bound for our single DM production benchmarks with $M = 1 \, {\rm TeV}$.}
 \label{tab:mom_disp}
\end{table}

\subsection{Matter power spectrum}
\label{sec:trans}

The linear matter power spectrum $P(k)$ encodes almost all the relevant information about the process of structure formation. The behavior of the FIMP power spectrum at small scales, with respect to the standard $P\unt{\Lambda CDM}(k)$, characterizes the warmness of the DM candidate. The computation of $P(k)$ takes into account the full form of the considered PSD and not just its first two moments (namely the mean and momentum dispersion): this quantity is the most reliable source to obtain bounds on non-thermal relics from structure formation. 

We define the relative squared ``transfer function''\footnote{Following the conventions adopted in the literature, we define the squared ``transfer function'' as the ratio of the FIMP linear matter power spectrum with respect to the $\Lambda$CDM one.}
\be
\mathcal{T}^2(k)\equiv \frac{P(k)}{P\unt{\Lambda CDM}(k)} \ ,
\ee
which encodes information about the small-scale power suppression for our non-thermal relic with respect to the perfectly cold thermal DM. Indeed, Lyman-$\alpha$ bounds are usually expressed in terms of the limiting transfer function for thermal WDM $\mathcal{T}\unt{WDM}^2(k)$. 

Here, we compute the transfer function for every point in parameter space using a Boltzmann-solver code optimized for non-CDM. We employ the \ttt{CLASS} code~\cite{Blas_2011,Lesgourgues_2011} to compute the linear matter power spectrum of our FIMP. Details about our determination of the matter power spectrum and the transfer function via \ttt{CLASS} can be found in App.~\ref{app:deltaA}. 

\begin{figure}
\centering
\includegraphics[width=1\textwidth]{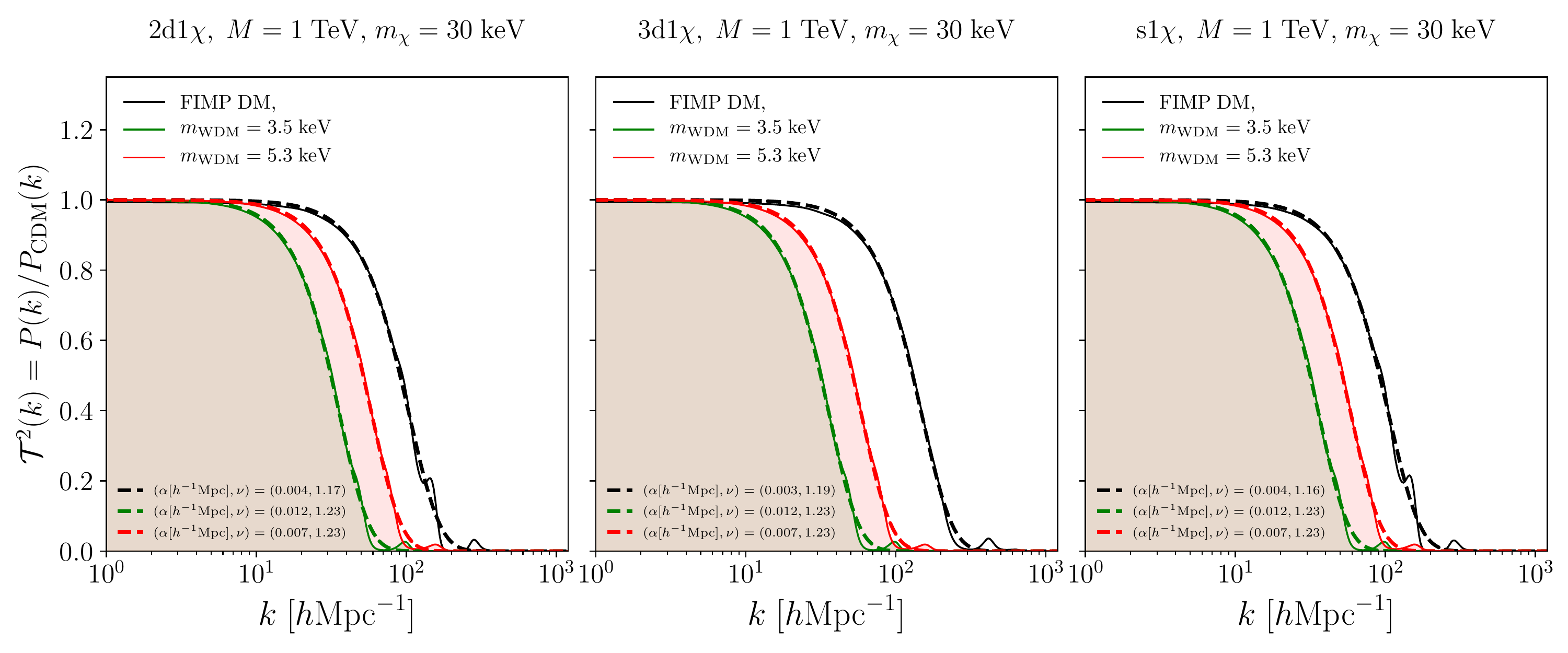}
\caption{Squared transfer functions for single DM production benchmarks with $M = 1 \, {\rm TeV}$ and $m_\chi=30$ keV (black lines). We superimpose the ones for WDM models with $m\unt{WDM}=3.5$ keV (green lines) and $m\unt{WDM}=5.3$ keV (red lines), and the fit with Eq.~\eqref{eq:trans_fit} (dashed lines). Shaded regions correspond to modes at which one could not have a further power suppression.}
\label{fig:T2k_bench}
\end{figure}

We fit our squared transfer functions with the expression 
\be\label{eq:trans_fit}
\trans^2(k)=[1+(\alpha k)^{2\nu}]^{-10/\nu} \ ,
\ee
where $\alpha$ are $\nu$ are fit parameters. This generalizes the form in Ref.~\cite{Bode:2000gq} for thermal WDM. Ref.~\cite{Murgia:2017lwo} provides a more general fit form with three parameters, $\trans^2(k)=[1+(\alpha k)^{\beta}]^{2\gamma}$; we find that Eq.~\eqref{eq:trans_fit} works well within our framework. We show in Fig.~\ref{fig:T2k_bench} the squared transfer functions corresponding to our three single DM production benchmarks with $M = 1 \, {\rm TeV}$ and $m_\chi=30$ keV. We show in the same figures the limiting transfer functions for the conservative and stringent WDM bounds, corresponding to $m\unt{WDM}=3.5$ keV and $m\unt{WDM}=5.3$ keV, together with their fits.

In order to provide bounds on the FIMP parameter space, we compare the resulting transfer functions to the limiting ones from the thermal WDM models from Lyman-$\alpha$ forest data~\cite{Irsic:2017ixq}. We perform this comparison by adopting the two different criteria described in what follows. As we will see explicitly in the next section, these two criteria provide bounds that are essentially identical.

\subsubsection*{The $k_{1/2}$ criterion}

The simplest criterion is a direct comparison between the reference modes at which there is a significative suppression of power with respect to the standard CDM model. Usually, this mode is chosen to be the one at which the squared transfer function drops by a factor of two, i.e. the so-called \textit{half-mode} $k_{1/2}$
\be
\trans^2(k_{1/2})\equiv\frac{1}{2} \ .
\ee
If the half-mode of the FIMP DM is smaller than the half-mode of a chosen reference WDM model, the considered FIMP scenario is excluded. The physical meaning of this criterion is that the small-scale cut-off of the FIMP model happens at scales which are too large and can compromise the Lyman-$\alpha$ data. From Eq. \eqref{eq:trans_fit} we can obtain the half-mode as a function of the fit parameters $\alpha$ and $\nu$
\begin{equation}
 k_{1/2} = \frac{1}{\alpha}\bigg[\bigg(\frac{1}{2}\bigg)^{-\frac{\nu}{10}}-1\bigg]^{\frac{1}{2	\nu}} \ .
 \end{equation} 
Notice how the half-mode is mostly set by $\alpha$. We exclude parameter space points where the half-mode is lower than the associated value corresponding to WDM, $k_{1/2}< k_{1/2}^{\rm WDM}$, and we consider both the conservative and the stringent values 
\be
k_{1/2}^{\rm WDM} =  
\begin{dcases} 
32.2 \; h/{\rm Mpc} & \hspace{2cm} m\unt{WDM} = 3.5 \; \mathrm{keV}  \\
52.5 \; h/{\rm Mpc} & \hspace{2cm} m\unt{WDM} = 5.3 \; \mathrm{keV}
\end{dcases} \ .
\ee

\subsubsection*{The $\delta A$ criterion}

The second criterion we consider employs a ``one-dimensional'' squared transfer function
\begin{equation}
 \mathcal{R}^2(k) \equiv \frac{P\upt{1D}(k)}{P\upt{1D}\unt{\Lambda CDM}(k)} \ ,
 \end{equation} 
where $P\upt{1D}(k)$ is the one dimensional projection of the linear matter spectrum defined as
\begin{equation}\label{eq:P1d}
P\upt{1D}(k) =\frac{1}{2\pi}\int_k^\infty dk'\,k'P(k') \ .
\end{equation}
The power suppression of the FIMP model with respect to the standard $\Lambda$CDM one is estimated through the following quantity~\cite{Murgia:2017lwo,Schneider:2016uqi}
\begin{equation}
\delta A=1-\frac{1}{(k\unt{max}-k\unt{min})}\int_{k\unt{min}}^{k\unt{max}} dk\, \mathcal{R}^2(k) \ ,
\label{eq:deltaA}
\end{equation}
where $k\unt{min}$ and $k\unt{max}$ are the minimum and maximum scale probed by the Lyman-$\alpha$ survey. This estimator is the mean deviation of the suppression of power along a line of sight due to FIMPs with respect the standard $\Lambda$CDM in the scales probed by the survey under consideration. If we refer to the analysis in Ref~\cite{Irsic:2017ixq}, the authors exploited the MIKE/HIRES+XQ-100 combined dataset which explored scales $k\in [0.5,20]$ $h/$Mpc. 

We exclude parameter space points where $\delta A > \delta A\unt{WDM}$. The calculations of both $\delta A$ as well as $\delta A\unt{WDM}$ depend on the UV cutoff $k\unt{lim}$ defined as the upper integration extremum in the integral in Eq. \eqref{eq:P1d}; while $P\upt{1D}(k)$ is formally defined as an integral over all possibile wave-numbers larger than $k$, we can only compute it numerically up to a finite value $k\unt{lim}$. We choose $k\unt{lim}=1.2\times 10^3\;h/$Mpc which gives the maximum allowed values
\be
\delta A\unt{WDM} =  
\begin{dcases} 
0.446  & \hspace{2cm} m\unt{WDM} = 3.5 \; \mathrm{keV}  \\
0.304 & \hspace{2cm} m\unt{WDM} = 5.3 \; \mathrm{keV}
\end{dcases} \ .
\ee
Nevertheless, the mass bounds extracted from this criterion do not depend on the cutoff we choose, see App.~\ref{app:deltaA} for further details.  

\subsection{Milky Way Satellites}

Cosmological $N$-body simulations predict a very large number of subhalos within the Milky Way (MW) virial radius. This is one of the few good reasons to introduce a WDM component into the cosmic budget. These subhalos are large enough to host a baryon fraction so that there should be many satellite galaxies around the MW. However these objects are not observed. Although there are other reasons why the predicted subhalos could not host a relevant baryon fraction for the respective satellite to be seen (e.g. complex baryon physics), it seems that a cut-off in the $\Lambda$CDM power spectrum could solve the problem. 

\begin{figure}
\centering
\includegraphics[width=0.75\textwidth]{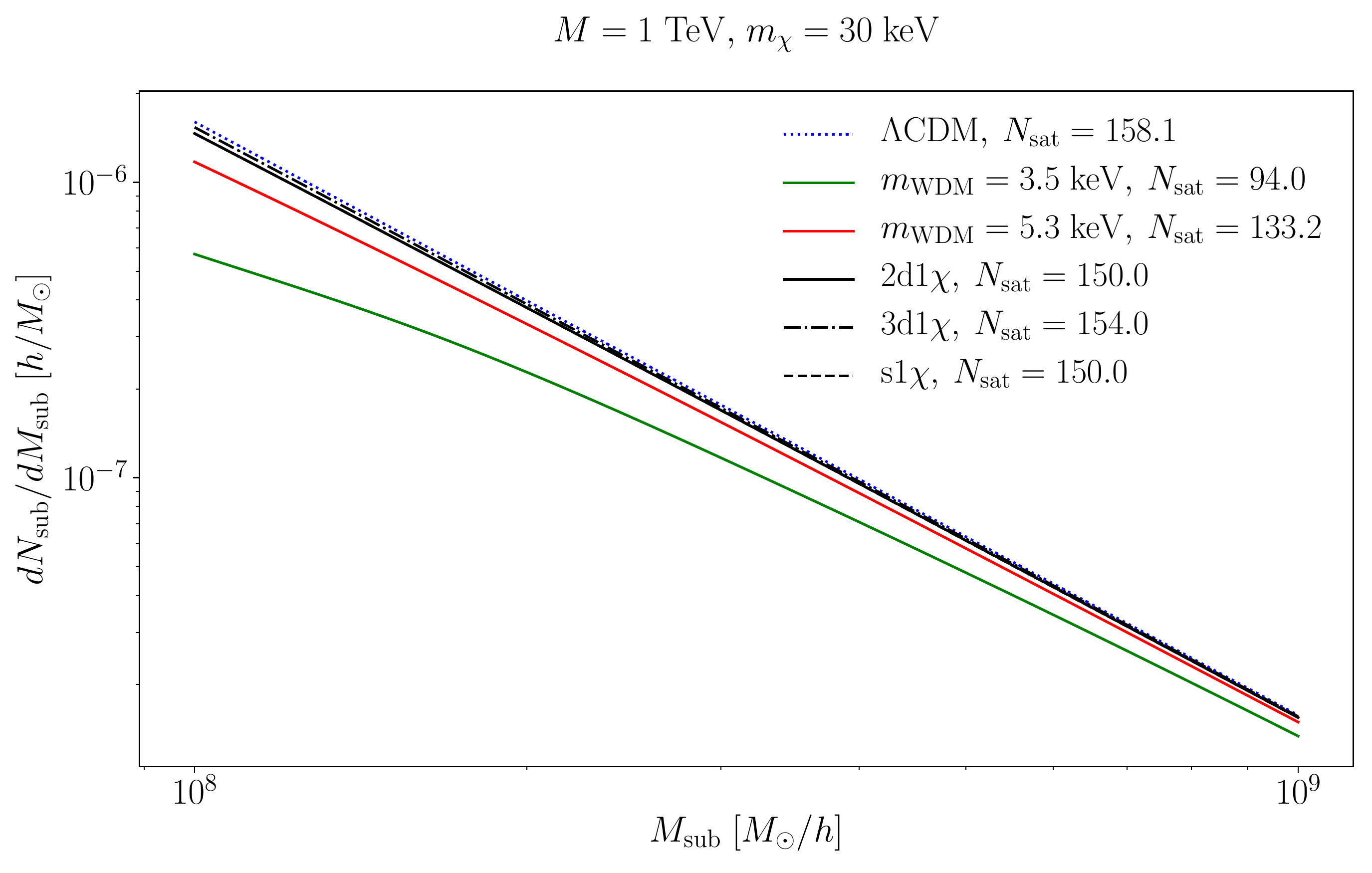}
\caption{Differential number of Milky Way satellites as a function of the mass. We show results for $\Lambda$CDM (dotted blue) as well as for WMD with masses set to the limiting conservative (green) and stringent (red) values. Black lines correspond to our single DM production benchmarks: two-body decays (solid), three-body decays (dot-dashed) and scatterings (dashed).}
\label{fig:Nsat}
\end{figure}

Thus MW satellite count provides a powerful and independent tool to constrain DM properties with respect to the aforementioned Lyman-$\alpha$ constraints. We follow the approach of Refs.~\cite{Schneider:2016uqi,Polisensky:2010rw} and multiply the number of observed MW satellites $N\unt{SDSS}=15$ observed by the SDSS by a factor of 3.5 to account for the limited sky coverage and add the known 11 MW satellites for a total of $N\unt{sat}=63$ estimated satellites. A more conservative estimate is $N\unt{sat}=57$~\cite{Murgia:2017lwo}. The constraint on the FIMP model comes from the comparison of the number of subhalos $N\unt{sub}$ predicted in the considered model and $N\unt{sat}$. A model is excluded if $N\unt{sub}<N\unt{sat}$. The number of subhalos is estimated through the following formula \cite{Schneider:2016uqi,Schneider:2014rda,Murgia:2017lwo}
\begin{equation}\label{eq:subhalos}
\frac{dN\unt{sub}}{dM\unt{sub}}=\frac{1}{C_n}\frac{1}{6\pi^2R\unt{sub}^3}\bigg(\frac{M\unt{halo}}{M\unt{sub}^2}\bigg)\frac{P(1/R\unt{sub})}{\sqrt{2\pi(S\unt{sub}-S\unt{halo})}} \ .
\end{equation}
Here $M\unt{sub}$ and $S\unt{sub}$ are the mass and the variance of a subhalo of radius $R\unt{sub}$, while $M\unt{halo}$ and  $S\unt{halo}$ are the mass and the variance of the main halo of radius $R\unt{halo}$. These quantities are defined as follows:
\begin{equation}
S_J= \frac{1}{2\pi^2}\int_0^{1/R_J}dk\,k^2P(k),\hspace{1cm}M_J=\frac{4\pi}{3}\Omega_m\rho\unt{cr}(2.5 R_J)^3\hspace{1cm} (J=\text{sub, halo})  \ .
\end{equation}
Here $C_n = 44.5$ since the host MW halo is defined as delimited by a density threshold of 200 times the background matter density, $\Omega_m=0.315(7)$~\cite{Zyla:2020zbs} the matter density parameter. We assume the mass of the MW halo to be $M\unt{halo}=1.7\times 10^{12}\:M_\odot/h$ following Ref.~\cite{Lovell:2013ola}, and from it we derive $R\unt{halo}=0.6673$ Mpc$/h$. Considering subhalos of mass $M\unt{sub} \geq 10^8\;M_\odot/h$, we obtain the predicted number of subhalos numerically integrating Eq.~\eqref{eq:subhalos}. In figure \ref{fig:Nsat}, we show the predicted differential number of Milky Way satellites as a function of the mass for different models and for fixed benchmark values for particles' masses.

%%%%%%%%%%%%%%%%%%%%%%%%%%%%%%%%%%%%%%%%%%%%%%%%%%%%%%%%%
%%%%%%%%%%%%%%%%%%%%%%%% RESULTS %%%%%%%%%%%%%%%%%%%%%%%%%%%
%%%%%%%%%%%%%%%%%%%%%%%%%%%%%%%%%%%%%%%%%%%%%%%%%%%%%%%%%

\section{Results}
\label{sec:results}

With the tools introduced in the previous section in hand, we explore the FIMP parameter space for the different production channels shown in Fig.~\ref{fig:setup}. We divide the discussion into two parts. First, we investigate the scenario where FIMPs account for the entire observed DM abundance, and this is the case where the mass bounds are more severe. We relax this assumption and we consider a mixed FIMP/CDM scenario where FIMPs constitute only a fraction $F < 1$ of the total amount of DM. Mass bounds are significantly weaker in this second case, and they disappear for small enough $F$.

\subsection{FIMP Dark Matter}
\label{sec:FIMPDM}

We begin our investigation of FIMP DM by considering the benchmarks shown in Fig.~\ref{fig:benchmarks}. Bath particles for these cases are either heavy with mass $M$, and in particular heavier than the FIMP, or massless. We focus on single DM production because changing the number of DM particles in the final state does not affect substantially the PSD; this is manifest from the equations in Fig.~\ref{fig:gCE} where we notice how the main change is just a multiplicative factor.

\begin{figure}
\centering
\includegraphics[width=1\textwidth]{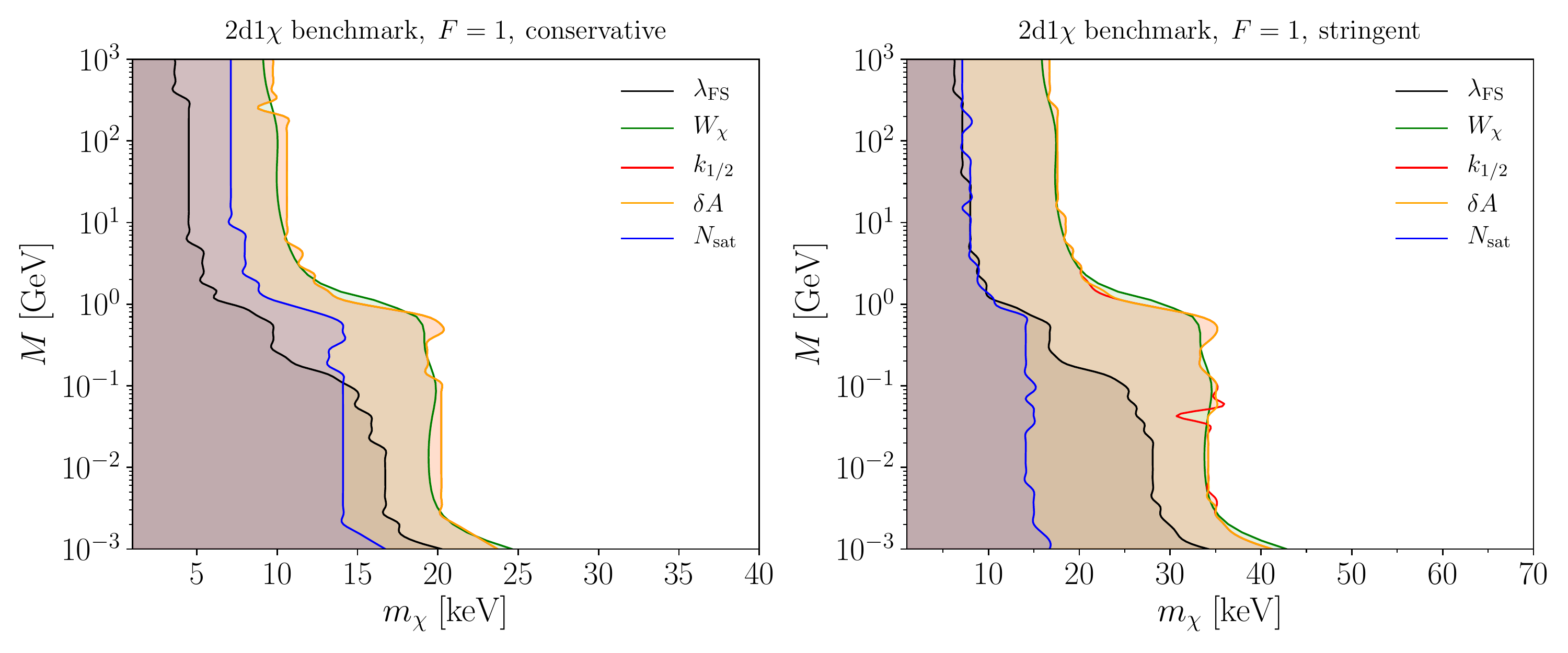}
\includegraphics[width=1\textwidth]{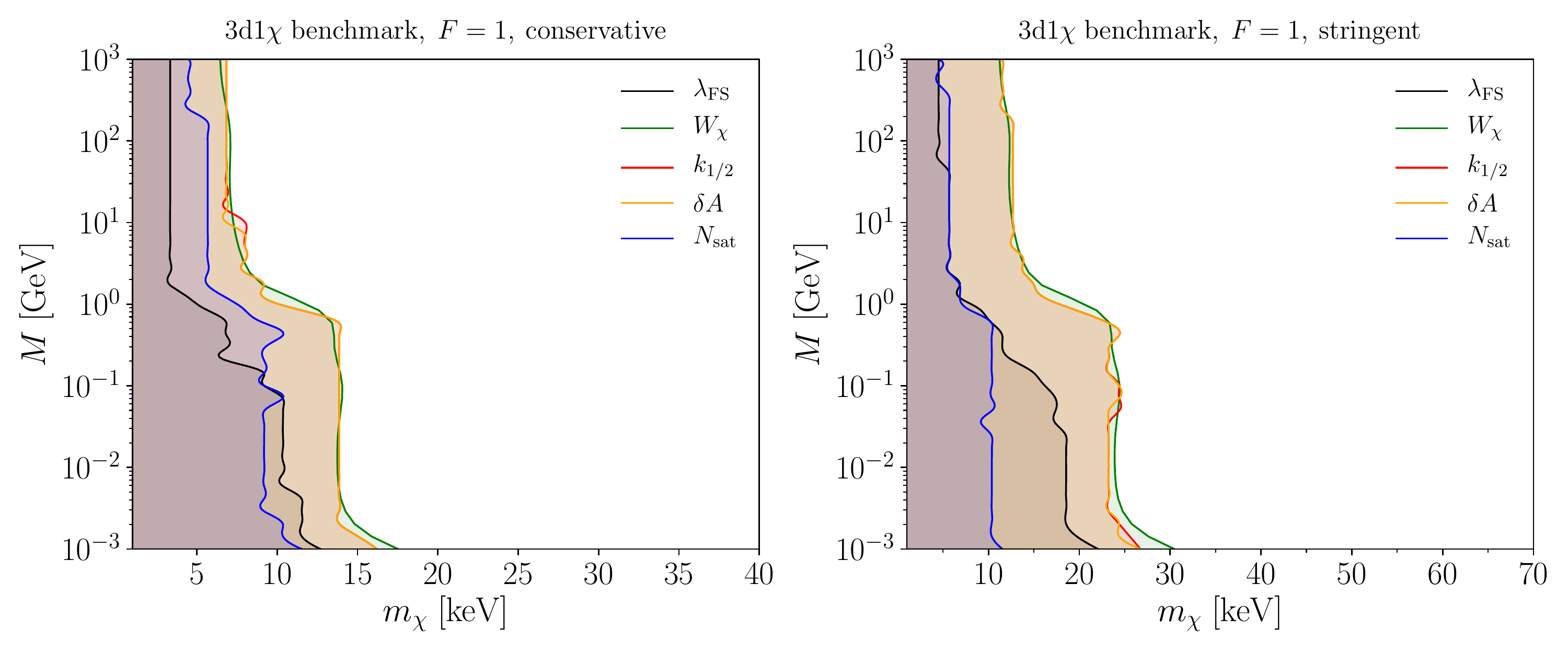}
\includegraphics[width=1\textwidth]{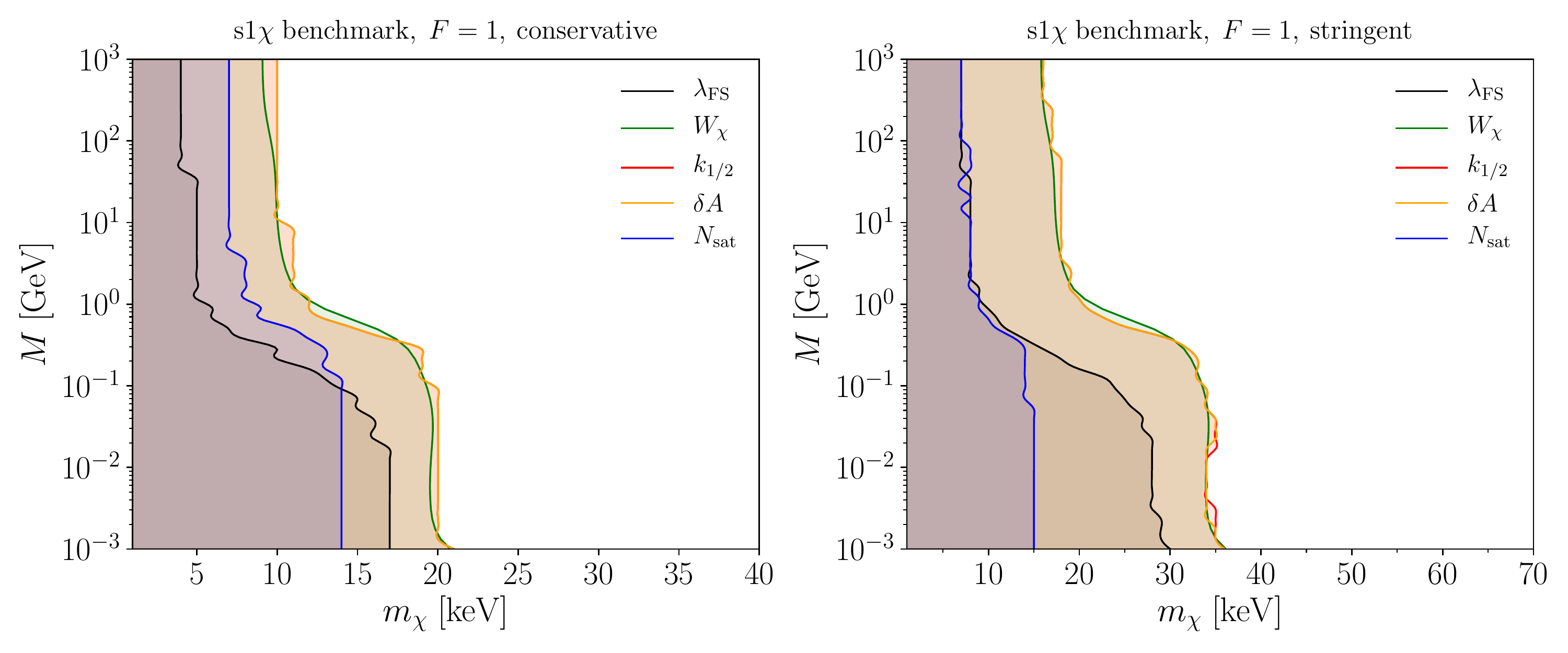}
\caption{Bounds on the FIMP mass for the single DM production benchmarks defined in Fig.~\ref{fig:benchmarks} as a function of the heaviest particle mass $M$. Here, the FIMP relic density accounts for the observed DM ($F = 1$).}
\label{fig:M_mX}
\end{figure}

Fig.~\ref{fig:M_mX} summarizes our mass bounds for the three different production channels we investigate: two-body decays (top panels), three-body decays (middle panels), and scatterings (bottom panels). We consider both the conservative (left panels) and the stringent (right panels) bounds on the WDM mass. In each panel, we show constraints by all the criteria listed in Sec.~\ref{sec:warmness}: free-streaming horizon (black lines), warmness quantity (green lines), half-mode $k_{1/2}$ (red lines), $\delta A$ (orange lines) and Milky Way satellites (blue lines). 

We consider a wide range for the overall freeze-in mass scale $M$. On one hand, we take $M$ above the MeV scale because we do not want new relativistic bath particles in thermal equilibrium at the time of BBN. On the other hand, we stop our plots for $M$ around the TeV scale because nothing changes for higher values; this is just a consequence of the fact that $\gst$ and $\gsts$ do not change above the Fermi scale and therefore the red-shift of DM particles after production, quantified by Eq.~\eqref{eq:Tchi}, is unaffected. Things would be different if one considers beyond the SM frameworks, such as supersymmetric theories, with several additional bath particles with a mass larger than the weak scale. 

As already observed before, three-body decays produce significantly colder FIMPs with respect to two-body decays and scatterings. Thus mass bounds are weaker for three-body decays. Once we look at the upper regions of each panel of Fig.~\ref{fig:M_mX}, the ones where $M = 1 \, {\rm TeV}$, we recover the bounds provided in Tabs.~\ref{tab:lambdaFS} and \ref{tab:mom_disp}. If we move to lower values of the overall mass scale $M$, we find that the bounds are quite stronger. The physics behind this is clear: FIMPs produced at later times have a ``temperature'', in the sense of Eq.~\eqref{eq:Tchi}, closer to the one of the thermal bath and therefore are less cold. Warmer FIMPs are more constrained. 

We see in each panel of Fig.~\ref{fig:M_mX} how the Milky Was satellites criterion is the less stringent, followed by the one using the free-streaming horizon $\lambda\unt{FS}$. The most stringent constraints come from the analysis of the transfer function with the two criteria illustrated in Sec.~\ref{sec:trans} and the analytical estimate provided by the warmness quantity $W_\chi$; the associated bounds are very similar. The mass bounds obtained by considering the ``one-dimensional'' squared transfer function, and quantifying the suppression $\delta A$ via Eq.~\eqref{eq:deltaA}, are the most reliable because they account for the transfer function in the momentum range probed by the Lyman-$\alpha$ survey. From now on, we will only impose the $\delta A$ criterion and derive the associated bounds.

\begin{figure}
\centering
\includegraphics[width=1\textwidth]{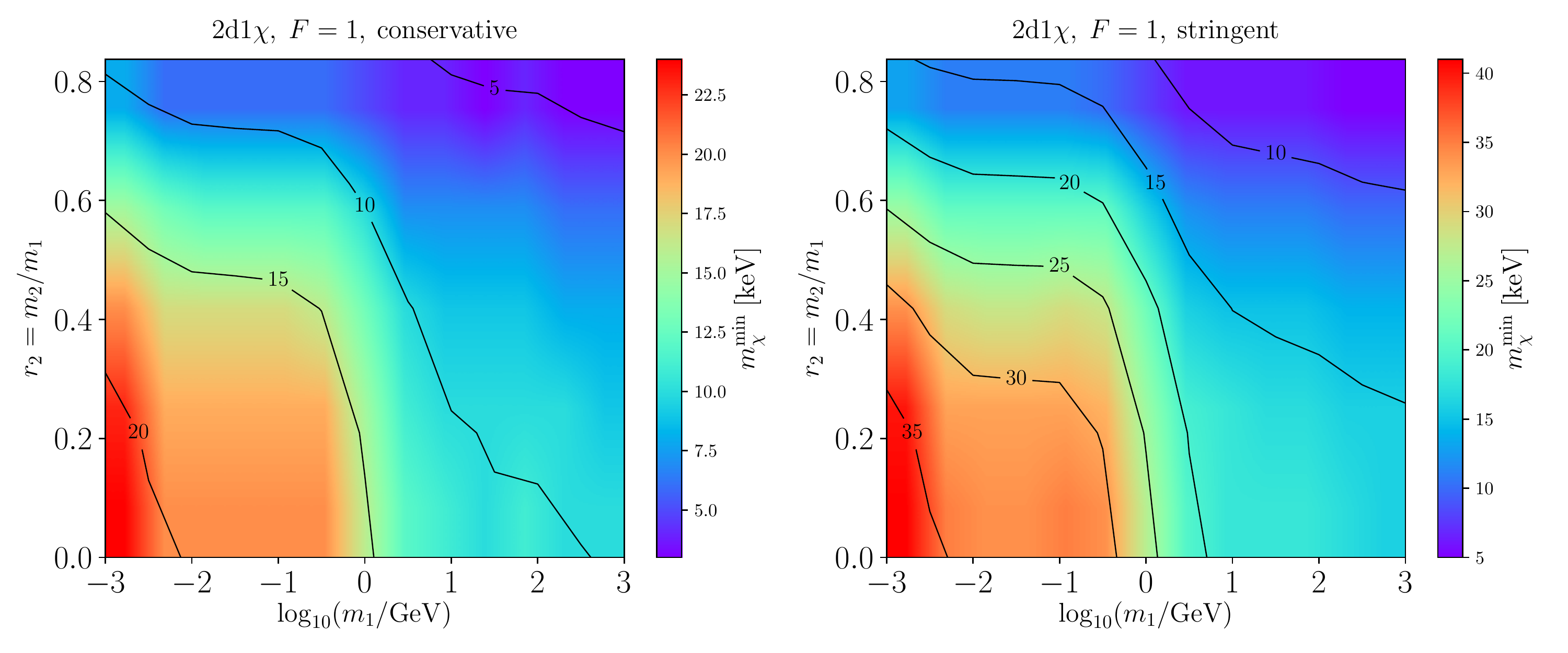}
\includegraphics[width=1\textwidth]{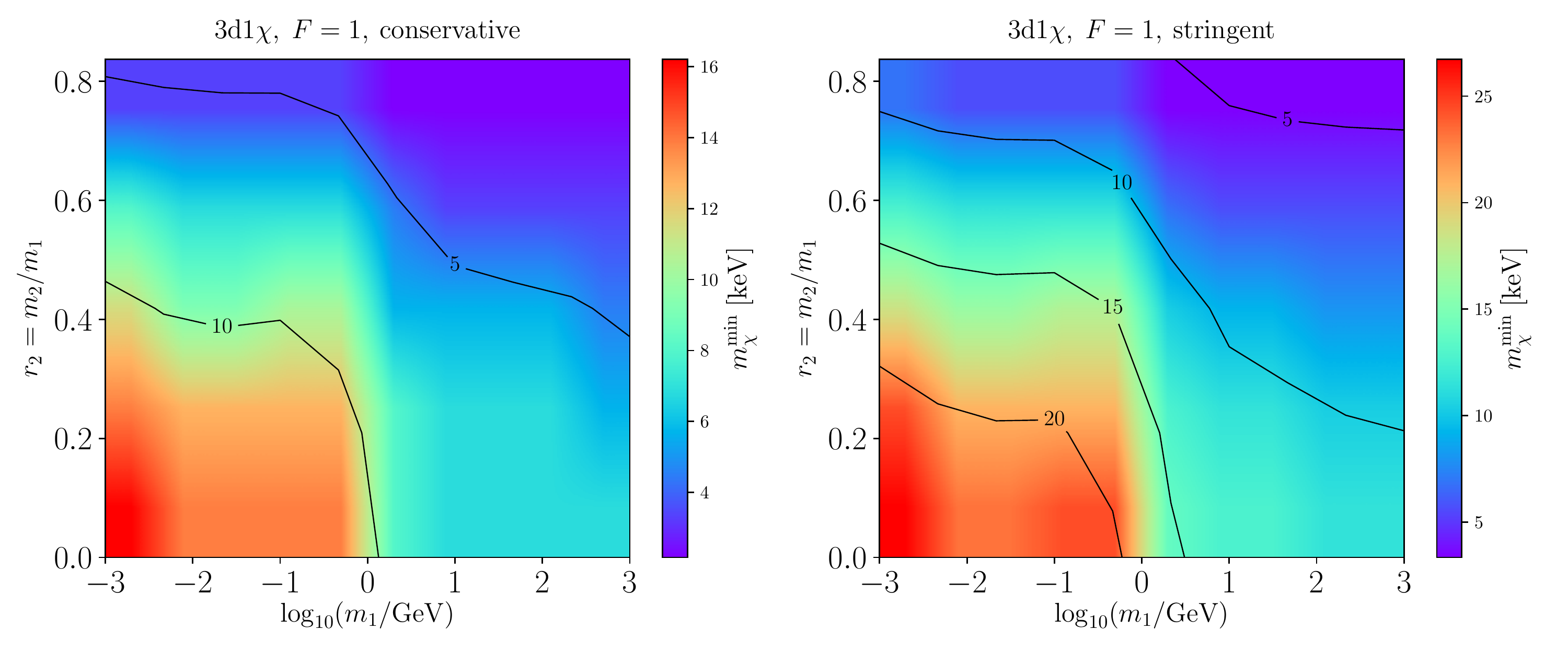} 
\includegraphics[width=1\textwidth]{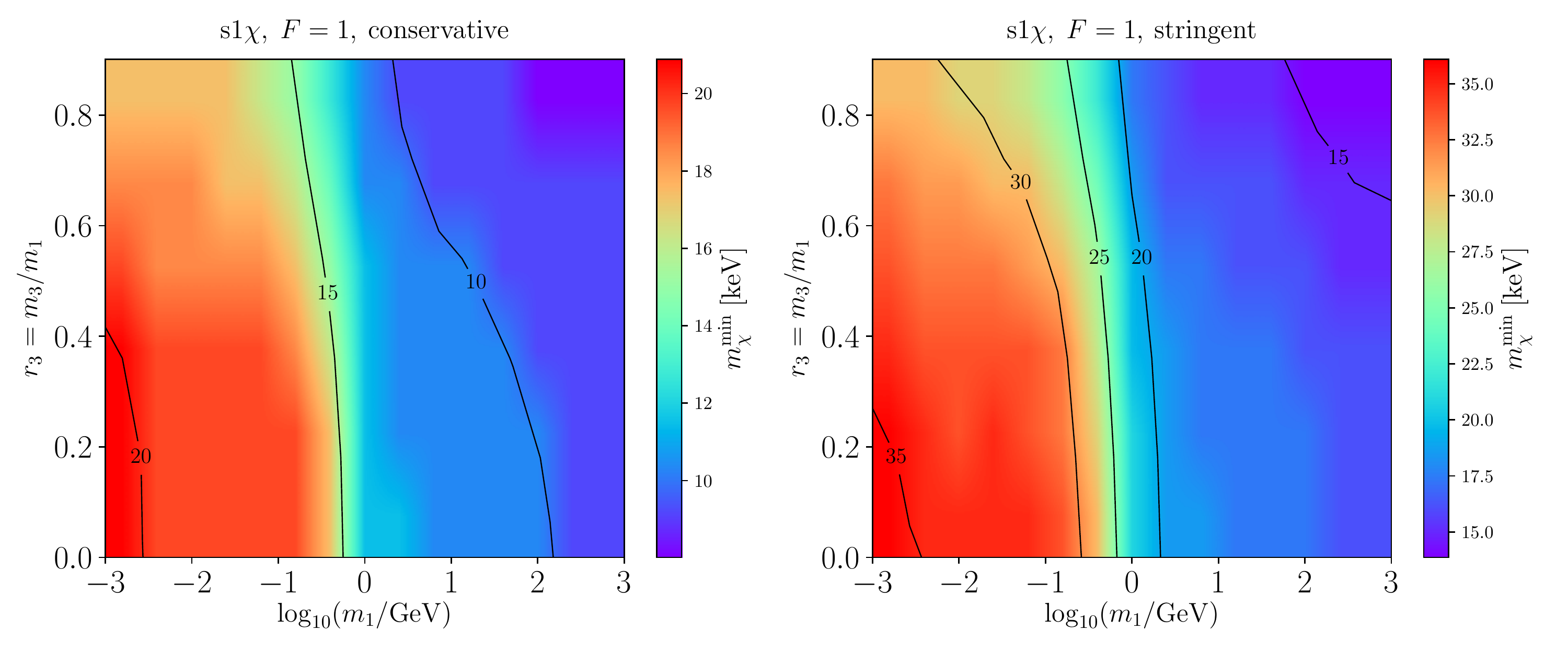}
\caption{Bounds on the DM mass for non minimal thermal bath mass spectra. For the benchmarks defined in Fig.~\ref{fig:benchmarks}, where the overall mass scale for freeze-in is $M$, we include one massive bath particle in the final state with mass $m_2 = r_2 \, m_1$ (decays) or $m_3=r_3m_1$ (scatterings). Here, the FIMP relic density accounts for the observed DM ($F = 1$).}
\label{fig:M_r}
\end{figure}

We go beyond the benchmarks in Fig.~\ref{fig:benchmarks} and we consider less minimal spectra. Always focusing on the same three channels for single DM production, we allow this time for two different mass scales in the process: an initial state heavy mass $m_1$ and a final state mass $m_2$ (for decays) or $m_3$ (for collisions). We define the ratio between the two masses in the process, $r_2 = m_2 / m_1$ and $r_3 = m_3 / m_1$, and we take it always smaller than one. Working in the $(m_1, r_2)$ and $(m_1, r_3)$ planes, we draw in Fig.~\ref{fig:M_r} isocontours for the minimum FIMP mass allowed by the $\delta A$ criterion. For two body-decay, we take $m_2$ the mass of the other final state bath particles. We do the same for three-body decays and we allow for one massive bath particle in the final state. Thus once the ratio $r_2$ gets close to one, decays are kinematically forbidden. As already observed for the benchmarks, three-body decays produce colder FIMPs and therefore are subject to weaker mass bounds. Regardless of what kind of decay we consider, bounds are relaxed for larger values of the mass ratio $r_{2}$; this is intuitive since there is less phase space available in the final state and FIMPs are produced with less kinetic energy. This effect was also observed in Refs.~\cite{Heeck:2017xbu,Kamada:2019kpe}. Finally, we show the results for scatterings where $m_3$ is the mass of the final state bath particle. Interestingly, there is a milder dependence on $r_3$ this time. Here the total energy available in the process is not only the mass of $m_1$ but the whole center of mass energy, which depends on the other initial state particle $B_2$. Therefore if $m_3$ increases, we have a milder reduction of the DM kinetic energy with respect to the decays. 

\subsection{Subdominant FIMP Component}
\label{sec:CDMWDM}

\begin{figure}
\centering
\includegraphics[width=1\textwidth]{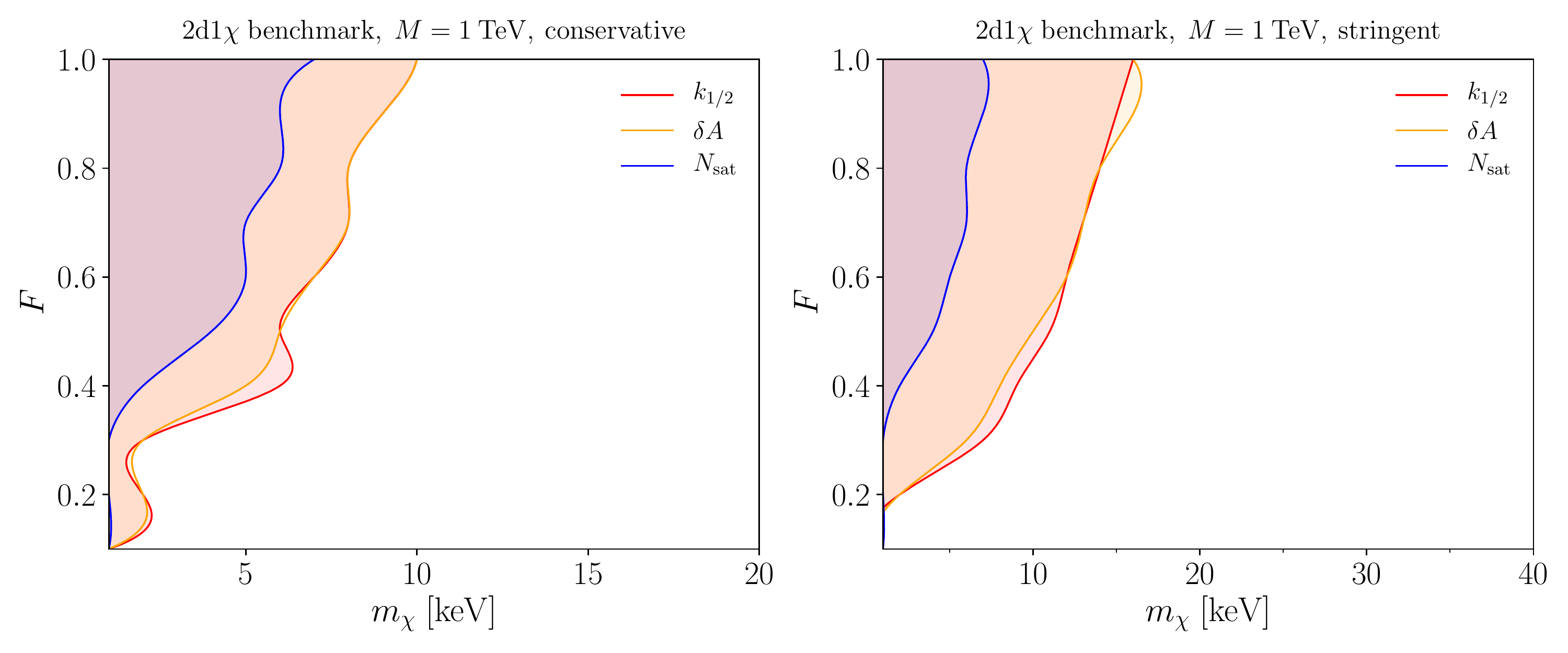}
\includegraphics[width=1\textwidth]{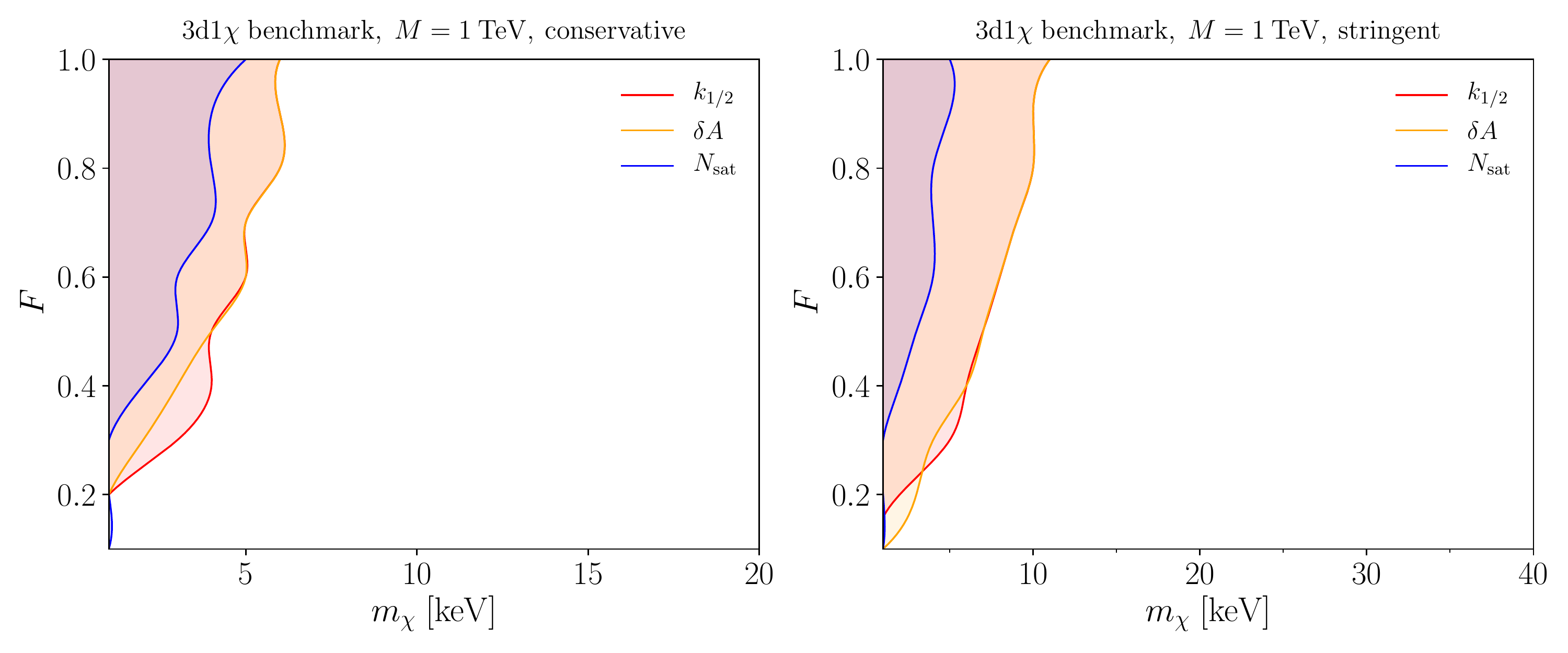}
\includegraphics[width=1\textwidth]{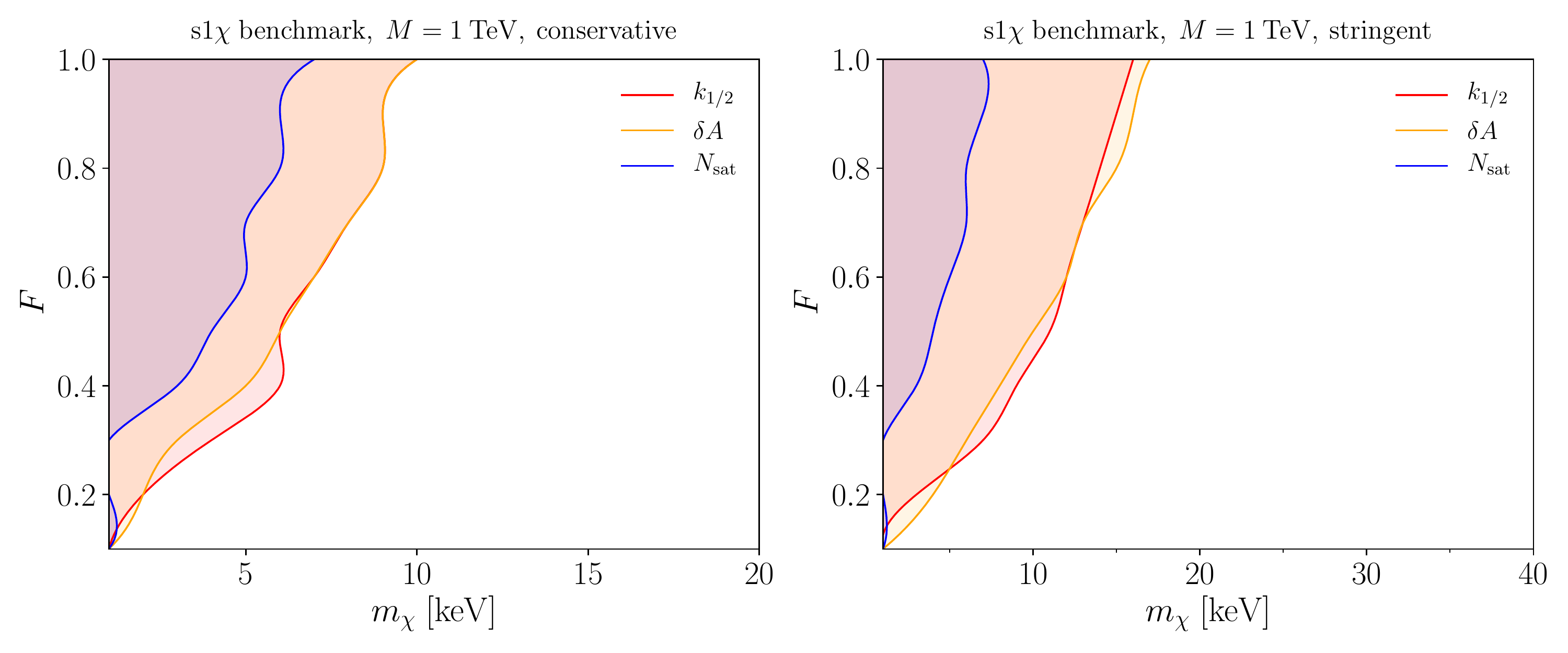}
\caption{Bounds on the DM mass for a mixed FIMP/CDM scenario where FIMPs account for a fraction $F < 1$ of the observed DM abundance. Benchmarks defined in Fig.~\ref{fig:benchmarks}.}
\label{fig:F_mX}
\end{figure}

We complete our investigation of the parameter space by considering a mixed FIMP/CDM scenario where FIMPs constitute only a fraction $F < 1$ of the total DM density. Once such a fraction gets very close to one we expect to recover the bounds just discussed, whereas they should disappear if $F$ is small enough. We quantify this in Fig.~\ref{fig:F_mX} where we show the usual three benchmarks for single DM production, and we show both conservative (left panels) and stringent (right panels) bounds on the FIMP mass as a function of $F$. 

We illustrate bounds for the two criteria based on the transfer function, $k_{1/2}$ and $\delta A$, and the one concerning Milky Way satellites. Not surprisingly, the most severe constraints come from the transfer function criteria also in this case. As usual, three-body decays are subject to less severe bounds due to colder final state FIMPs at the production time. Mass bounds are still above the keV scale for a fractional FIMP component as low as $F \simeq 0.1$ and they go away if the fraction is even smaller.

\section{Conclusions}
\label{sec:conclusions}

If FIMPs are behind the particle identity of DM, the only residual dark sector degrees of freedom that we have around today are particles with extremely tiny couplings to SM fields. Clearly, searching for FIMPs is a very challenging business and typical experimental rates at any conventional DM search are too low to yield any observable signal.

In this work, we considered FIMPs in the mass region where they can give astrophysical signals in cosmological structures at small length scales. Considering several topologies for DM production via decays and scatterings, we derived the resulting PSD after integrating the correspondent Boltzmann equation. With this in hand, we investigated the suppression of the matter power spectrum at small scales. Our study relied on the analysis performed in Ref.~\cite{Irsic:2017ixq} for WDM where the authors provided two mass bounds: $m\unt{WDM}>3.5$ keV (conservative) and $m\unt{WDM} > 5.3$ keV (stringent). We applied different criteria to put mass bounds on FIMPs, and the most reliable one (the $\delta A$ criterion~\cite{Murgia:2017lwo,Schneider:2016uqi}) turned out to be also the most severe. 

We summarize our results in Tab.~\ref{tab:results}. For FIMP production via two-body decays of a mother particle with mass around the TeV scale, FIMP mass bounds range from $m\unt{2d1\chi} \gtrsim 9.2$ keV (conservative) to $m\unt{2d1\chi} \gtrsim 16.0$ keV (stringent). These bounds get stronger if the decaying particle is lighter. Three-body decays produce colder final states, and for the decay of a TeV mass scale particles the bounds range from $m\unt{3d1\chi} \gtrsim 6.4$ keV (conservative) to $m\unt{3d1\chi} \gtrsim 11.1$ keV (stringent). Finally, bounds from scattering are  identical to the ones for two-body decays. It is worth keeping in mind how all our bounds were derived under the assumption of a constant squared matrix element for the transition. This is exact for two-body decays, and not necessarily the case for scattering although it is a reasonable approximation for IR-dominated freeze-in production. Once one considers a specific microscopic theory, mass bounds for two-body decays and scattering may differ slightly. In the last part of the paper, we considered a mixed FIMP+CDM scenario where FIMPs provide a sub-dominant DM component. Mass bounds are consequently weaker, and they disappear once we consider a fraction $F \lesssim 0.1$.

\begin{table}
\centering
\begin{tabular}{?c?c?c?c?}\Xhline{2\arrayrulewidth} 
Topology & $M$&conservative $m_\chi\upt{min}$  [keV] &stringent $m_\chi\upt{min}$  [keV]\\\Xhline{2\arrayrulewidth} 
\multirow{3}{*}{2d1$\chi$}& $1$ TeV&9.2 & 16.0 \\
& $1$ GeV&15.3 & 26.1 \\
& $1$ MeV &23.4 & 40.9 \\\Xhline{2\arrayrulewidth} 
\multirow{3}{*}{3d1$\chi$}&   1 TeV &6.4 & 11.1 \\
& $1$ GeV &9.9 & 14.0 \\
& $1$ MeV &15.4 & 26.8 \\\Xhline{2\arrayrulewidth} 
\multirow{3}{*}{s1$\chi$} &  1 TeV& 9.2 & 16.0 \\
& $1$ GeV &11.7 & 20.2 \\
& $1$ MeV &20.4 & 35.7 \\
 \Xhline{2\arrayrulewidth} 
\end{tabular}\caption{Summary of our analysis: FIMP mass bounds for the single DM production benchmarks defined in Fig.~\ref{fig:benchmarks} and for different freeze-in mass scale $M$. }
 \label{tab:results}
\end{table}

Our investigation relied strongly upon the WDM study in Ref.~\cite{Irsic:2017ixq}, and consistently with it we provided both stringent and conservative bounds. For both cases, we applied five different criteria (see, e.g., the five lines in Fig.~\ref{fig:M_mX}) with the goal of comparing the most reliable method (the ``area criterion'' $\delta A$) with all the ones adopted in the literature. Those methods include: free streaming length comparison to a reference value or rescaling with average comoving momentum~\cite{Shaposhnikov:2006xi,Biswas:2016iyh,Heeck:2017xbu}, momentum dispersion matching to map Lyman-$\alpha$ constraints of WDM to frozen-in DM~\cite{Kamada:2019kpe} and three-dimensional matter power spectrum transfer function criteria~\cite{Bae:2017dpt,Boulebnane:2017fxw,Huo:2019bjf,Baumholzer:2019twf,Dvorkin:2020xga}. Our model-independent results are in agreement with bounds for specific models by the aforementioned literature, taking into account the different reference bounds on the WDM mass that can vary in the references listed above. We stress two novel features of our work: the first bounds on DM produced via three-body decays, and the investigation of sub-dominant FIMP scenarios. Furthermore, we employ for the first time the ``area criterion'' to derive mass bounds on FIMPs.

We assumed an early universe dominated by a thermal bath of radiation at the time of FIMP production. This is a reasonable extrapolation of what we know was valid at the time of BBN, but it is worth keeping in mind that it is not supported by any observation. One possible direction to explore in the future is to perform the same analysis with a different cosmological background. Among several plausible options~\cite{Allahverdi:2020bys}, a motivated one is freeze-in during inflationary reheating when the universe undergoes an early matter domination phase~\cite{Co:2015pka,Roszkowski:2015psa,Evans:2016zau,Belanger:2018sti}. Alternatively, one can investigate freeze-in for fast-expanding universes where the Hubble rate scales with the higher power of the temperature with respect to the case of radiation~\cite{Redmond:2017tja,DEramo:2017ecx,Visinelli:2017qga,Biswas:2018iny}, as for example during the kination phase in theories of quintessence~\cite{Caldwell:1997ii,Sahni:1999gb}.

Following a different path, one can employ our methodology to study the phenomenology of FIMP warm DM within specific microscopic realizations. If we insist on the renormalizability of FIMP couplings to the visible world, in order to ensure IR domination, we have several options for freeze-in via both decays and scattering. For production via decays, if the decaying bath particles is colored then DM production must happen necessarily at high temperatures given the current collider bounds. Natural particle candidates for this scenario are supersymmetric squarks decaying to gravitinos~\cite{Cheung:2011nn}. The mother particle can be lighter if it is not colored, but still not lighter than the weak scale if it carries electroweak gauge quantum numbers. Examples include supersymmetric sleptons~\cite{Ibarra:2008kn,Junius:2019dci}, DFSZ axinos~\cite{Covi:1999ty,Covi:2001nw,Co:2016fln} and the singlet-doublet model~\cite{Calibbi:2018fqf,No:2019gvl}. Finally, the dark photon as a mediator between the FIMP and the SM is a motivated candidate if one wants to have FIMP production via scattering~\cite{Chu:2011be,Essig:2015cda,Chang:2019xva}.

We leave these interesting future directions to forthcoming work. 

\paragraph{Note added.} While finalizing our paper, Ref.~\cite{Ballesteros:2020adh} appeared on the arXiv which also studies the matter power spectrum of non-thermal DM candidates and Lyman-$\alpha$ forest constraints. 

\paragraph{Acknowledgments.} The authors thank L. Lopez-Honorez, S. Matarrese, R. Murgia, D. Teresi, and M. Viel for useful discussions. We acknowledge \textit{Cloud Veneto} for granting access to their computational resources. The work of F.D. is supported by the research grants: ``The Dark Universe: A Synergic Multi-messenger Approach'' number 2017X7X85K under the program PRIN 2017 funded by the Ministero dell'Istruzione, Universit\`a e della Ricerca (MIUR); ``New Theoretical Tools for Axion Cosmology'' under the Supporting TAlent in ReSearch@University of Padova (STARS@UNIPD); ``New Theoretical Tools to Look at the Invisible Universe'' funded by the University of Padua. F.D. is also supported by Istituto Nazionale di Fisica Nucleare (INFN) through the Theoretical Astroparticle Physics (TAsP) project. F.D. acknowledges support from the European Union's Horizon 2020 research and innovation programme under the Marie Sk\l odowska-Curie grant agreement No 860881-HIDDeN. The work of A.L. is supported by the Deutsche Forschungsgemeinschaft under Germany’s Excellence Strategy - EXC 2121 Quantum Universe - 390833306.

\appendix

\section{Notation, conventions and useful results}
\label{app:useful}

We set the notation and conventions adopted in our work and we collect useful results.

\subsection*{Four-momenta and Lorentz invariant phase space}

Bath particles $B_i$ participate in the production of FIMPs via decay and scattering processes. Each degree of freedom has a four-momentum whose components are the energy and the spatial momentum. Throughout our paper, we employ uppercase characters to denote Lorentz four-vectors, and we use the correspondent lowercase character to express the modulus of the associated three-vector. Moreover, we use different symbols for bath particles and DM.  

For particles belonging to the primordial bath we have the four-momenta
\begin{align}
K_i^\mu = & \, (E_i, \vec{k}_i) \ , \\
E_i = & \, \sqrt{k_i^2 + m_i^2} \ ,
\end{align}
where energy and spatial momentum are related via a dispersion relation with $m_i$ the mass of the bath particle $B_i$. Likewise, for DM particles we have
\begin{align}
P_i^\mu = & \, (E_{\chi i}, \vec{p}_i) \ , \\
E_{\chi i} = & \, \sqrt{p_i^2 + m_\chi^2} \ .
\end{align}
It is convenient to introduce the Lorentz invariant phase space (LIPS) of initial and final state particles 
\begin{align}
\label{eq:LIPS1} d\mathcal{K}_i = & \, g_i\frac{d^3k_i}{2E_i \, (2\pi)^3} \ , \\
\label{eq:LIPS2} {d\Pi}_i = & \, g_\chi\frac{d^3p_i}{2E_{\chi i} \, (2\pi)^3} \ ,
\end{align}
where $g_i$ and $g_\chi$ are the numbers of internal degrees of freedom (colors, spins, etc.) of bath particles $B_i$ and FIMPs $\chi$, respectively.

\subsection*{FRW Cosmology}

The cosmological background for FIMP production in our work is a Friedmann-Robertson-Walker (FRW) expanding universe described by the metric
\be
ds^2= dt^2 - a(t)^2 \delta_{ij} dx^i dx^j \ .
\ee
Here, $t$ is the cosmic time and $x^i$ are comoving spatial coordinates ($i,j=1,2,3$). Physical distances grow with the expansion proportionally to the scale factor $a(t)$.
  
The expansion rate is quantified by the Hubble parameter defined as $H \equiv \dot{a} / a$. Here and thereafter, the dot denotes the derivative with respect to the cosmic time $t$. The total energy density of the universe $\rho$ sets the functional time dependence of the Hubble parameter through the Friedmann equation
\be
H = \frac{\sqrt{\rho}}{\sqrt{3} M_{\rm Pl}} \ ,
\ee
where we use the reduced Planck mass, $\Mpl = \left(8 \pi G_N\right)^{-1/2} = 2.4 \times 10^{18} \, {\rm GeV}$. In our analysis, we assume that a gas of relativistic particles in thermal equilibrium at temperature $T$ dominates the energy budget during freeze-in production. This is the extrapolation of the BBN snapshot that we get when the universe was approximately one second old. The energy density of the radiation bath reads
\be
\rho = \frac{\pi^2}{30} \gst(T) T^4 \ ,
\ee 
where $\gst(T)$ accounts for the effective number of relativistic degrees of freedom. The Hubble parameter as a function of the temperature explicitly reads
\be
H(T) = \frac{\pi \, \gst^{1/2}(T)}{3 \sqrt{10}} \frac{T^2}{\Mpl} \ .
\ee

\begin{figure}
\centering
\includegraphics[width=0.9\textwidth]{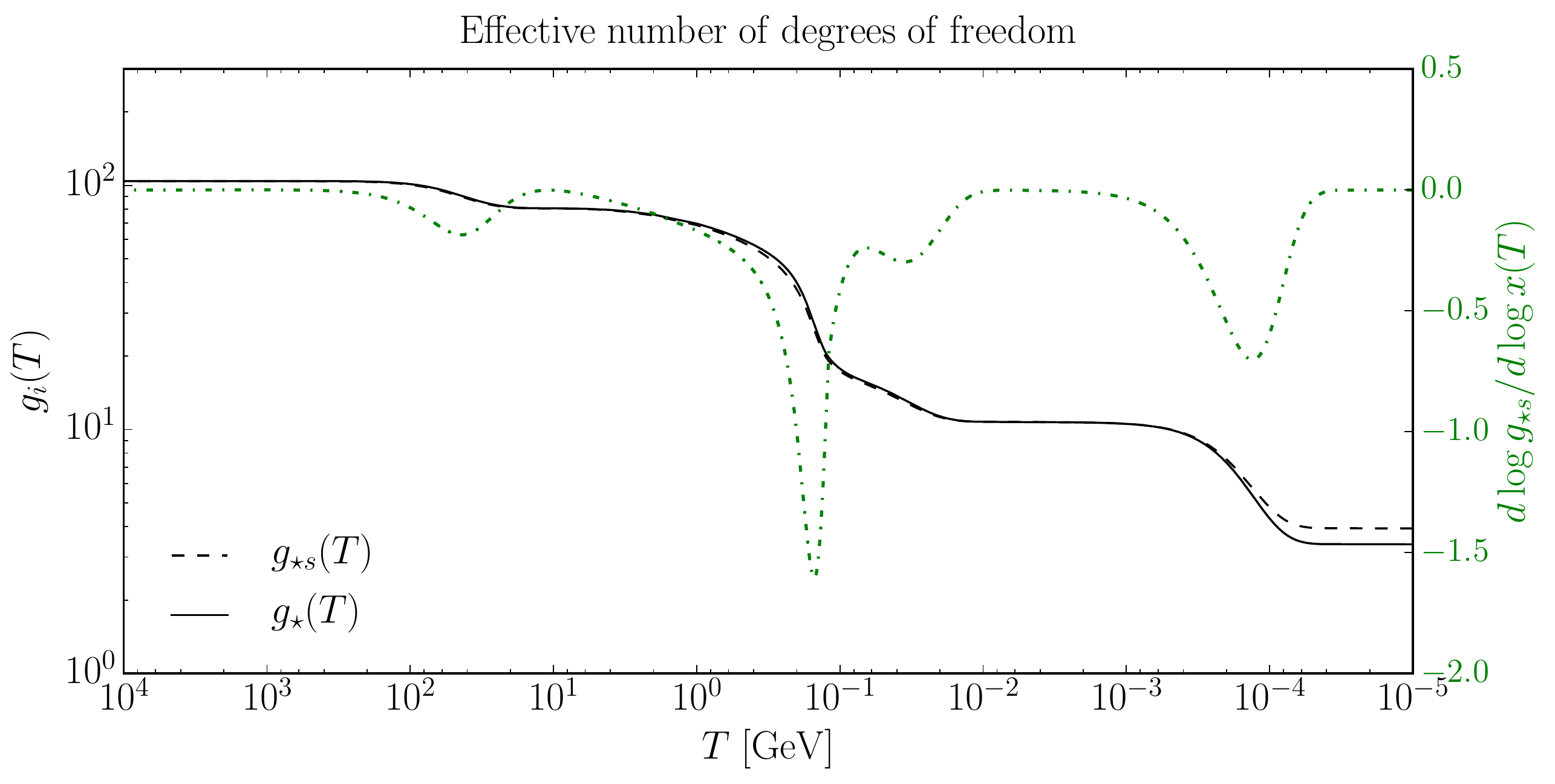}
\caption{Effective number of relativistic degrees of freedom throughout the expansion history. In green (right y-axis) we show the term $d\log \gsts(T)/d\log x $ (with $x = M / T$).}
\label{fig:gstgsts}
\end{figure}

Another useful quantity describing the expanding universe is the entropy density of the radiation bath
\be
s = \frac{2 \pi^2}{45} \gsts(T) T^3 \ ,
\ee
with $\gsts(T)$ the effective number of entropic relativistic degrees of freedom. We use $\gst(T)$ and $\gsts(T)$ given in Ref.~\cite{Saikawa:2018rcs}, which we reproduce in Fig.~\ref{fig:gstgsts}, where the authors found an analytical fit to lattice simulations for a careful treatment of the QCD phase transition (QCDPT).

\subsection*{Time vs temperature derivative}

The expansion of a radiation dominated universe is adiabatic and therefore the entropy in a comoving volume, $S = s a^3$, is a conserved quantity. This leads to the useful relation
\be
\frac{d T}{d t} = - \frac{H \, T}{1 + \frac{1}{3} \frac{d \log \gsts}{d \log T}}  \ .
\label{eq:Tvst}
\ee
We use the above expression to trade time derivatives with temperature derivatives. The derivative of a generic function of time $\xi(t)$ with respect to the cosmic time $t$ itself can be expressed as follows
\be
\dev{\xi}{t} = -\frac{H}{1 + \frac{1}{3} \frac{d \log \gsts}{d \log T}} \, \frac{d \xi}{d \log T} \ .
\label{eq:t_to_T}
\ee
The logarithm derivative term, also shown in Fig.~\ref{fig:gstgsts}, achieves its maximum values around the QCDPT but remains a small correction for almost the entire cosmological history.

It is often convenient to employ dimensionless variables when solving Boltzmann equations numerically. Instead of using the bath temperature $T$, we introduce the dimensionless ``time variable'' $x \equiv M / T$ where $M$ is a mass scale and its choice is purely conventional. It is advantageous to set it to the value of the heaviest particle mass involved in the process. Regardless of the specific choice, we have the following identity among derivatives
\be \label{eq:T_to_x}
\frac{d \xi}{d\log T} = - \frac{d \xi}{d\log x}  \  .
\ee

\subsection*{Phase space distributions}

The PSD $f_i$ denotes the phase space occupation number of a given species labeled by the index $i$. In the most general case, we have the functional dependence $f_i(X_i, K_i)$: the PSD depends on physical coordinates $X_i$ and four-momenta $K_i$. We are using physical distances and spatial momenta and therefore Lorentz indexes are raised and lowered with the Minkowski metric tensor; this is achieved by absorbing the scale factor into the spatial component of the four-vectors. We provide a general relativistic normalization for the PSD through the relation
\be
dN_i = 2 \, dS_\mu \, K_i^{\mu} d^4 K_i\; f_i(X_i,K_i) \delta(K_i^2 - m_i^2) \ .
\ee
Here, the Dirac delta function ensures the on-shellness condition, and $dN_i$ is the number of particle worldlines crossing the infinitesimal space-like surface orthogonal to $dS_\mu$. 

As we explain in Sec.~\ref{sec:PSFI}, homogeneity and isotropy of the FRW metric allow us to reduce the number of independent variables in the PSD, $f_i = f_i(t,k_i(t))$. Within our conventions, $k_i \equiv |\vk_i|$ and we use the notation $f_i = f_i(k_i)$.

Thermal bath particles $B_i$ are described by Bose-Einstein (BE) or Fermi-Dirac (FD) equilibrium PSDs. Quantum degeneracy effects in the early universe give small corrections, and the PSD is effectively well described by a Maxwell-Boltzmann (MB). These equilibrium distributions explicitly read
\be
f_i\upt{eq}(k_i) = \left\{ \begin{array}{lllllll}
\left[\exp(E_i / T ) - 1\right]^{-1} & & & & & & \text{BE} \\
\left[\exp(E_i / T ) + 1\right]^{-1} & & & & & & \text{FD} \\
 \exp( - E_i / T ) & & & & & & \text{MB}
\end{array}
\right. \ ,
\ee
where $T$ is the temperature of the thermal bath and $E_i^2= k^2_i+m_i^2$. 

\subsection*{Dark matter measured relic density}

Throughout our analysis, we solve the Boltzmann equation in momentum space and we determine the PSD due to freeze-in production. FIMPs never achieve thermal equilibrium, and the solution for the PSD is always proportional to the squared matrix element integrated over the phase space of final states. This in turn implies that the result  is proportional to the squared coupling mediating FIMP interactions with the thermal bath. In order to derive the shape of the PSD, we do not actually need to know such a coupling since it is only a multiplicative factor. We can fix its value if we impose the relic density constraint. 

The measured DM abundance is reported via the dimensionless combination~\cite{Aghanim:2018eyx}
\be
\Omega_{\rm DM} h^2 = \frac{\rho_{\rm DM}}{\rho_{\rm cr} / h^2} = 0.1200(12) \ .
\ee
The critical density is defined as $\rho_{\rm cr} \equiv 3 H_0^2 M_{\rm Pl}^2$, and if we express the current value of the Hubble parameter as $H_0 = 100 \, h \, {\rm km} / ({\rm s} \; {\rm Mpc})$ we have the numerical value~\cite{Zyla:2020zbs}
\be
\rho_{\rm cr} / h^2 = 1.053672(24) \times 10^{-5} \, {\rm GeV} \, \text{cm}^{-3} \  .
\ee 

Once we find the PDS $f_\chi$, we can compare with the observed DM relic density as follows. First, we integrate it over the phase space to find the number density
\be
n_\chi \equiv g_\chi  \int \frac{d^3 p}{(2\pi)^3} f_\chi(p) \ .
\ee
This is not the most convenient variable because after freeze-in production it still decreases as $n_\chi \propto a^{-3}$ as a consequence of the Hubble expansion. We normalize it with another quantity scaling with the expansion as $a^{-3}$, namely the entropy density $s$ (the entropy in a comoving volume $s a^3$ is conserved). In so doing we define the FIMP comoving density
\be
Y_\chi \equiv \frac{n_\chi}{s} \ .
\ee

The FIMP relic density today results in $\rho_\chi  = m_\chi n_\chi = m_\chi Y_\chi  s_0$, with the present entropy density~\cite{Zyla:2020zbs}
\be
s_0 = 2891.2 \, {\rm cm}^{-3} \ .
\ee
If we express this result in terms of the critical density we find
\be
\Omega_\chi h^2 = \frac{\rho_\chi}{\rho_{\rm cr} / h^2} = \frac{m_\chi Y_\chi s_0}{\rho_{\rm cr} / h^2}  \ .
\ee

We conclude this appendix with a useful expression to quantify the FIMP relic density. For a general mixed FIMP/CDM scenario where a FIMP $\chi$ contributes to a fraction $F$ of the DM relic density, $F \equiv \Omega_\chi / \Omega_{\rm DM}$, we need to satisfy the constraint
\be
m_\chi Y_\chi \simeq F \times 0.44 \, {\rm eV}  \ .
\ee

\section{General collision operator}
\label{app:Cderivation} 

The PSD variation due to the space-time geometry is accounted for by the Liouville operator, the collision operator accounts for interactions and it is independent on the metric. Here, we provide a derivation of the collision operator in Eq.~\eqref{eq:Cfgeneral} for a Minkowski flat spacetime. 

We consider a finite region of space, a box of volume $V$. The time variation of the total number $N_\chi$ of $\chi$ particles inside the box results in
\be
\frac{d N_\chi}{d t} = g_\chi \int \frac{V d^3 p}{(2\pi)^3} \frac{d f_\chi(p)}{dt} = 2 V \int d\Pi_\chi C[f_\chi(p)] \ .
\label{eq:NN}
\ee
The first equality follows from the PSD definition whereas in the second equality we use Eq.~\eqref{eq:BEgeneral} and we identify the Lorentz invariant phase space as defined in Eq.~\eqref{eq:LIPS2}.

In this appendix, we consider the direct (production) process in Eq.~\eqref{eq:genprocess} leading to  $n$ particles $\chi$ in the final state; the analysis for the inverse process is analogous. The variation of the number of $\chi$ particles inside the box per unit time results in
\be
\frac{d N_\chi}{dt} = n \int  \prod_{i=1}^{\ell} \frac{V d^3 k_i}{(2\pi)^3} f_i(k_i) \, \frac{d w}{dt} \ ,
\label{eq:dNchidt}
\ee
where we integrate the differential rate $d w / dt$ over all possible initial state momenta. 

The transition probability for the process is the square of the S-matrix element between initial and final states. We consider one-particle states with a Lorentz invariant normalization, see e.g. Ref.~\cite{Maggiore:2005qv}, and the S-matrix element reads
\be
S_{\text{fi}} = \delta_{\text{fi}}+i(2\pi)^4\delta^{(4)}(P_\text{i}-P_\text{f})
\, \mathcal{M}_{\text{fi}} \, (2 EV)^{-1/2} \prod_{i=2}^{n}(2 E_{\chi i} V)^{-1/2} \prod_{i=1}^{\ell+m}(2E_i V)^{-1/2} \ ,
\label{eq:Sfi}
\ee
with $P_\text{i}=\sum_{i=1}^\ell K_i$ and $P_\text{f}=\sum_{i=\ell+1}^{\ell+m} K_i + P + \sum_{i=2}^{n} P_i$. The products run over the energies of the DM particles different from the one under consideration (if any) and the bath particles, respectively. The matrix element $\mathcal{M}_{\text{fi}}$ depends on the microscopic theory. We squared the S-matrix element, and for different initial and final states we find
\be
|S_{\text{fi}}|^2 = (2\pi)^4\delta^{(4)}(P\unt{i}-P\unt{f}) \, V dt \, |\mathcal{M}_{\text{fi}}|^2 
(2 E  V)^{-1}  \prod_{i=2}^{n}(2 E_{\chi i} V)^{-1} \prod_{i=1}^{\ell+m}(2E_i V)^{-1} \ ,
\ee
where we regularize time and space via the relation $(2\pi)^4 \delta^{(4)}(0)= V dt$. Notice here $dt$ is a macroscopic time with respect to the timescale at which processes happen.

The interaction rate results from the sum over all possible final states. This means that we have to sum over all final state internal degrees of freedom (spin, colors, etc.) as well as final state momenta. The latter corresponds to the sum over the possible discrete values of the momenta allowed in the box of volume $V$
\be
\sum_{\vp} \sum_{\vp_i} \sum_{\vk_i} \simeq \int  \frac{V d^3p}{(2\pi)^3} \; \prod_{i=2}^{n} \frac{V d^3p_i}{(2\pi)^3} \; \prod_{i=\ell+1}^{\ell+m} \frac{V d^3k_i}{(2\pi)^3}  \ ,
\ee
with the right-hand side valid in the large volume limit. 

The squared matrix element in Eq.~\eqref{eq:Sfi} is for a given configuration of initial and final state internal degrees of freedom (spin, colors, etc.). It is convenient to introduce the squared matrix element \textit{averaged} over both initial and final states
\be
\obar{|\mathcal{M}|^2} \equiv \frac{\sum |\mathcal{M}_{\text{fi}}|^2}{g_\chi^n \times \prod_{i=1}^{\ell+m} g_i}  \ .
\ee
The sum runs over all possible internal degrees of freedom configurations, and we account for all possible outcomes of the process and all possible initial states. 

We can finally identify the differential interaction rate $dw / dt$. For a given initial state momenta configuration, and summing over all internal degrees of freedom, we find
\be
\begin{split}
 \frac{d w}{dt} = V  \, \prod_{i=1}^{\ell} \frac{g_i}{2 E_i V} \, & \, \int d \Pi_\chi \prod_{i=2}^{n} d \Pi_i \prod_{i=\ell+1}^{\ell+m} d\mathcal{K}_i \;
(2\pi)^4\delta^{(4)}(P\unt{i}-P\unt{f}) \ \obar{|\mathcal{M}|^2}  \times \\ &  \qquad (1\pm f_\chi(p)) \prod_{i=2}^{n}  (1\pm f_\chi(p_i)) \prod_{i=\ell + 1}^{\ell+m}(1\pm f_i(k_i)) \ .
\end{split}
\ee
In the second row of the above equation, we account for quantum correction: phase space cells for final states may be already occupied, and fermions and bosons are less and more likely to occupy the same states, respectively. This leads to Pauli-blocking ($-$ sign) and Bose-enhancement ($+$ sign) phenomena. We plug this rate into Eq.~\eqref{eq:dNchidt} and we find 
\be
\begin{split}
\frac{d N_\chi}{dt} = n V & \, \int d \Pi_\chi \prod_{i=2}^{n} d \Pi_i \prod_{i=1}^{\ell+m} d\mathcal{K}_i \;
(2\pi)^4\delta^{(4)}(P\unt{i}-P\unt{f}) \ \obar{|\mathcal{M}|^2}  \times \\ &  \qquad \prod_{i=1}^{\ell} f_i(k_i) (1\pm f_\chi(p)) \prod_{i=2}^{n}  (1\pm f_\chi(p_i)) \prod_{i=\ell + 1}^{\ell+m}(1\pm f_i(k_i))  \ .
\end{split}
\ee
We compare this result with the relation in Eq.~\eqref{eq:NN} and we find the collision operator corresponding to the production process. The analysis for the inverse process is analogous, and once we identify both contributions we recover the expression in Eq.~\eqref{eq:Cfgeneral}. 

\section{Collision terms for our topologies}
\label{app:Ctopologies}

Once one focuses on freeze-in production, the collision operator is actually not an operator but rather a function of the DM momentum and the bath temperature. We denote this \textit{collision term}, defined in Eq.~\eqref{eq:collterm}, with the symbol $\mathcal{C}$. In this appendix, we provide computational details for the collision terms for the three main topologies considered in this work, corresponding to the three rows of Fig. \ref{fig:setup}. We allow for all possible statistics for bath particles in the initial state (FD, BE, MB). These general expressions can be used to justify the approximations employed in this work where we assume instead the MB statistics for all particles involved, and they can also be applied to specific microscopic scenarios.

\subsection*{Two-body decays}

For two-body decays, we consider both single and double production
\be
B_1 \rightarrow \left\{\begin{array}{l}
B_2 + \chi \\
\chi + \chi 
\end{array} \right. \ .
\ee
Our convention for the four-momenta is the same as in Eq.~\eqref{eq:genprocess}. We can write the collision term in a general form accounting for both cases
\be
\mathcal{C}_2(T,p)  = \frac{n}{2}\int d\mathcal{K}_1 d\mathcal{Q}_2 (2\pi)^4 \delta^{(4)}(K_1 - P - Q_2) \obar{|\mathcal{M}_{2}|^2} f_1(k_1) \ ,
\ee
with $n$ the number of DM particles in the final state. The integrated over four-momentum in the final state, which we dub $Q_2=(\mathcal{E}_2,\vq_2)$, is equal to $K_2$ or $P_2$ for single or double production, respectively. This final state particle is on-shell via the constraint $\mathcal{E}_2 = \sqrt{q_2^2 + M_2^2}$ where $M_2$ is either $m_2$ (single production) or $m_\chi$ (double production). The associated phase space integration over the LIPS $d\mathcal{Q}_2$ follows accordingly. Two-body decays are monochromatic and the squared matrix element $\obar{|\mathcal{M}_{2}|^2}$, which we always take as averaged over both initial and final degrees of freedom, is a constant and independent of any momentum. 

We integrate over $d^3k_1$ by using the three-dimensional Dirac delta ensuring conservation of spatial momentum and we find
\be
\mathcal{C}_2(T,p)  = n \frac{g_1 g_{\mathcal{Q}_2}}{32 \pi^2}  \obar{|\mathcal{M}_{2}|^2} \int \frac{d^3 q_2}{\mathcal{E}_2} \frac{f_1(k_1)}{E_1} \delta(E_1 - E - \mathcal{E}_2)   \ ,
\ee
with $g_{\mathcal{Q}_2}$ equal to $g_2$ or $g_\chi$ for single or double production, respectively. The four-momentum of the decaying particle is fixed by spatial momentum conservation
\begin{align}
\label{eq:k1DD} k_1 = & \, \sqrt{(\vp + \vq_2)^2} = \sqrt{p^2 + q_2^2 + 2 p q_2 \cos\theta} \ , \\
E_1 = & \, \sqrt{m_1^2+ k_1^2} = \sqrt{m_1^2 + p^2 + q_2^2 + 2 p q_2 \cos\theta} \ ,
\end{align}
with $\theta$ the angle between vectors $\vp$ and $\vq_2$. 

We employ polar coordinates for the $d^3 q_2$ integration, and we orient the polar axis along the fixed direction of the vector $\vp$. The integration over the azimuthal angle is straightforward and we can write the collision term in the form
\be
\mathcal{C}_2(T,p)  = n \frac{g_1 g_{\mathcal{Q}_2}}{16 \pi p}  \obar{|\mathcal{M}_{2}|^2} \int_{M_2}^\infty d \mathcal{E}_2 \int_{-1}^{+1} d \cos\theta
\, f_1(k_1) \, \delta(\cos\theta - \cos\theta_\star) \ .
\ee
The value of the polar angle $\theta_\star$ satisfying energy conservation reads 
\be
\cos\theta_\star = \frac{(E + \mathcal{E}_2)^2 - m_1^2 -p^2 -q_2^2}{2 p q_2} \ .
\label{eq:thetastar}
\ee
As a consequence of the Dirac delta function, the integral over $d \cos\theta$ is non-vanishing only if $ \cos\theta_\star \in[-1, 1]$. This puts constraints on the possible values of $\mathcal{E}_2$ we can integrate over. We find that the integration over $d \mathcal{E}_2$ has support in the range $[\mathcal{E}_2^-,\mathcal{E}_2^+]$ where we define
\be\label{eq:E2+-}
\mathcal{E}_2^\pm = \sqrt{M_2^2+(q_2^\pm)^2} \ ,
\ee
and
\begin{align}
\label{eq:q2+} q_2^+ = & \, \frac{p(m_1^2-M_2^2-m_\chi^2)+\sqrt{(p^2+m_\chi^2)\lambda (m_1^2,M_2,m_\chi)}}{2m_\chi^2} \ , \\
\label{eq:q2-} q_2^- = & \, \frac{\left|p(m_1^2-M_2^2-m_\chi^2)-\sqrt{(p^2+m_\chi^2)\lambda (m_1^2,M_2,m_\chi)}\right|}{2m_\chi^2} \ .
\end{align}
Here, we find it convenient to express the results in terms of the function $\lambda$ defined as
\be\label{eq:lambda}
\lambda(x,y,z)\equiv[x-(y+z)^2][x-(y-z)^2] \ .
\ee
In the limit of light DM, relevant to our analysis, we find $q_2^+\rightarrow \infty$ while $q_2^-$ remains finite. 

We write the final result in the compact form 
\be
 \mathcal{C}_2(T,p)  =  n \frac{g_1 g_{\mathcal{Q}_2}}{16 \pi p}  \obar{|\mathcal{M}_{2}|^2} \int_{\mathcal{E}_2^-}^{\mathcal{E}_2^+} d\mathcal{E}_2 \;  f_1(k_{1\star})  \ ,
\label{eq:Cdec_gen}
\ee
where the PSD is understood to be evaluated for $k_1$ given in Eq.~\eqref{eq:k1DD} and with $\theta = \theta_\star$
\be
k_{1\star} \equiv \sqrt{(E + \mathcal{E}_2)^2 - m_1^2} \ . 
\ee
This general result is valid for any PSD $f_{1}$. If we assume a thermal distribution for $B_1$, as it is the case for our analysis, we can perform the integral analytically
\be\label{eq:Cdecayseq}
 \mathcal{C}_2(T,p)  =  n \frac{g_1 g_{\mathcal{Q}_2}}{16 \pi p} \obar{|\mathcal{M}_{2}|^2} \, T \, \times \, \begin{dcases}
- \log\bigg[\frac{1 - \exp\left[ - (\mathcal{E}_2^-+E)/T \right]}{1 - \exp\left[ -(\mathcal{E}^+_2+E)/T \right]}\bigg] & \text{BE} \\
\log\bigg[\frac{1+ \exp\left[ - (\mathcal{E}_2^-+E)/T \right]}{1+ \exp\left[ -(\mathcal{E}^+_2+E)/T \right]}\bigg] & \text{FD} \\
\exp[-(\mathcal{E}_2^-+E)/T] - \exp[- (\mathcal{E}^+_2+E)/T] & \text{MB}
\end{dcases} \ .
\ee

\subsection*{Three-body decays}

For three-body decays we have single and multiple production
\be
B_1 \rightarrow \begin{cases} B_2+B_3+\chi\\
B_2+\chi+\chi\\
\chi+\chi+\chi \end{cases} \ ,
\ee
and the general collision term reads
\be
 { \mathcal{C}_3(T,p)}  =\frac{ n}{2  } \int d\mathcal{K}_1 d\mathcal{Q}_2 d\mathcal{Q}_3 (2\pi)^4 \delta^{(4)}(K_1-Q_2-Q_3-P)\obar{ |\mathcal{M}_{3}|^2}f_1(k_1) \ .
\ee
As above, $n$ is the number of DM particles produced and $Q_i$ is equal to $K_i$ or $P_i$ ($i=2,3$).

We consider the LISPs of final state particles
\be
d\mathcal{Q}_2 d\mathcal{Q}_3 (2\pi)^4 \delta^{(4)}(K_1-Q_2-Q_3-P) = \frac{g_{\mathcal{Q}_2} g_{\mathcal{Q}_3}}{16 \pi^2} \frac{d^3 q_3}{\mathcal{E}_2 \mathcal{E}_3} \delta(E_1 - \mathcal{E}_2 - \mathcal{E}_3  - E) \ ,
\label{eq:LISP3bodydecay}
\ee
where the second equality follows from integrating over $d^3 q_2$ through the three-dimensional Dirac delta function imposing spatial momentum conservation. As a consequence, the energy $\mathcal{E}_2$ is understood to be evaluated on-shell with the spatial momentum $\vq_2 = \vk_1 - \vq_3 - \vp$. The expression in Eq.~\eqref{eq:LISP3bodydecay} is Lorentz invariant, and we exploit such an invariance to perform the remaining integrals. First, we introduce the Mandelstam variables 
\begin{align}
s \equiv & \, (Q_2+ Q_3)^2 = (K_1-P)^2 \ , \\
t \equiv  & \, (Q_3+ P)^2 = (K_1-Q_2)^2 \ .
\end{align}
We work in the center of momentum frame of $Q_2$ and $Q_3$ where the particles $2$ and $3$ have equal and opposite spatial momenta with modulus
\be
\hat{p}_{23} = \frac{\sqrt{\lambda(s,M_2,M_3)}}{2\sqrt{s}} \ .
\ee
 In the same frame, $B_1$ and the DM particle $\chi$ under consideration have spatial momenta along the same direction and with the same modulus
\be
\hat{p}_{1\chi} = \frac{\sqrt{\lambda(s,m_1,m_\chi)}}{2\sqrt{s}} \ .
\ee
Here, the function $\lambda$ is the same as in Eq.~\eqref{eq:lambda}. The Dirac delta function in Eq.~\eqref{eq:LISP3bodydecay} can be expressed as follows
\be
\delta(E_1 - \mathcal{E}_2 - \mathcal{E}_3  - E) = \delta\left(\sqrt{s} - \sqrt{p_{23}^2 + M_2^2} - \sqrt{p_{23}^2 + M_3^2}\right) = 
\frac{\mathcal{E}_2 \mathcal{E}_3}{\hat{p}_{23} \sqrt{s}} \delta(p_{23} - \hat{p}_{23}) \ .
\label{eq:DiracDelta}
\ee
As done for two-body decays, we employ polar coordinates for the integration over $d^3 q_3$ and we choose the direction of the polar axis along $\vp$. The integration over the azimuthal angle is straightforward, whereas the integration over the modulus of the momentum can be perfomed with the delta function in Eq.~\eqref{eq:DiracDelta}. The expression in Eq.~\eqref{eq:LISP3bodydecay} becomes
\be
d\mathcal{Q}_2 d\mathcal{Q}_3 (2\pi)^4 \delta^{(4)}(K_1-Q_2-Q_3-P) = \frac{g_{\mathcal{Q}_2} g_{\mathcal{Q}_3}}{8 \pi} \frac{\hat{p}_{23}}{\sqrt{s}} d \cos\theta_{3\chi}   \ .
\ee
Finally, we trade the integration over $d \cos\theta_{3\chi}$ with the one over $dt$ by using the Mandelstam variable evaluated in this frame
\be
t = M_3^2 + m_\chi^2 + 2 \left( \sqrt{\hat{p}_{23}^2 + M_3^2} \sqrt{\hat{p}_{1\chi}^2 + m_\chi^2} -  \hat{p}_{23} \hat{p}_{1\chi} \cos\theta_{3\chi} \right) \ .
\label{eq:t3dec}
\ee
Thus we manage to write the expression in Eq.~\eqref{eq:LISP3bodydecay} in a manifest Lorentz invariant way
\be
d\mathcal{Q}_2 d\mathcal{Q}_3 (2\pi)^4 \delta^{(4)}(K_1-Q_2-Q_3-P) = - \frac{g_{\mathcal{Q}_2} g_{\mathcal{Q}_3}}{16 \pi} \frac{dt}{\hat{p}_{1 \chi} \sqrt{s}}   \ .
\ee

Likewise, we evaluate the LIPS of the decaying particle in the FRW frame
\be
d\mathcal{K}_1 = g_1 \frac{d^3 k_1}{(2 \pi)^3 2 E_1} = \frac{g_1}{8 \pi^2} k_1 dE_1 d\cos\theta_{1\chi} = - \frac{g_1}{16 \pi^2} \frac{dE_1 ds}{p} \ ,
\ee
where we employ polar coordinates for the vector $\vk_1$ with the direction of $\vp$ as the polar axis. Similarly to what done above for $d t$, we trade the $d \cos \theta_{1\chi}$ integration with the one over $ds$. 

Thus the full integration measure reads
\be
d\mathcal{K}_1 d\mathcal{Q}_2 d\mathcal{Q}_3  (2\pi)^4 \delta^{(4)}(K_1-Q_2-Q_3-P)=  
\frac{g_1 g_{\mathcal{Q}_2} g_{\mathcal{Q}_3}}{256\pi^3}\frac{dt\, ds \, dE_1}{p \, \hat{p}_{1\chi} \, \sqrt{s}} \ ,
\ee
and the collision term for three-body decays results in
\be 
\algn{\mathcal{C}_3(T,p)  = &{}\; n\frac{ g_1 g_{\mathcal{Q}_2}g_{\mathcal{Q}_3}}{256 \pi^3 p} \int_{s\unt{min}}^{s\unt{max}}  \frac{ds}{\sqrt{\lambda(s,m_1,m_\chi)}} \int_{E^-_1(s)}^{E^+_1(s)} dE_1 f_1(k_{1}) \\ &\hspace{2cm}\times \int_{t\unt{min}(s)}^{t\unt{max}(s)}   {dt}  \obar{|\mathcal{M}_{3}|^2}(s,t)\ .} 
\label{eq:C3decE1}
\ee

We still need to determine the allowed range for the integration variables, and we begin with the integration over $ds$. The minimum value of $s$ is $(M_2+M_3)^2$, and it corresponds to the configuration when particles with momentum $Q_2$ and $Q_3$ are at rest in their center of momentum frame. The maximum value is obtained when $E=m_{\chi}$, namely when $\chi$ is produced at rest. Thus we have the integration range
\be\label{eq:s3dec}
s\in \left[(M_2+M_3)^2,(m_1-m_\chi)^2\right].
\ee
In order to determine the range for the integration over $dE_1$, we evaluate the Mandelstam variable $s$ in the FRW frame
\be
s = m_1^2 + m_\chi^2 - 2 (E_1 E - k_1 p \, \cos\theta_{1\chi\star}) \ .
\ee
The emission angle must satisfy the constraint $\cos\theta_{1\chi\star}\in[-1,1]$ and therefore the extrema of the integration range, $[ E_1^-(s), E_1^+(s)]$, are functions of $s$
\be\label{eq:E1dec3}
 E_1^\pm(s) = \sqrt{m_1^2+(k_1^\pm(s))^2},
\ee
where
\be\label{eq:k1dec3+} 
k_1^+(s)= \frac{p(m_\chi^2+m_1^2-s)+\sqrt{(p^2+m_\chi^2)\lambda(s,m_1,m_\chi)}}{2m_\chi^2} \ ,
\ee
\be\label{eq:k1dec3-} 
 k_1^-(s) = \frac{\left|p(m_\chi^2+m_1^2-s)-\sqrt{(p^2+m_\chi^2)\lambda(s,m_1,m_\chi)}\right|}{2m_\chi^2}  \ .
\ee
In the light DM limit, relevant for our study, we have $k_1^+\rightarrow \infty$ while $k_1^-$ stays finite. Finally, the variable $t$ in the center of mass frame of 2 and 3 is given in Eq.~\eqref{eq:t3dec}, and it is fully determined as a function of $s$ apart from $\cos\theta_{3\chi}$. Therefore the maximum and minimum are obtained respectively for $\cos\theta_{3\chi}=-1$ and $\cos\theta_{3\chi}=+1$. 

The collision term in Eq.~\eqref{eq:C3decE1} is completely general and it is valid for any PSD of the decaying particle. If we assume that $B_1$ is in thermal equilibrium, as it is the case for freeze-in production, we can perform the integral over $E_1$ analytically 
\be\label{eq:C3decayseq}
   \mathcal{C}_3(T,p)  =n\frac{  g_1 g_{\mathcal{Q}_2}g_{\mathcal{Q}_3} }{256\pi^3 }\frac{T}{p} \int_{s\unt{min}}^{s\unt{max}} ds \frac{h\unt{eq}(s)}{\sqrt{\lambda(s,m_1,m_\chi)}}\int_{t\unt{min}(s)}^{t\unt{max}(s)}   {dt}  \obar{|\mathcal{M}_{3}|^2}(s,t)  \ ,
\ee
where
\be\label{eq:heq}
h\unt{eq}(s)=\frac{1}{T}\int_{E^-_1(s)}^{E^+_1(s)} dE_1 f_1(k_{1})  =\begin{dcases}
-\log\bigg[\frac{1-\exp[-{E_1^-}/{T}]}{1- \exp[-{E_1^+}/{T}]}\bigg]& \text{BE} \\
+\log\bigg[\frac{1+ \exp[-{E_1^-}/{T}]}{1+\exp[-{E_1^+}/{T}]}\bigg]  & \text{FD} \\
 \exp[-{E_1^-}{T}]-\exp[- {E_1^+}/{T}]  & \text{MB}
\end{dcases}\ .
\ee

\subsection*{Scatterings}

We conclude this appendix with the collision term for FIMP production via binary collisions of bath particles. We consider both single and double production
\be
B_1 +B_2\rightarrow \begin{cases} B_3+\chi\\
 \chi+\chi \end{cases} \ ,
\ee
and the general collision term reads 
\be \algn{
  \mathcal{C}_s(T,p) =&{}\frac{n}{2} \int d\mathcal{K}_1 d\mathcal{K}_2 d\mathcal{Q}_3 (2\pi)^4 \delta^{(4)}(K_1+K_2-Q_3-P) \obar{|\mathcal{M}_{s}|^2}  f_1(k_1)f_2(k_2) \ , }
\ee
where $n$ is the number of DM particles in the final state. 

As done for the previous case, we introduce the Mandelstam variables
\be
\algn{ s=&{} (K_1+K_2)^2 =  (Q_3+P)^2,\\
		t=&(K_1-P)^2= (Q_3-K_2)^2 \ ,}
\ee
and we exploit Lorentz invariance of the LISPs to perform the integrations conveniently. In the center of momentum frame of the collision, initial state particles have opposite momenta with equal modulus
\be
\hat{k}_{12} = \frac{\sqrt{\lambda(s,m_1,m_2)}}{2\sqrt{s}}\ .
\ee 
Likewise, the momentum of final state particles results in
\be
\hat{p}_{3\chi} = \frac{\sqrt{\lambda(s,M_3,m_\chi)}}{2\sqrt{s}}\ .
\ee 

The procedure to perform the integrations is analogous to the one already adopted for three-body decays. For binary collisions we have
\be
d\mathcal{K}_1 d\mathcal{K}_2 d\mathcal{Q}_3 (2\pi)^4 \delta^{(4)}(K_1+K_2-Q_3-P) = 
\frac{g_1 g_2 g_{\mathcal{Q}_3}}{256\pi^3}\frac{dtdsd\mathcal{E}_3}{p \, \hat{p}_{3\chi}\sqrt{s}} \ ,
\ee
and the associated collision term is given by
\be
\algn{
 \mathcal{C}_s(T,p) =&{}\; n\frac{g_1g_2g_{\mathcal{Q}_3}}{256 \pi^3 p} \int_{s\unt{min}}^{s\unt{max}}\frac{ds}{\sqrt{\lambda (s,M_3,m_\chi)}}   \int_{\mathcal{E}_3^-(s)}^{\mathcal{E}^+_3(s)} d\mathcal{E}_3 \\ &\hspace{2cm}\times \int_{t\unt{min}(s)}^{t\unt{max}(s)} {dt}\obar{   |\mathcal{M}_{s}|^2}(s,t)f_1(k_{1\star})f_2(k_{2\star})\ .}
\label{eq:Csca_gen}
\ee

The minimum value of $s$ is obtained if particles have zero momentum in the center of mass in the initial or final state, hence the support for the integration over $ds$ reads
\be\label{eq:ssca}
s \ = \left[\max \left\{(m_{1}+m_{2})^2,(M_3+m_\chi)^2\right\}, +\infty\right[\ .
\ee
The integration over $\mathcal{E}_3$ is constrained by the values of $s$ in the FRW frame
\be
s=M_3^2+m_\chi^2+2(\mathcal{E}_3 E-q_3p\cos\theta_{3\chi\star}) \ , 
\ee
allowing $\cos\theta_{3\chi\star}\in [-1,1]$. We get that the integration over $d\mathcal{E}_3$ has support $[\mathcal{E}_3^-,\mathcal{E}_3^+]$ with 
 \be\label{eq:E3sca}
 \mathcal{E}_3^\pm(s) = \sqrt{M_3^2+(q_3^\pm(s))^2},
\ee
and
\be\label{eq:q3sca+} 
q_3^+(s)= \frac{p(s-M_3^2-m_\chi^2)+\sqrt{(p^2+m_\chi^2)\lambda(s,M_3,m_\chi)}}{2m_\chi^2}  \ , \\
\ee
\be\label{eq:q3sca-} 
 q_3^-(s) = \frac{\left|p(s-M_3^2-m_\chi^2)-\sqrt{(p^2+m_\chi^2)\lambda(s,M_3,m_\chi)}\right|}{2m_\chi^2} \ .
\ee
Again, for very light DM, $q_3^+\rightarrow \infty$, while $q_3^-$ remains finite. The variable $t$ in the center of mass frame is given by
\be\label{eq:tsca}
t=m_{1}^2+m_{\chi}^2-2\left(\sqrt{m_1^2+\hat{k}_{12}^2 }\sqrt{m_\chi^2+\hat{p}_{3\chi}^2 } - \cos\theta_{1\chi}\hat{k}_{12} \hat{p}_{3\chi}\right) \ ,
\ee
where all variables involved but $\cos\theta_{1\chi}$ are known functions of $s$. Then the maximum and minimum values of $t$ are found setting  $\cos\theta_{1 \chi}=-1$ and  $\cos\theta_{1\chi}=+1$, respectively. Finally, the PSDs of initial state particles are evaluated for arguments consistent with four-momentum conservation. Here, $k_{1\star}=k_{1\star}(s,t,p,\mathcal{E}_3)$ is the solution of the system of four equations in the FRW frame with other three unknown angles $\theta_{1\chi\star},\theta_{2\chi\star},\theta_{3\chi\star}$
\be 
\begin{dcases}
s=m_1^2+m_2^2+2\left(\sqrt{k_{1\star}^2+m_1^2}\sqrt{k_{2\star}^2+m_2^2}-\cos(\theta_{1\star}+\theta_{2\star})k_{1\star}k_{2\star}\right)\\
t=m_1^2+m_\chi^2-2\left(E\sqrt{k_{1\star}^2+m_1^2}-\cos\theta_{1\chi\star}pk_{1\star}\right)\\
%s+t=m_1^2+M_3^2+2\left(E\sqrt{k_{2\star}+m_2^2}+\cos\theta_{1\chi\star}pk_{2\star}\right)\\
k_{1\star}\cos\theta_{1\chi\star}+k_{2\star}\cos\theta_{2\chi\star}=q_3\cos\theta_{3\chi\star}+p\\
k_{1\star}\sin\theta_{1\chi\star}-k_{2\star}\sin\theta_{2\chi\star}=q_3\sin\theta_{3\chi\star}
\end{dcases}\ .
\ee
Energy conservation sets the value of $k_{2\star}$ as follows
\be 
k_{2\star}=\sqrt{\left(E+\mathcal{E}_3-\sqrt{k_{1\star}^2+m_1^2}\right)^2-m_2^2} \ .
\ee

The expression in Eq.~\eqref{eq:Csca_gen} is general and valid for any distribution for initial state particles. If we assume equilibrium MB distributions for $B_1$ and $B_2$, we can perform at least the integral over $\mathcal{E}_3$ analytically. We can exploit the conservation of energy
\be
f_1\upt{eq}(k_{1\star})f_2\upt{eq}(k_{2\star})  = \exp[-(E+\mathcal{E}_{3})/T]\ ,
\ee
to obtain
\be\label{eq:Cscaeq}\algn{
 \mathcal{C}_s(T,p)   = &{}n\frac{ g_1g_2g_{\mathcal{Q}_3}}{256 \pi^3 }\frac{Te^{-E/T}}{p }\int_{s\unt{min}}^{s\unt{max}} \frac{ds}{\sqrt{\lambda(s,M_3,m_\chi)}}    \bigg\{\exp[-\mathcal{E}_3^-/T]-\exp[-\mathcal{E}^+_3/T]\bigg\}\\ &\hspace{2cm}\times\int_{t\unt{min}(s)}^{t\unt{max}(s)}  {dt}  \obar{ |\mathcal{M}_{s}|^2}(s,t) \ .}
\ee
 
\section{Analytical solutions for the PSD}
\label{app:PSD_an}

In this appendix we provide analytical estimates for the collision terms above as function of the dimensionless time variable $x \equiv M/T$ and the dimensionless comoving momentum $q$ defined in Eq. (\ref{eq:q}) under some assumptions on the time when freeze-in happens and on the mass spectrum of the particles involved. Then these collision terms can be integrated to give the DM PSD $f_\chi(q)$, given by Eq. (\ref{eq:PSD}) after a choice of the Hubble parameter function $H(x)$.  All the assumptions we will make are reasonable as one can see in concrete microscopic models of DM. We assume
 \begin{enumerate}
 \item[\textbf{1)}] that FIMP production happens during radiation domination so that the Hubble parameter is given by Eq.~\eqref{eq:Hubblevsx};
 \item[\textbf{2)}] that all the particles are described by a MB statistics;
 \item[\textbf{3)}] that the relevant mass scale for the FIMP-production process is $m_1$ so $x=m_1/T$;
 \item[\textbf{4)}] that the matrix element responsible for the processes can be approximated as constants and then replaced with physically meaningful observables such as the decay rate $\Gamma_1$ or the scattering cross section evaluated at the relevant mass scale $\sigma\upt{FI}$, as done in Sec. \ref{sec:PSD}.
 \item[\textbf{5)}] that the effective number of relativistic degrees of freedom $\gst$ and $\gsts$ do not depend on temperature. So they can be computed at the scale of production $m_1.$ This also implies that $q=p/T$
\item[\textbf{6)}] to work in the very light DM limit in which $m_\chi \ll m_1$, implying $E=p=qm_1/x$.
 \end{enumerate}
 
\subsection*{Two-body decays}
We expand the results of Eqs.~\eqref{eq:q2+} and \eqref{eq:q2-} in the light DM limit. For single production we have $q_2=k_2$ and
\be \label{k2+_appr}
k_2^+=   p\bigg(\frac{m_1^2-m_2^2}{m_\chi^2}-\frac{m_1^2}{m_1^2-m_2^2}\bigg)+\frac{m_1^2-m_2^2}{4p} +\mathcal{O}(m_\chi^2)\ ,
\ee
\be \label{k2-_appr}
k_2^- =  \left| p\bigg(\frac{m_2^2}{m_1^2-m_2^2}\bigg)-\frac{m_1^2-m_2^2}{4p}\right|	+\mathcal{O}(m_\chi^2)  \ .
\ee
Otherwise, for double production $q_2=p_2$ and
\be\label{p2+_appr} 
p_2^+= \;  p\bigg(\frac{m_1^2}{m_\chi^2}-2\bigg)+\frac{m_1^2}{4p} +\mathcal{O}(m_\chi^2)\ , 
\ee
\be\label{p2-_appr}
 p_2^- = \; \frac{m_1^2}{4p}+\mathcal{O}(m_\chi^2)  \ . 
\ee
From Eq. \eqref{eq:E2+-}, we obtain that, under our approximations, 
\be
\mathcal{E}_2^+= \infty\ ,
\ee
\be
\mathcal{E}_2^-=\frac{m_1x}{4q}(1-r_2^2)+\frac{qm_1}{x}\frac{r_2^2}{1-r_2^2}\ ,
\ee
with $r_2=0$ for double production and $r_2=m_2/m_1$ for single production. Then, we get an analytical expression for Eq.~\eqref{eq:CdecayseqMB} 
\be\label{eq:Cdec_an}
\frac{g_\chi \mathcal{C}(x,q)}{E} \simeq  n\frac{g_1\Gamma_1}{y_{\mathcal{Q}_2\chi}} \frac{x}{q^2}\exp\bigg\{-x^2\frac{1-r_2^2}{4q}-\frac{q}{1-r_2^2} \bigg\}\ .
\ee 
Inserting this expression in Eq.~\eqref{eq:PSD} we obtain the  following approximation for the PSD
\be
g_\chi f_\chi(q)\simeq 6\sqrt{\frac{10}{\pi}}n \frac{g_1\Gamma_1 \Mpl}{y_{\mathcal{Q}_2\chi}m_1^2\gst(m_1)^{1/2}} (1-r_2)^{-3/2}\frac{1}{\sqrt{q}}\exp\bigg\{-\frac{q}{1-r_2^2}\bigg\}\ .
\ee
The average comoving momentum is $\mean{q}=5(1-r_2^2)/2$ and the comoving momentum dispersion is $\sigma_q=\sqrt{35}(1-r_2^2)/2$.
The comoving FIMP energy density is hence given by
\be
m_\chi Y_\chi ^\infty \simeq  m_\chi \times \frac{9}{\pi^2}\sqrt{\frac{5}{8}}n\frac{g_1\Gamma_1 \Mpl}{y_{\mathcal{Q}_2\chi}m_1^2\gst(m_1)^{1/2}}\frac{T_0^3}{s_0}(1-r_2^2) \ .
\ee
We can reproduce the DM relic density with the scalings
\be
m_\chi Y_\chi ^\infty \simeq 0.44 F \bigg(\frac{m_\chi}{100\text{ keV}}\bigg) \bigg(\frac{ng_1 \Gamma_1/y_{\mathcal{Q}_2\chi}}{10^{7.85}\text{ s}^{-1}}\bigg)\bigg(\fracinv{m_1}{1\text{ TeV}}\bigg)^2 \bigg(\frac{106.75}{\gst(m_1)}\bigg)^{1/2} \ .
\ee

\subsection*{Three-body decays}
For three-body decays we further assume
\begin{enumerate}
\item[\textbf{7.a)}] that all particles but $B_1$ and $\chi$ are massless in the cases of three-body decays.
\end{enumerate}
If we expand Eqs.~\eqref{eq:q3sca+} and~\eqref{eq:q3sca-} for $m_\chi \ll m_1,\sqrt{s}$, we obtain
\be 
k_1^+(s)=   p\bigg(\frac{ m_1^2-s}{m_\chi^2}-\frac{s}{m_1^2-s}\bigg)+\frac{m_1^2-s}{4p} +\mathcal{O}(m_\chi^2) \ ,
\ee
\be 
 k_1^-(s) =   \bigg| p\bigg(\frac{m_1^2}{m_1^2-s}\bigg)-\frac{m_1^2-s}{4p}\bigg|+\mathcal{O}(m_\chi^2) \ .
\ee
Defining $\varsigma\equiv s/m_1^2$, from Eq.~\ref{eq:E1dec3} in our approximations,
\be
\algn{E_1^+=&{}\:\infty \ , \\
E_1^-=&\;\frac{m_1}{4qx}\bigg(\frac{4q^2+x^2(1-\varsigma)^2}{1-\varsigma} \bigg)\equiv \frac{m_1}{x}\mathcal{F}(\varsigma,x,q) \ ,  \\
\lambda^{1/2}(\sqrt{s},0,0)=&\:s= \,m_1^2 \varsigma \ ,  \\
s\unt{min}=&\:0 \ ,  \\		
s\unt{max}=&\:m_1^2\ .}
\ee
Then Eq. (\ref{eq:C3decayseqMB}) becomes
\be
\frac{g_\chi \mathcal{C}(x,q)}{E}\simeq n  \frac{g_1{\Gamma_1}}{y_{\mathcal{Q}_2\mathcal{Q}_3\chi}}\frac{x}{q^2}\int_0^1 d\varsigma\exp\bigg\{-\mathcal{F}(\varsigma,x,q)\bigg\} \ .
\ee
We approximate the integral with the saddle point method, expanding the argument of the exponential at second order about the minimum value $\varsigma=1-2q/x$ of $\mathcal{F}(\varsigma,x,q)$ and extending the domain of integration from $[0,1]$ to $]-\infty,\infty[$, to obtain a gaussian integral. We obtain
\be\label{C3dec_an}
\frac{g_\chi \mathcal{C}(x,q)}{E} \simeq  2n \frac{ g_1\Gamma_1}{y_{\mathcal{Q}_2\mathcal{Q}_3\chi}}  \frac{1}{q}\sqrt{\frac{2\pi}{x}} e^{-q-x} \ .
\ee
Performing the integration over $dx$ in Eq.~\eqref{eq:PSD} we get the PSD 
\be 
g_\chi f_\chi(q)\simeq 6\sqrt{5}n   \frac{g_1\Gamma_1 \Mpl}{y_{\mathcal{Q}_2\mathcal{Q}_3\chi}m_1^2\gst(m_1)^{1/2}} \frac{1}{q}e^{-q}  \ .
\ee
The average comoving momentum is $\mean{q}=2$ and the comoving momentum dispersion $\sigma_q=\sqrt{6}$. 

The comoving FIMP energy density is hence given by
\be
m_\chi Y_\chi ^\infty \simeq  m_\chi \times\frac{3\sqrt{5}}{\pi^2}n   \frac{g_1\Gamma_1 \Mpl}{y_{\mathcal{Q}_2\mathcal{Q}_3\chi}m_1^2\gst(m_1)^{1/2}} \frac{T_0^3}{s_0} \ . 
\ee
We can reproduce the DM relic density with the scalings
\be
m_\chi Y_\chi ^\infty \simeq 0.44 F \bigg(\frac{m_\chi}{100\text{ keV}}\bigg) \bigg(\frac{n   g_1\Gamma_1/y_{\mathcal{Q}_2\mathcal{Q}_3\chi}}{10^{7.88}\text{ s}^{-1}}\bigg)\bigg(\fracinv{m_1}{1\text{ TeV}}\bigg)^2 \bigg(\frac{106.75}{\gst(m_1)}\bigg)^{1/2} \ .
\ee

\subsection*{Scatterings}
We further assume
\begin{enumerate}
\item[\textbf{7.b)}] that if present, $B_3$ is massless in the case of scatterings.
\end{enumerate}
We expand Eqs.~\eqref{eq:q3sca+} and \eqref{eq:q3sca-} for light DM. For single production $q_3=k_3$
\be\label{eq:q3+_appr} 
k_3^+(s)=   p\bigg(\frac{s-m_3^2}{m_\chi^2}-\frac{s}{s-m_3^2}\bigg)+\frac{s-m_3^2}{4p} +\mathcal{O}(m_\chi^2),\,
\ee
\be\label{eq:q3-_appr} 
 k_3^-(s) = \bigg|p\bigg(\frac{m_3^2}{s-m_3^2}\bigg)-\frac{s-m_3^2}{4p}\bigg|+\mathcal{O}(m_\chi^2)  \ .
\ee
Instead for double production, $q_3=p_2$ we have the following expressions
\be\label{eq:p3+_appr} 
p_2^+(s)= p\bigg(\frac{s}{m_\chi^2}-2\bigg)+\frac{s}{4p} +\mathcal{O}(m_\chi^2)\ ,
\ee
\be\label{eq:p3-_appr} 
 p_2^-(s) =\frac{s}{4p}+\mathcal{O}(m_\chi^2) \ . 
\ee
We choose the mass of $B_2$ to be either equal to the one of $B_1$ ($r_2=1$) or 0 ($r_2=0$). Defining $\varsigma=s/m_1^2$ and $r_2=m_2/m_1$, we get
\be
\algn{\mathcal{E}_2^+=&{}\:\infty \ , \\
\mathcal{E}_2^-=&\;\frac{{m_1} \varsigma x}{4q}\ ,	  \\
\lambda^{1/2}(\sqrt{s},m_1,m_2)=&\:\sqrt{(s-(1-r_2)^2)(s-(1+r_2)^2)}= \begin{cases}\,m_1^2 (\varsigma-1) & \text{ if } r_2=0 \\ m_1^2 \sqrt{\varsigma(\varsigma-4)} & \text{ if } r_2=1 \end{cases} \ ,\\
s\unt{min}=&\:m_1^2(1+r_2)^2 \ .}
\ee
Then  Eq. \eqref{eq:Cscaeq} becomes
\be
\frac{g_\chi \mathcal{C}(x,q)}{E} \simeq\frac{n }{16 \pi^2 }   \frac{g_1 g_2 m_1^3\sigma\upt{FI}_{\mathcal{Q}_3\chi}}{y_{\mathcal{Q}_3\chi}}  x\frac{e^{-q}}{q^2}\int_{(1+r_2)^2}^{\infty} \frac{d\varsigma}{\varsigma} \exp\bigg[-\frac{\varsigma x^2}{4q}\bigg] \times \begin{cases}\, \varsigma-1 & \text{ if } r_2=0 \\  \sqrt{\varsigma(\varsigma-4)} & \text{ if } r_2=1 \end{cases}  \ ,
\ee
and, finally,
\be \label{eq:Csca_an}
\frac{g_\chi \mathcal{C}(x,q)}{E} \simeq \frac{n }{16\pi^2 }  \frac{g_1 g_2 m_1^3\sigma\upt{FI}_{\mathcal{Q}_3\chi}}{y_{\mathcal{Q}_3\chi}^s}  x \frac{e^{-q}}{q^2}\times \begin{dcases}\, \frac{4q}{x^2}\exp\left[-\frac{ x^2}{4q}\right] +\text{Ei}\left[\frac{-x^2}{4q}\right]& \text{ if } r_2=0 \\  2\sqrt{\pi}G^{2,0}_{1,2}\left(\frac{x^2}{q}\bigg|\{\{\},\{\frac{1}{2}\}\},\{\{-1,0\},\{\}\}\right) & \text{ if } r_2=1 \end{dcases}   \ ,
\ee
where Ei$(x)$ is the exponential integral function and 
\[G_{p,q}^{m,n}\left(z|\{\{a_1,\dots ,a_n\},\{a_{n+1},\dots a_p\}\},\{\{b_1,\dots, b_m\},\{b_{m+1}\dots, b_q\}\}\right)\] is the Mejer $G$ function. Inserting this expression in Eq.~\eqref{eq:PSD} we obtain
\be 
g_\chi f_\chi(q)\simeq  \frac{n }{16\pi^2 }    \frac{g_1g_2m_1\Mpl\sigma\upt{FI}_{\mathcal{Q}_3\chi} }{y_{\mathcal{Q}_3\chi}^s\gst(m_1)^{1/2}} \frac{1}{\sqrt{q}}e^{-q}\times   \begin{dcases}\, 8\sqrt{\dfrac{10}{\pi}}& \text{ if } r_2=0 \\   3\sqrt{\dfrac{5\pi}{2}}& \text{ if } r_2=1 \end{dcases}   \ .
\ee
The average comoving momentum is $\mean{q}=5/2$ and the comoving momentum dispersion $\sigma_q= {\sqrt{35}}/{2}$. The comoving FIMP energy density is hence given by
\be
m_\chi Y_\chi ^\infty \simeq m_\chi \times  \frac{n }{16\pi^2 }   \frac{g_1g_2m_1\Mpl\sigma\upt{FI}_{\mathcal{Q}_3\chi} }{y_{\mathcal{Q}_3\chi}^s\gst(m_1)^{1/2}} \frac{T_0^3}{s_0} \times   \begin{dcases}\,  \dfrac{3\sqrt{10}}{\pi^2}& \text{ if } r_2=0 \\   \dfrac{9}{8\pi}\sqrt{\dfrac{5\pi}{2}}& \text{ if } r_2=1 \end{dcases}  \ .
\ee
We can reproduce the DM relic density with the scalings
\be
m_\chi Y_\chi ^\infty \simeq 0.44 F \bigg(\frac{m_\chi}{100\text{ keV}}\bigg)\bigg(\fracinv{m_1}{1\text{ TeV}}\bigg)^2 \bigg(\frac{106.75}{\gst(m_1)}\bigg)^{1/2} \times\begin{dcases} \bigg(\dfrac{n  g_1 g_2 \sigma\upt{FI}_{\mathcal{Q}_3\chi}/y_{\mathcal{Q}_3} }{10^{-50.7}\text{cm}^{2}}\bigg)&\text{ if } r_2=0 \\  \bigg(\dfrac{n  g_1 g_2 \sigma\upt{FI}_{\mathcal{Q}_3\chi}/y_{\mathcal{Q}_3} }{10^{-50.4}\text{cm}^{2}}\bigg)& \text{ if } r_2=1 \end{dcases}  \ .
\ee

\section{Matter power spectrum computation}
\label{app:deltaA}

In this Appendix, we provide technical details about how we compute the linear matter power spectrum with \ttt{CLASS}~\cite{Blas_2011,Lesgourgues_2011} for CDM, WDM and FIMP DM models. We also give details about the implementation of the $\delta A$ criterion and we comment on its robustness. 

\subsection*{Computation of $P(k)$ with \ttt{CLASS}}

We utilize \ttt{CLASSv2.9.3} implemented in a \ttt{python3.7} code. The cosmological parameters for the inizialization of \ttt{CLASS} to compute the linear matter power spectrum of the standard $\Lambda$CDM are the ones set by default in the program, and are taken from the most recent results published by the Planck collaboration \cite{Aghanim:2018eyx}. The inizialization of \ttt{CLASS} from our code is done with the command
\begin{mylisting}
LambdaCDM = Class()
LambdaCDM.set({'omega_cdm':0.12038,
	           'output':'mPk',          
	           'P_k_max_1/Mpc':klim,
	            })
\end{mylisting}
The only parameter we tune is \ttt{klim} which is the highest mode at which our power spectrum is computed. We choose \ttt{klim=1.2e3} to capture effects on fair small scales.
 
In order to compute the linear matter power spectrum in a model involving non-CDM candidate such as WDM or FIMP DM, further input parameter are needed. First we modify the \ttt{background\_ncdm\_distribution} function inside the \ttt{background.c} module of \ttt{CLASS} to include a very general expression for the PSD
\begin{mylisting}
 double A0=param[0];		
 double A1=param[1];		
 double A2=param[2];		
 double A3=param[3];		
 double A4=param[4];		
 double A5=param[5];		
	*f0 = A0*(pow(q,2+A1)+A3)*pow(exp(A2*q+A4*q*q)+A5,-1)/q/q;	#gXfX(q)
\end{mylisting}
This corresponds to the very general formula
\begin{equation}
f(q)= 	A_0\frac{q^{2+A_1}+A_3}{q^2}\frac{1}{\exp(A_2q+A_4q^2)+A_5}
\end{equation}
that includes the cases of WDM PSD $f_{\mathrm{WDM}}(q)=2/(e^q+1)$ for $A_0=2$, $A_1=A_3=0$, $A_2=A_5=1$ and our fit function for the FIMP PSD $f_\chi(q)=\mathcal{N}_Fq^b e^{-cb}$ for $A_0=\mathcal{N}_F$, $A_1=b$, $A_2=c$, $A_3=A_4=A_5=0$. 

The initialization for the computation of a WDM power spectrum for a mass \ttt{mWDM} is done with the following command
\begin{mylisting}
LambdaWDM = Class()
LambdaWDM.set({	'omega_cdm':0.0,
	           'output':'mPk',          
	           'P_k_max_1/Mpc':klim,	
	           'N_ncdm' : 1,    #number of ncdm species
	           'ncdm_psd_parameters': "2, 0.0, 1.0, 0.0, 0.0, 1.0",				
	           'm_ncdm' : mWDM , #eV
	           'T_ncdm' : ( 93.14/mWDM*0.12038)**(1./3.)*0.71611, #units of T0
	           'omega_ncdm' : 0.12038,				
	           }) 
\end{mylisting}
If one specifies the WDM relic density the PSD is automatically normalized to reproduce this result. The important parameter here is \ttt{T\_ncdm} which is chosen to be the WDM temperature defined in $T_\mathrm{WDM}/T_0$
Eq.~\eqref{eq:TWDM}. 

For a FIMP model we have an analogous initialization command
\begin{mylisting}
LambdaFIMP = Class()
LambdaFIMP.set({	'omega_cdm':(1-F)*0.12038,
	           	'output':'mPk',          
	           	'P_k_max_1/Mpc':klim,	
	           	'N_ncdm' : 1,    #number of ncdm species
	           	'ncdm_psd_parameters': str(N)+", "+str(b)+", "\
	      	                 		     +str(c)+", "+"0.0"+", "\
	           		  	          		 +"0.0"+", "+"0.0",
	            'm_ncdm' : mX , #eV
	           	'T_ncdm' : TX(T0,M)/T0, #units of T0
	           	'omega_ncdm' : F*0.12038,				
	           	}) 
\end{mylisting}
where we left dependence on the FIMP DM fraction \ttt{F}, the FIMP mass \ttt{mX}. Now \ttt{T\_ncdm} is obtained from Eq. \eqref{eq:Tchi} as $T_\chi/T_0$.

\subsection*{Implementation of the $\delta A$ criterion}

As long as \ttt{klim} is sufficiently larger than the value of $k_{1/2}$ for our FIMP and WDM models we do not have problems in computing power spectra and transfer functions. However, if one wants to impose the $\delta A$ criterion then the three-dimensional power spectra need to be projected into one-dimensional ones. Numerically, one has to put a UV cutoff into the definition of the one-dimensional power spectrum
\begin{equation}\label{eq:P1d}
P\upt{1D}(k) =\frac{1}{2\pi}\int_k^\infty dk'\,k'P(k')  \simeq \frac{1}{2\pi}\int_k^{k_{\mathrm{lim}}} dk'\,k'P(k') \ .
\end{equation}

\begin{figure}
\centering
\includegraphics[width=0.6\textwidth]{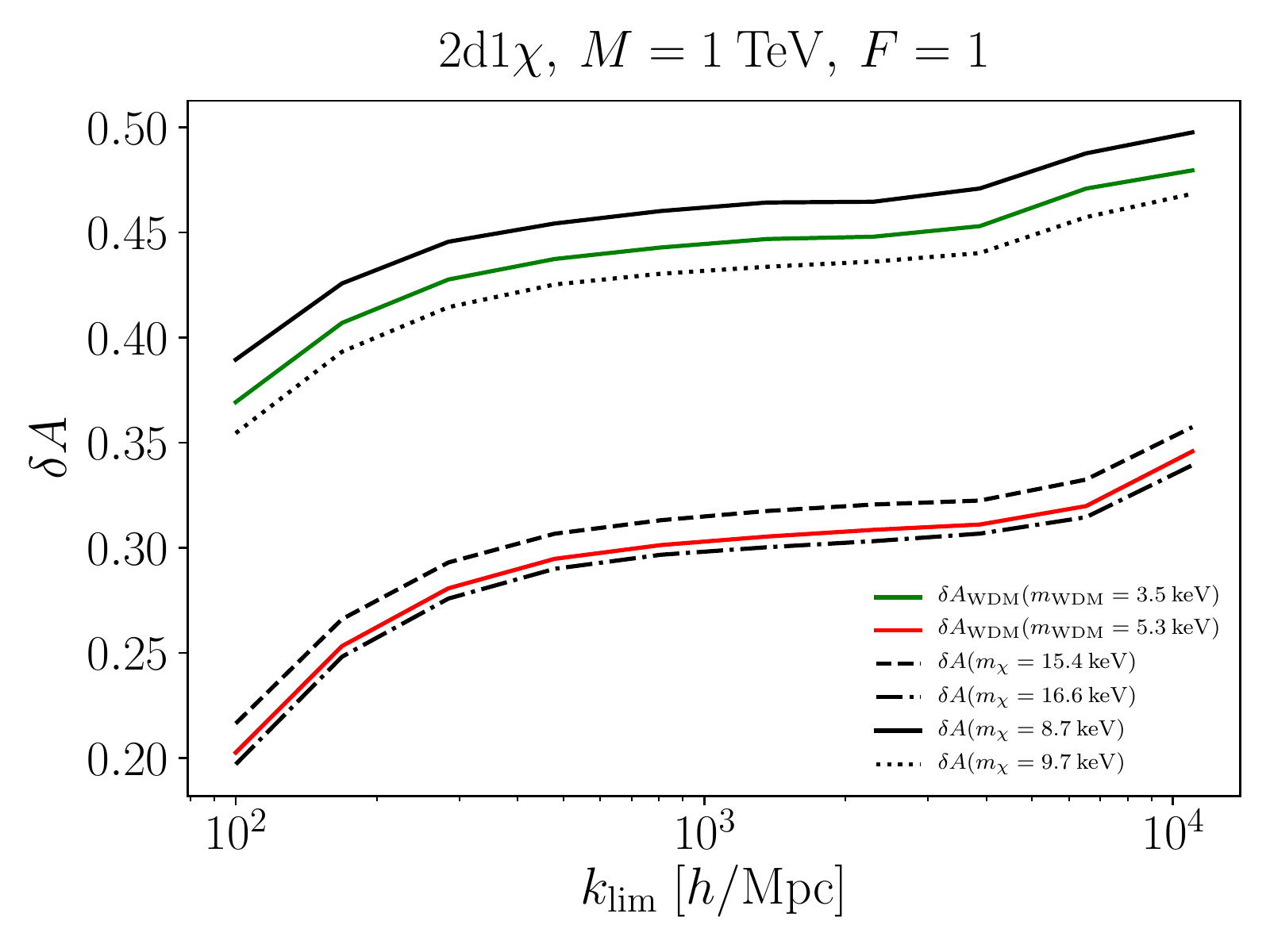}
\includegraphics[width=0.6\textwidth]{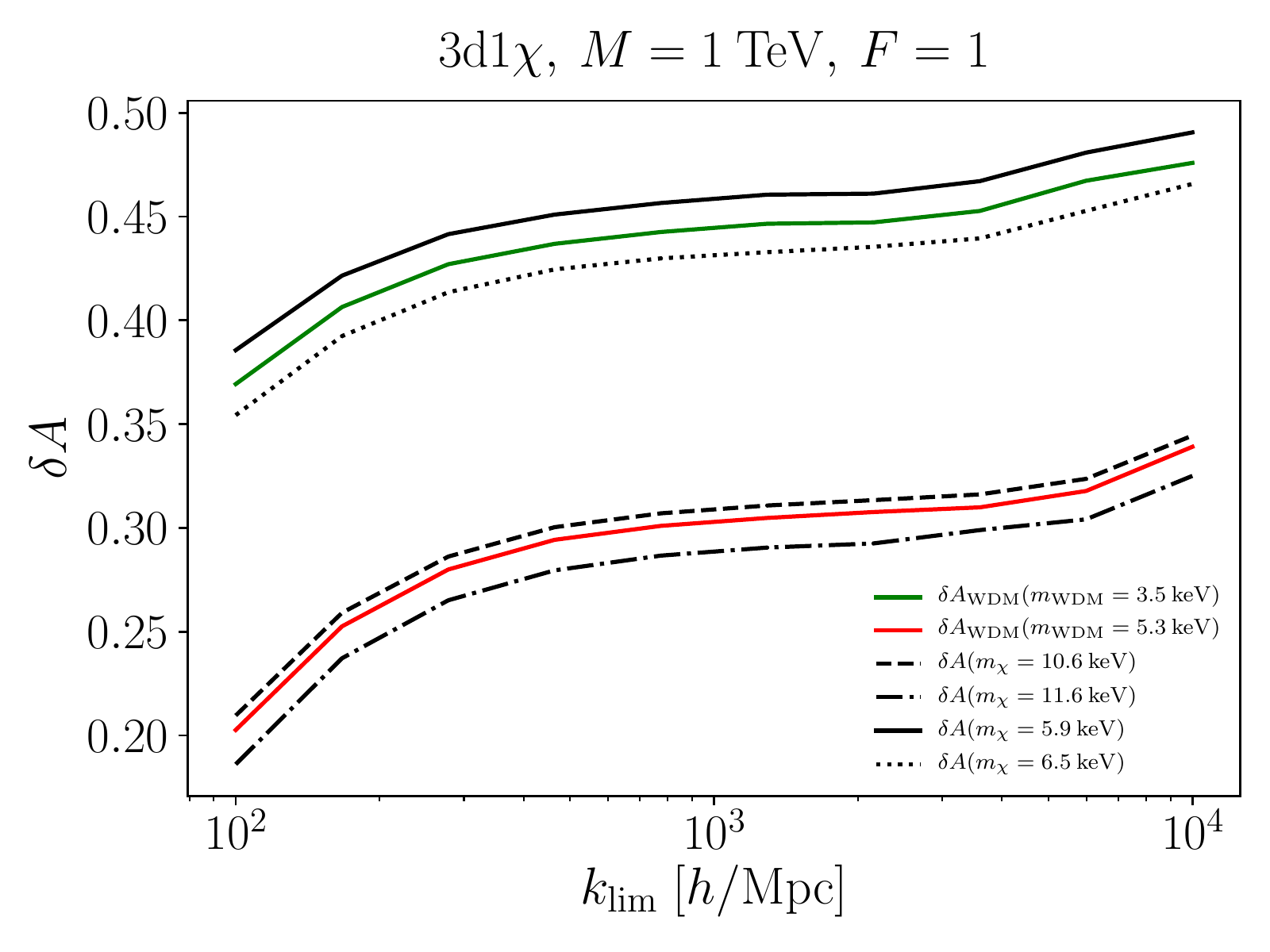}
\includegraphics[width=0.6\textwidth]{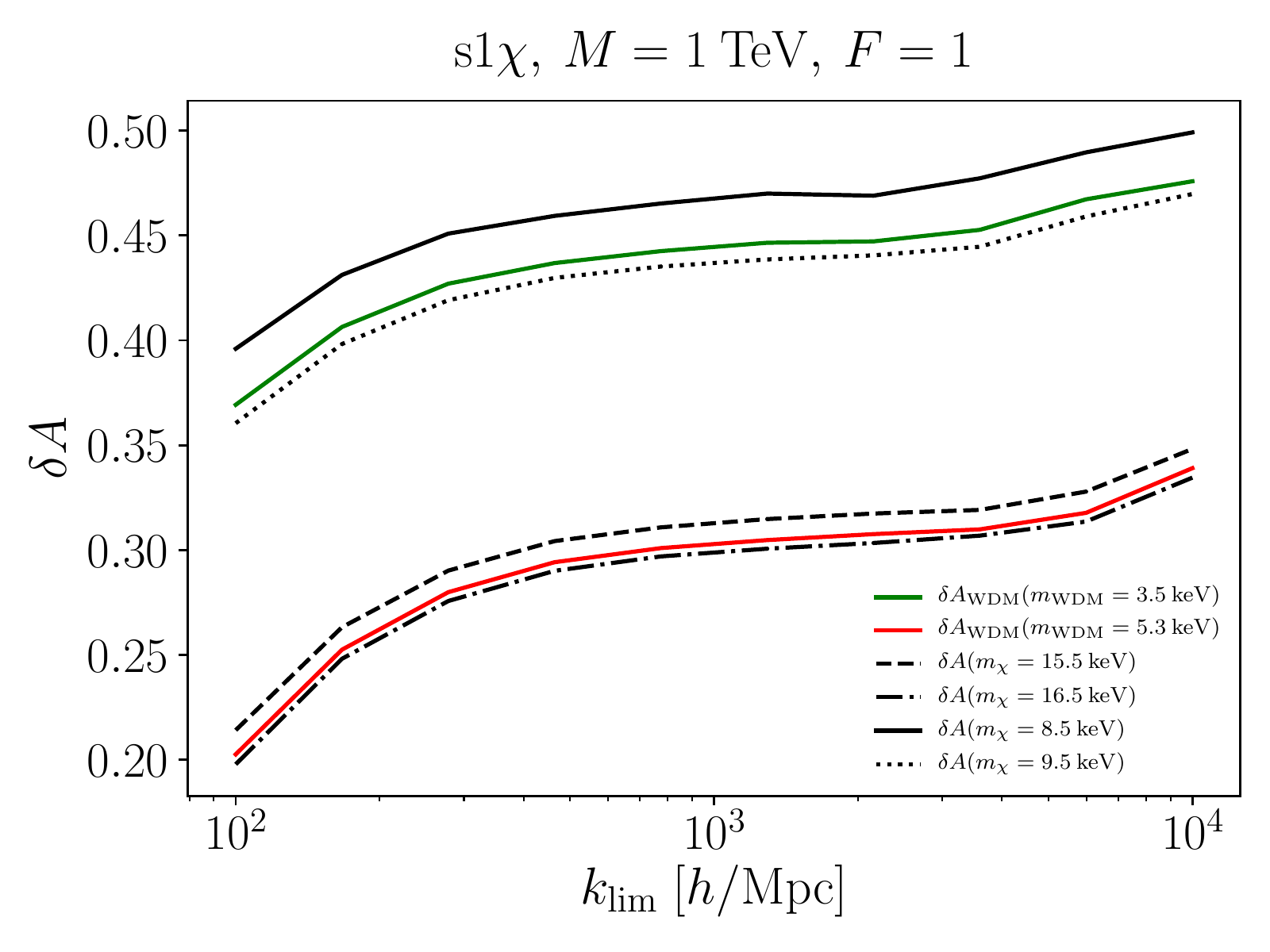}
\caption{Plots of $\delta A$ for WDM (conservative and stringent bounds) and for FIMP benchmarks for different values of the power spectrum cutoff $k_{\mathrm{lim}}$.}
\label{fig:delta_A}
\end{figure}

The choice of the UV cutoff impacts the resulting value of $\delta A$ via Eq. \eqref{eq:deltaA}, and we show this dependence in Fig. \ref{fig:delta_A}. However, the curves correspondent to the conservative and stringent WDM mass bounds as well as the ones for FIMPs, for various values of $m_\chi$ and fixing $M=1$ TeV, have the same shape independently on the choice of the cutoff. This guarantees that the FIMP mass bounds we derive with the $\delta A$ criterion \textit{do not} depend on the cutoff $k_\mathrm{lim}$ we choose.

\bibliographystyle{JHEP}
\bibliography{FIMPmass}

\end{document}